\documentclass[preprint,superscriptaddress,secnumarabic,amssymb, nobibnotes, aps, pre,floatfix]{revtex4-2}

\usepackage{bm}
\usepackage{setspace}
\usepackage{tikz}
\usetikzlibrary{automata,positioning}
\usepackage{graphicx}
\usepackage{epstopdf, epsfig}
\usepackage{makecell}
\usepackage{bm}
\usepackage{mathrsfs}
\usepackage{amsmath}
\usepackage{array}

\newcommand{\bcdot}{\boldsymbol{\cdot}}

\newcommand\NavSto{Nav\-ier--Sto\-kes}
\newcommand\threed{three-di\-men\-sion\-al}

\newcommand\twod{two-di\-men\-sion\-al}

\newcommand\qtwod{quasi-two-di\-men\-sion\-al}

\newcommand\TS{TS}
\newcommand\Rey{\textit{Re}}  
\newcommand\eg{e.g.}  

\newcommand{\vect}[1]{\bm{#1}}

\newcommand{\pde}[2]{\frac{\partial #1}{\partial #2}}

\newcommand\Fig{Figure} 
\newcommand\Figs{Figures} 
\newcommand\fig{Fig.}  
\newcommand\figs{Figs.}  
\newcommand\tbl{Table}  
\newcommand\tbls{Tables}  

\newcommand\alphaCrit{\alpha_{\mathrm{S},\mathrm{crit}}}
\newcommand\alphaOpt{\alpha_\mathrm{opt}}
\newcommand\deltaS{\delta_\mathrm{S}}
\newcommand\tauOpt{\tau_\mathrm{opt}}
\newcommand\dUP{\mathrm{d}}
\newcommand\ELD{E_\mathrm{D}}
\newcommand\EUD{E_{\mathrm{D},2}}
\newcommand\Ezero{E_0}
\newcommand\Euv{E}
\newcommand\Ev{E_\mathrm{v}}
\newcommand\EzeroR{E_{0}}
\newcommand\Gmax{G_\mathrm{max}}
\newcommand\HsubD{L_{y}}
\newcommand\WlenD{L_{x}}
\newcommand\wlenD{l_{x}}
\newcommand\lxopt{l_{x,\mathrm{opt}}}
\newcommand\Np{N_\mathrm{p}}
\newcommand\ReyS{\Rey_\mathrm{S}}

\newcommand\ReyCrit{\Rey_{\mathrm{S},\mathrm{crit}}}
\newcommand\rrc{r_\mathrm{c}}

\newcommand\Nel{N_\mathrm{el}}

\graphicspath{ {Figures/} }

\begin{document}
\title{Subcritical route to turbulence via the Orr mechanism in a quasi-two-dimensional boundary layer}%
\author{Christopher J. Camobreco}%
\email{christopher.camobreco@monash.edu}
\affiliation{Department of Mechanical and Aerospace Engineering, Monash University, VIC 3800, Australia}
\author{Alban Poth{\'e}rat}%
\email{alban.potherat@coventry.ac.uk}
\affiliation{Fluid and Complex Systems Research Centre, Coventry University, Coventry CV15FB, UK}
\author{Gregory J. Sheard}%
\email{greg.sheard@monash.edu}
\affiliation{Department of Mechanical and Aerospace Engineering, Monash University, VIC 3800, Australia}
\date{\today}%
\begin{abstract}
The link to the online abstract of this manuscript, accepted in Phys.~Rev.~Fluids, is {https://journals.aps.org/prfluids/accepted/32074S4aH8b1c608e19768b42571f9001086a3f44}.

A subcritical route to turbulence via purely \qtwod\ mechanisms, for a \qtwod\ system composed of an isolated exponential boundary layer, is numerically investigated. Exponential boundary layers are highly stable, and are expected to form on the walls of liquid metal coolant ducts within magnetic confinement fusion reactors. Subcritical transitions were detected only at weakly subcritical Reynolds numbers (at most $\approx 70$\% below critical). Furthermore, the likelihood of transition was very sensitive to both the perturbation structure and initial energy. Only the \qtwod\ Tollmien--Schlichting wave disturbance, attained by either linear or nonlinear optimisation, was able to initiate the transition process, by means of the Orr mechanism. The lower initial energy bound sufficient to trigger transition was found to be independent of the domain length. However, longer domains were able to increase the upper energy bound, via the merging of repetitions of the Tollmien--Schlichting wave. This broadens the range of initial energies able to exhibit transitional behaviour. Although the eventual relaminarization of all turbulent states was observed, this was also greatly delayed in longer domains. The maximum nonlinear gains achieved were orders of magnitude larger than the maximum linear gains (with the same initial perturbations), regardless if the initial energy was above or below the lower energy bound. Nonlinearity provided a second stage of energy growth by an arching of the conventional Tollmien--Schlichting wave structure. A streamwise independent structure, able to efficiently store perturbation energy, also formed.
\end{abstract}
\maketitle
\section{Introduction}\label{sec:introduction}
There is significant interest in understanding transitions to quasi-two-dimensional (Q2D) turbulence, given the wide range of natural and industrial flows which exhibit quasi-two-dimensionality. These include magnetohydodynamic (MHD), shallow channel and atmospheric flows \citep{Lindborg1999atmospheric, Potherat2011shallow}. The conditions under which 3D MHD turbulence becomes quasi-two dimensional, and the appearance of three-dimensionality in Q2D MHD turbulence have been clarified \citep{Sommeria1982why,thess2007_jfm,klein2010_prl,Potherat2014why}. However, a clear subcritical path to Q2D turbulence from a Q2D laminar state has not been identified. The aim of the present work is thus to establish a purely Q2D subcritical route to turbulence. This is motivated by the design of coolant ducts in magnetic confinement fusion reactors, where pervading field strengths range between $4$--$10$ T \citep{Smolentsev2008characterization, Kluber2019numerical}.  Understanding transition in coolant ducts is important for ensuring sufficient heat transfer at the plasma-facing (Shercliff) wall \citep{Barleon2000heat, Burr2000turbulent, Cassels2016heat, Mistrangelo2009influence, Mistrangelo2014buoyant} and to establish the feasibility of self-cooled reactor designs \citep{Smolentsev2008characterization}. Limits on maximum pressure gradient \citep{Barleon2000heat, Smolentsev2010considerations, Mistrangelo2011magnetohydrodynamic} and pumping efficiency \citep{Hussam2012enhancing, Cassels2016heat, Hamid2016combining, Hamid2016heat} motivate seeking the most efficient route to turbulence. However, quasi-two-dimensional turbulence is unlikely to arise in blankets via strongly three-dimensional turbulence \citep{Smolentsev2008characterization}. Thus, this work limits itself only to the use of an initial two-dimensional perturbation; secondary excitations with \threed\ random noise are not applied. 

Transitions in MHD flows have previously been initiated by a perturbation comprising either two \threed\ oblique-waves or a \twod\ initial field with \threed\ random noise \citep{Krasnov2004numerical, Krasnov2008optimal}, which are routes prohibited in Q2D systems. Using these techniques, for Hartmann channel flow, \cite{Krasnov2004numerical} found excellent agreement with the critical Reynolds numbers at which transition was observed experimentally \citep{Moresco2004experimental}, observing a strongly \threed\ subcritical transition. Although less energetic perturbations generated more growth, they did not sufficiently modulate the base flow. The perturbations which attained the highest maximum energy, regardless of initial energy, were most likely to incite transition. Complicating matters at high field strengths, \threed\ noise relaminarized the flow, instead of triggering transition. 

To assess subcritical transitions in Q2D MHD flows, the SM82 model \citep{Sommeria1982why} is applied, as realistic magnetic confinement field strengths ($4$--$10$ T) are currently beyond the capability of three-dimensional numerics. The SM82 model governs the evolution of a velocity field averaged along uniform magnetic field lines. In the limit of quasi-static Q2D MHD, the magnetic field is imposed and the Lorentz force dominates all other forces. The bulk flow is two-dimensional, with thin Hartmann layers formed along walls perpendicular to field lines. In the SM82 model, the presence of Hartmann layers is modelled with linear friction on the average flow. The validity of the SM82 approximation is well supported in the \qtwod\ limit \citep{Muck2000flows, Potherat2000effective, Dousset2008numerical, Kanaris2013numerical}. Departure from the \twod\ average has been observed in regions of strong viscosity or inertia. \cite{Potherat2000effective} demonstrates errors less than $10\%$ between \qtwod\ and laminar \threed\ Shercliff layers, which do not vanish, even in the asymptotic limit when the Lorentz force dominates. There is also excellent agreement at high magnetic field strengths \citep{Cassels2019from3D} between the linear transient growth of full \threed\ simulations, and Q2D simulations based on the SM82 model. 

The linear stability and linear transient growth of duct flows under strong magnetic fields are determined solely by boundary layer dynamics \citep{Potherat2007quasi,Vo2017linear}. Direct numerical simulations depict instabilities isolated to the Shercliff layers, on walls parallel to the magnetic field \citep{Krasnov2010optimal, Cassels2019from3D}. As such, an exponential boundary layer in isolation is considered. The isolated \qtwod\ boundary layer profile is identical to an asymptotic suction boundary layer \citep{Roberts1967introduction}, where friction replaces wall suction. The analogy has been highlighted in \cite{Levin2005transition}, by performing a change of variables, such that the wall suction boundary condition becomes impermeable. This introduces an additional term in the governing equations for the transformed velocity, of the form $-(\partial \vect{u}/\partial y)/\Rey$. Comparatively, the friction term in the SM82 model is $-\vect{u}/\Rey$. However, as the underlying exponential boundary layer remains the same, both flows are very stable \citep{Roberts1967introduction, Albrecht2006stability}.

\begin{figure}
\begin{center}
\addtolength{\extrarowheight}{-10pt}
\addtolength{\tabcolsep}{-2pt}
\begin{tabular}{ ll }
\makecell{ \\  \vspace{9mm} \rotatebox{90}{\footnotesize{$\int \hat{u}^2 \mathrm{d}\Omega$}}} & \makecell{\includegraphics[width=0.458\textwidth]{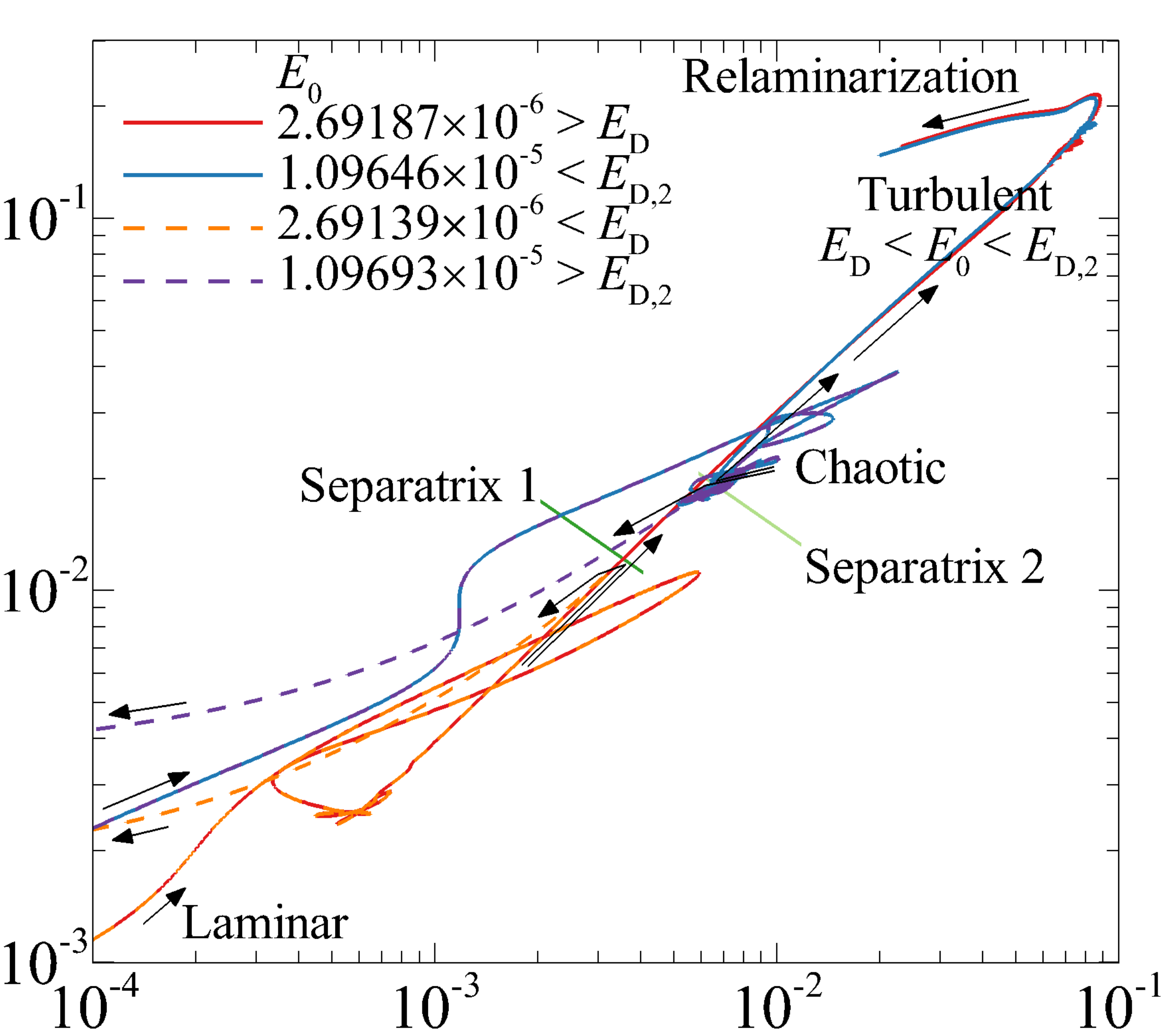}}  \\
 & \hspace{32mm} \footnotesize{$\int \hat{v}^2 \mathrm{d}\Omega$}  \\
\end{tabular}
\addtolength{\tabcolsep}{+2pt}
\addtolength{\extrarowheight}{+10pt}
\end{center}
    \caption{
    A state space representation of the problem. Four cases are considered, two with initial energies $E_0$ just below and above the minimum initial energy sufficient to cross separatrix 1 ($\ELD$) and two with $E_0$ just above and below the maximum initial energy sufficient to cross separatrix 2 ($\EUD$). An initial energy $\ELD<E_0<\EUD$ either crosses separatrix 1 (red curve crosses solid dark green line) or avoids crossing separatrix 2 (blue curve eventually avoids solid  light green line) to transition to turbulence. Eventually the turbulent state relaminarizes.   
    }
    \label{fig:state_space}
\end{figure}

Nonlinear optimisation and edge tracking algorithms have been widely used to assess subcritical turbulent transitions in hydrodynamic pipe \citep{Pringle2012minimal, Kerswell2014optimization}, plane Couette \cite{Duguet2009localized, Duguet2013minimal} and plane Poiseuille flows \cite{Farano2016subcritical, Zammert2019transition}, as well as in Blasius \cite{Duguet2012self, Cherubini2011minimal, Beneitez2019edge, Vavaliaris2020optimal} and asymptotic suction \cite{Khapko2014complexity,Cherubini2015nonlinear} boundary layers. A fundamental part of this process involves searching the state space for seperatrices, which divide the basins of attraction of the laminar fixed point and turbulent state \cite{Khapko2014complexity}.  The minimal seed is then the nonlinearly optimised perturbation with the smallest initial energy that is able to cross the separatrix \cite{Pringle2012minimal}. Separatrix 1 is henceforth defined as a segment of the laminar-turbulent basin boundary where the minimal seed crosses. Hydrodynamic studies of three-dimensional turbulent transitions have determined that the laminar-turbulent basin boundary is the `edge' of a stable manifold. At a saddle node (the edge state) an unstable solution crosses \cite{Budanur2020upper,Khapko2014complexity}. However, such an unstable solution is not necessarily the minimal seed \cite{Duguet2013minimal} as the seperatrix can be closer to the fixed laminar point elsewhere in the state space. This discussion  is aided by \fig\ \ref{fig:state_space}, which depicts two initial conditions with slightly different initial energies. One perturbation has an initial energy $E_0<\ELD$ and returns back to the laminar state without crossing separatrix 1, such that $\ELD$ is the minimum initial energy sufficient to cross separatrix 1. The case with $E_0>\ELD$ continues on to the turbulent attractor. An upper bound on the edge state was also identified by \cite{Budanur2020upper}. It stemmed from additional dissipation generated by distortion of overly energised initial seeds. This segment of the laminar-turbulent boundary is henceforth defined as separatrix 2. The perturbation with initial energy $E_0>\EUD$ crosses seperatrix 2, missing the trajectory toward the turbulent attractor, such that $\EUD$ is the maximum initial energy sufficient to avoid separatrix 2. The perturbation with $E_0<\EUD$ reaches the turbulent attractor, following an almost identical trajectory to the turbulent state as the perturbation with $E_0>E_D$. After remaining in the basin of the turbulent attractor for some time, relaminarization occurs.


 
 

Nonlinear optimisation has also been used to demonstrate that nonlinear transient growth occurs solely via the collaboration of multiple linear transient growth mechanisms \citep{Kerswell2014optimization}. This cannot occur in \twod\ systems, as only the Orr mechanism is present. Thus, nonlinear optimisation effectively degenerates to linear optimisation. The \twod\ inviscid Orr mechanism is characterized by an initial perturbation that is tilted opposite to the mean shear \cite{Schmid2001stability}. Energy from the mean shear transiently grows the perturbation energy, as the base flow advects the structure into an upright position. Perturbation energy decays as the structure is further tilted into the mean shear, returning energy to the base flow \cite{Butler1992optimal}. Initially, this work compares linearly and nonlinearly optimised perturbations, which may form the minimal seeds for inciting subcritical turbulent transitions. 

Therefore, this paper considers:
\begin{itemize}
\item What roles linear transient growth (in particular, the Orr mechanism) and nonlinearity play in Q2D transition scenarios.
\item Whether distinct initial energies representing separatrix 1 and 2 on the laminar-turbulent boundary can be defined, as for 3D systems.
\item How sensitive transition is to the structure and wavelength of the initial field.
\end{itemize}

This paper proceeds as follows: the problem setup, \S~\ref{sec:prob_set}, establishes the Shercliff boundary layer domain and base flow. \S~\ref{sec:ltg} details the determination, validation and results of the linear transient growth analysis, as linear optimals form the initial seeds for nonlinear simulations. \S~\ref{sec:nltg} discusses and validates the approach for determining nonlinear optimals and compares the linear optimals to their nonlinear counterparts for small target times. \S~\ref{sec:nonlin} validates the nonlinear evolutions of linear optimals, for prescribed initial energies, and then considers the energies delineating transitional states, perturbation structures through growth and decay stages, and the effect of domain length. Conclusions are drawn in \S~\ref{sec:conc}.

\section{Problem setup and solution process}\label{sec:prob_set}
\subsection{Problem setup}\label{sec:pro_set}
An incompressible Newtonian fluid with density $\rho$, kinematic viscosity $\nu$ and electric conductivity $\sigma$ flows through a duct with rectangular cross-section of width $a$ ($z-$direction) and height $2L$ ($y-$direction). A uniform magnetic field $B\vect{e_z}$ is imposed. Quasi-two-dimensionality, based on the SM82 model \citep{Sommeria1982why,Potherat2000effective} is assumed. The revelant length scale is the Q2D Shercliff boundary layer thickness $\deltaS=L/H^{1/2}$, where the Hartmann friction parameter $H=L^2(2B/a)(\sigma/\rho\nu)^{1/2}$ \cite{Potherat2007quasi}. Normalizing lengths by $\deltaS$, velocities by maximum undisturbed duct velocity $U_0$, time $t$ by $\deltaS/U_0$ and pressure $p$ by $\rho U_0^2$, the governing momentum and mass conservation equations become
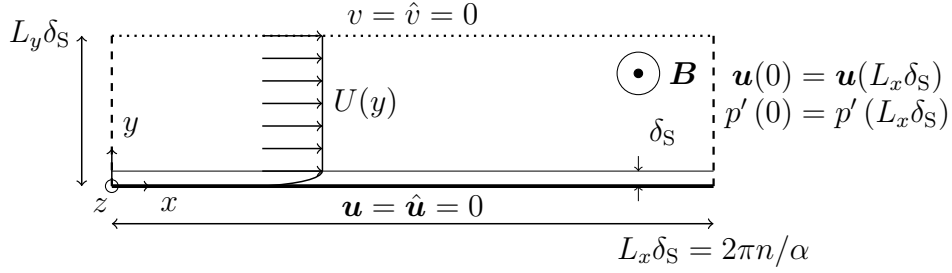
\begin{figure}
    \centering
\begin{tikzpicture}
\draw[dashed,line width = 0.3mm] (-4,-2) -- (-4,0);  
\draw[dotted,line width = 0.3mm] (-4,0)  -- (4,0); 
\draw[dashed,line width = 0.3mm] (4,0)   -- (4,-2);  
\draw[line width = 0.5mm] (-4,-2)  -- (4,-2);     
\draw[line width = 0.2mm,<->] (-4,-2.5) -- (4,-2.5) node[anchor=north] {$\WlenD\deltaS=2\pi n/\alpha$};
\draw[line width = 0.2mm,<->] (-4.4,-2) -- (-4.4,0) node[anchor=east] {$\HsubD\deltaS$};
\draw[line width = 0.2mm,->] (-4,-2) -- (-3.5,-2) node[anchor=north west] {$x$};
\draw[line width = 0.2mm,->] (-4,-2) -- (-4,-1.5) node[anchor=south west] {$y$};
\node at (-4,-2) [circle,draw=black,inner sep=0.6mm] {};
\node at (-4.15,-2.25) {$z$};
\draw[line width = 0.1mm] (-4,-1.8) -- (4,-1.8);
\draw[line width = 0.1mm,->] (3,-2.2) -- (3,-2);
\draw[line width = 0.1mm,<-] (3,-1.8) -- (3,-1.6) node[anchor=south west] {$\deltaS$};
\draw[line width = 0.2mm] (-1.2,-1.8) -- (-1.2,0);
\draw[line width = 0.2mm,->] (-2,0) -- (-1.2,0);
\draw[line width = 0.2mm,->] (-2,-0.3) -- (-1.2,-0.3);
\draw[line width = 0.2mm,->] (-2,-0.6) -- (-1.2,-0.6);
\draw[line width = 0.2mm,->] (-2,-0.9) -- (-1.2,-0.9)node[anchor=west] {$U(y)$};
\draw[line width = 0.2mm,->] (-2,-1.2) -- (-1.2,-1.2);
\draw[line width = 0.2mm,->] (-2,-1.5) -- (-1.2,-1.5);
\draw[line width = 0.2mm,->] (-2,-1.8) -- (-1.2,-1.8);
\draw[line width = 0.2mm] (-2,-2) .. controls (-1.5,-1.97) and (-1.22,-1.9) .. (-1.2,-1.8);
\node at (0,0.3) {$v=\hat{v}=0$};
\node at (0,-2.25) {$\vect{u}=\hat{\vect{u}}=0$};
\node at (5.65,-0.8) {\makecell{$\vect{u}(0)=\vect{u}(\WlenD\deltaS)$ \\ $p'\,(0)=p'\,(\WlenD\deltaS)$}};
\node at (-5.65,-0.8) {\hspace{23mm} };
\node at (3.0,-0.5) [circle,draw=black,fill=black,inner sep=0.4mm] {};
\node at (3.0,-0.5) [circle,draw=black,inner sep=2mm] {};
\node at (3.6,-0.5) {$\vect{B}$};
\end{tikzpicture}
    \caption{Schematic diagram of the sidewall domain with a characteristic length of the Shercliff boundary layer height $\deltaS$. The thick horizontal line represents an impermeable no-slip boundary. The dotted line represents a stress-free  parallel flow condition. The vertical dashed lines represent a periodicity constraint on velocity and fluctuating pressure. A uniform magnetic field is directed into the page. The out-of-plane Hartmann walls (the sources of linear friction) are not drawn.}
    \label{fig:prob_setup}
\end{figure}
%
\begin{equation} \label{eq:non_dim_m}
\pde{\vect{u}}{t} = -(\vect{u}\vect{\cdot}\vect{\nabla}_\perp)\vect{u} - \vect{\nabla}_\perp p + \frac{1}{\ReyS}\nabla_\perp^2\vect{u} - \frac{1}{\ReyS}\vect{u},
\end{equation}
\begin{equation} \label{eq:non_dim_c}
\vect{\nabla_\perp} \vect{\cdot} \vect{u} = 0,
\end{equation}
%
where $\vect{u}=(u,v)$ is the \qtwod\ velocity vector, representing the $z-$averaged field, and $\vect{\nabla}_\perp=(\partial_x,\partial_y)$ and $\nabla^2_\perp = \partial^2_x + \partial^2_y$ are the \qtwod\ gradient and vector Laplacian operators, respectively. The flow is governed by one dimensionless parameter, a Reynolds number based on the boundary layer thickness, $\ReyS = U_0\deltaS/\nu$. Hereafter, quantities are expressed in dimensionless form unless specified otherwise. The rightmost term in equation~(\ref{eq:non_dim_m}) is a linear friction term describing Hartmann braking from the two out-of-plane duct walls \citep{Sommeria1982why}. At $H \gg 100$, $\deltaS \ll L$ \citep{Cassels2019from3D, Potherat2007quasi}, such that the sidewall boundary layer that dictates transition behaviour is isolated. A domain extending from the sidewall a distance $\HsubD$  into the flow is considered, with streamwise-periodic length $\WlenD$, as depicted in \fig\ \ref{fig:prob_setup}. The streamwise length $\WlenD= n \wlenD$ spans $n$ integer repetitions of a flow structure having streamwise length $\wlenD = 2\pi/\alpha$ and streamwise wavenumber $\alpha$.

Instantaneous variables $(\vect{u},p)$ are decomposed into base $(\vect{U},P)$ and perturbation $(\hat{\vect{u}},\hat{p})$ components via small parameter $\epsilon$, as $\vect{u} = \vect{U} + \epsilon \hat{\vect{u}}$; $p = P + \epsilon \hat{p}$,
for use in linear transient growth analysis. The fully developed, time steady, parallel flow $\vect{U}=U(y)\vect{e_x}$, with boundary conditions $U(y=0)=0$, $U(y\rightarrow\infty)=1$, and a constant driving pressure gradient scaled to achieve a unit maximum velocity, is $\vect{U} = (1-\exp(-y),0)$.

\subsection{Solver}\label{sec:solver}
An in-house nodal spectral element solver temporally integrates equations (\ref{eq:non_dim_m}) and (\ref{eq:non_dim_c}) using a third order backward differencing scheme with operator splitting. The \twod\ Cartesian domain is discretized with quadrilateral spectral elements over which Gauss--Legendre--Lobatto nodes are placed. The \NavSto\ solver, with the inclusion of the friction term, has been previously introduced and validated \citep{Cassels2016heat, Cassels2019from3D, Hussam2012optimal, Sheard2009cylinders}. No-slip velocity boundary conditions are applied at the impermeable wall, $\vect{u}=\hat{\vect{u}}=0$, supplemented by high-order Neumann pressure boundary conditions \citep{Karniadakis1991high}. Pressure is decomposed into a constant pressure gradient, and a fluctuating component $p'$, and periodicity is imposed between the upstream and downstream boundaries on the velocity and fluctuating pressure. At the stress-free boundary a parallel flow condition $(v=\hat{v}=0)$ is strongly enforced. A constant flow rate condition is also enforced in nonlinear simulations, by appropriate adjustment of the flow rate after each time step.
%
%
%
\section{Linear transient growth}\label{sec:ltg}
\subsection{Formulation and validation}\label{sec:ltg_form}
At subcritical Reynolds numbers, all eigenmodes of the linear evolution operator decay. Thus, to begin establishing a subcrtical route to turbulent transitions, the linear initial value problem is considered. Linear growth is generated by the superposition of decaying non-orthogonal Orr--Sommerfeld modes \citep{Reddy1993pseudospectra, Trefethen1993hydrodynamic}. To interrogate the transient growth of a perturbation, total kinetic energy $E=(1/2)\int \hat{\vect{u}} \cdot \hat{\vect{u}} \, \mathrm{d}\Omega = (1/2)\left\lVert \hat{\vect{u}} \right\rVert$ is chosen to quantify growth, following \cite{Barkley2008direct, Blackburn2008convective}, where $\Omega$ represents the computational domain. The maximum possible linear transient growth is found by determining the initial condition for perturbation $\hat{\bm u}_\tau(t=0)$ maximizing $G=\left\lVert\hat{\bm u}(\tau)\right\rVert/\left\lVert\hat{\bm u}(0)\right\rVert$ via evolution to time $\tau$. For a given $\ReyS$, $\Gmax = \mathrm{max}\left(G(\tau,\alpha)\right)$ is sought, along with the optimal time horizon $\tauOpt$ and streamwise wavenumber $\alphaOpt$. Thereby $\lxopt=2\pi/\alphaOpt$. The analysis proceeds with integration of the linearised forward evolution equations
\begin{equation}\label{eq:lin_for}
\pde{\hat{\vect{u}}}{t} =  -(\hat{\vect{u}} \vect{\cdot} \vect{\nabla}_\perp ) \vect{U} -(\vect{U} \vect{\cdot} \vect{\nabla}_\perp ) \hat{\vect{u}} - \vect{\nabla}_\perp\hat{p} + \frac{1}{\ReyS}\nabla_\perp^2\hat{\vect{u}} - \frac{1}{\ReyS}\hat{\vect{u}},
\end{equation}
\begin{equation}\label{eq:lin_for_con}
\vect{\nabla}_\perp \vect{\cdot} \hat{\vect{u}} = 0
\end{equation}
from time $t=0$ to $t=\tau$. This is followed by backward time integration of the adjoint equations
\begin{equation}\label{eq:lin_adj}
\pde{\hat{\vect{u}}^\ddag}{t} = ( \vect{\nabla}_\perp \vect{U})^\mathrm{T} \vect{\cdot} \hat{\vect{u}}^\ddag   - (\vect{U} \vect{\cdot} \vect{\nabla}_\perp ) \hat{\vect{u}}^\ddag - \vect{\nabla}_\perp\hat{p}^\ddag - \frac{1}{\ReyS}\nabla_\perp^2\hat{\vect{u}}^\ddag - \frac{1}{\ReyS}\hat{\vect{u}}^\ddag,
\end{equation}
\begin{equation}\label{eq:lin_adj_con}
\vect{\nabla}_\perp \vect{\cdot} \hat{\vect{u}}^\ddag = 0
\end{equation}
for the Lagrange multiplier of the velocity perturbation $\hat{\vect{u}}^\ddag$, from $t=\tau$ to $t=0$. Boundary conditions $\hat{\vect{u}}=\hat{\vect{u}}^\ddag=0$ are applied at  the wall and $\hat{v}=\hat{v}^\ddag=0$ at the stress-free boundary. `Initial' conditions for forward and backward evolution are $\hat{\vect{u}}(0)=\hat{\vect{u}}^\ddag(0)$ and $\hat{\vect{u}}^\ddag(\tau)=\hat{\vect{u}}(\tau)$, respectively. $G$ is then the largest real eigenvalue of the operator representing the sequential action of forward then adjoint evolution \citep{Barkley2008direct, Blackburn2008convective}, obtained by a Krylov subspace scheme. The scheme iterates until a specified eigenvalue tolerance is reached. The corresponding eigenvector contains the optimal initial field (optimal for short).
\begin{table}
\begin{center}
\begin{tabular}{ ccccccc } 
\hline
$\Delta t$  &  $\Nel = 70$  & $|$\% Error$|$ & $\Nel = 98$  & $|$\% Error$|$  & $\Nel = 154$  & $|$\% Error$|$  \\
$2.5\times10^{-3}$   & 33.25571762 & $2.45\times10^{-1}$ & 33.36191967 & $2.59\times10^{-3}$ & 33.36189331 & $2.60\times10^{-3}$ \\
$1.25\times10^{-3}$   & 33.23149556 & $1.72\times10^{-1}$ & 33.36145641 & $1.20\times10^{-3}$ & 33.36142823 & $1.20\times10^{-3}$ \\
$6.25\times10^{-4}$   & 33.20232632 & $8.45\times10^{-2}$ & 33.36122729 & $5.15\times10^{-4}$ & 33.36119843 & $5.15\times10^{-4}$ \\
$3.125\times10^{-4}$  & 33.17957603 & $1.59\times10^{-2}$ & 33.36111304 & $1.73\times10^{-4}$ & 33.36108413 & $1.72\times10^{-4}$ \\
$1.5625\times10^{-4}$ & 33.17428683 & \makecell{0 \\ $5.60\times10^{-1}$} & 33.36105549 & \makecell{0 \\ $8.61\times10^{-5}$} & 33.36102678 & \makecell{0 \\ 0} \\
\hline
\end{tabular}
\caption{The real component of the leading eigenvalue, at $\ReyS=7.071\times10^3$, $\alpha=0.7071$ and $\tau=42.43$ (close to optimal), with domain height $\HsubD=14.14$ and polynomial order $\Np=15$ for various numbers of elements. Meshes with $1$, $2$ and $4$ elements per unit height ($\Nel = 70$, $98$ and $154$, respectively) within the first five units from the wall are compared. Absolute percentage errors are quoted for each mesh separately, relative to the smallest time step case, except the last row, which compares to the $\Nel = 154$ mesh. The eigenvalue convergence tolerance is $10^{-7}$.}
\label{tab:res_lin_dt_comp}
\end{center}
\end{table}

The mesh for computation of linear optimals has a region of high resolution near the wall, with sparse resolution further away. Element spacing is also sparse in the streamwise direction, as the variation must be sinusoidal (from linearity). Three key factors are considered when assessing accuracy, the number of elements in the wall normal direction, the temporal resolution and the domain height where the stress-free condition is applied, as shown in \tbls\ \ref{tab:res_lin_dt_comp} and \ref{tab:res_lin_dom_hght}. Based on the magnitude and behaviour of the errors, the highest near wall resolution ($\Nel=154$ mesh from \tbl\ \ref{tab:res_lin_dt_comp}) was selected, with $\Delta t=1.25\times10^{-3}$. Based on \tbl\ \ref{tab:res_lin_dom_hght}, $\HsubD=14.14$ is sufficient for determining the linear $\tauOpt$ and $\alphaOpt$. However, it was deemed pertinent to increase $\HsubD$ to $28.28$ and to recompute time and wavenumber optimised fields to initiate the nonlinear evolutions reported in \S~\ref{sec:nonlin}. This ensures that the parallel flow assumption remains valid if structures increase in height due to vortex merging. 



\begin{table}
\begin{center}
\begin{tabular}{ ccccccc } 
\hline
$\HsubD$  &  $7.071\times10^2$  & $|$\% Error$|$ & $7.071\times10^3$  & $|$\% Error$|$  & $7.071\times10^4$  & $|$\% Error$|$  \\
14.14 & 6.11779740087 & $3.14\times10^{-6}$ & 33.3619198126 & $2.66\times10^{-6}$ & 166.410928536 & $1.04\times10^{-3}$ \\
28.28 & 6.11779759275 & $7.63\times10^{-10}$ & 33.3619206992 & $7.05\times10^{-10}$ & 166.409189845 & $2.76\times10^{-9}$ \\
56.57 & 6.11779759280 & 0 & 33.3619206994 & 0 & 166.409189849 & 0 \\
\hline
\end{tabular}
\caption{The real component of the leading eigenvalue, varying the domain height, for various $\ReyS$. Initially, $\ReyS=7.071\times10^3$ at $\alpha=0.7071$ and $\tau=42.43$ was tested as part of a formal validation, $\Nel=154$ for $\HsubD=14.14$, $\Delta t=2.5\times10^{-3}$, $\Np=15$. The optimals at $\ReyS=7.071\times10^2$ and $7.071\times10^4$ were tested post validation, $\Nel=250$ for $\HsubD=14.14$, $\Delta t=1.25\times10^{-3}$, $\Np=13$. }
\label{tab:res_lin_dom_hght}
\end{center}
\end{table}



\subsection{Results}\label{sec:ltg_rsts}

\begin{figure}
\begin{center}
\addtolength{\extrarowheight}{-10pt}
\addtolength{\tabcolsep}{-2pt}
\begin{tabular}{ llll }
\makecell{\vspace{23mm} \footnotesize{(a)} \\  \vspace{32mm} \rotatebox{90}{\footnotesize{$\Gmax$}}} & \makecell{\includegraphics[width=0.458\textwidth]{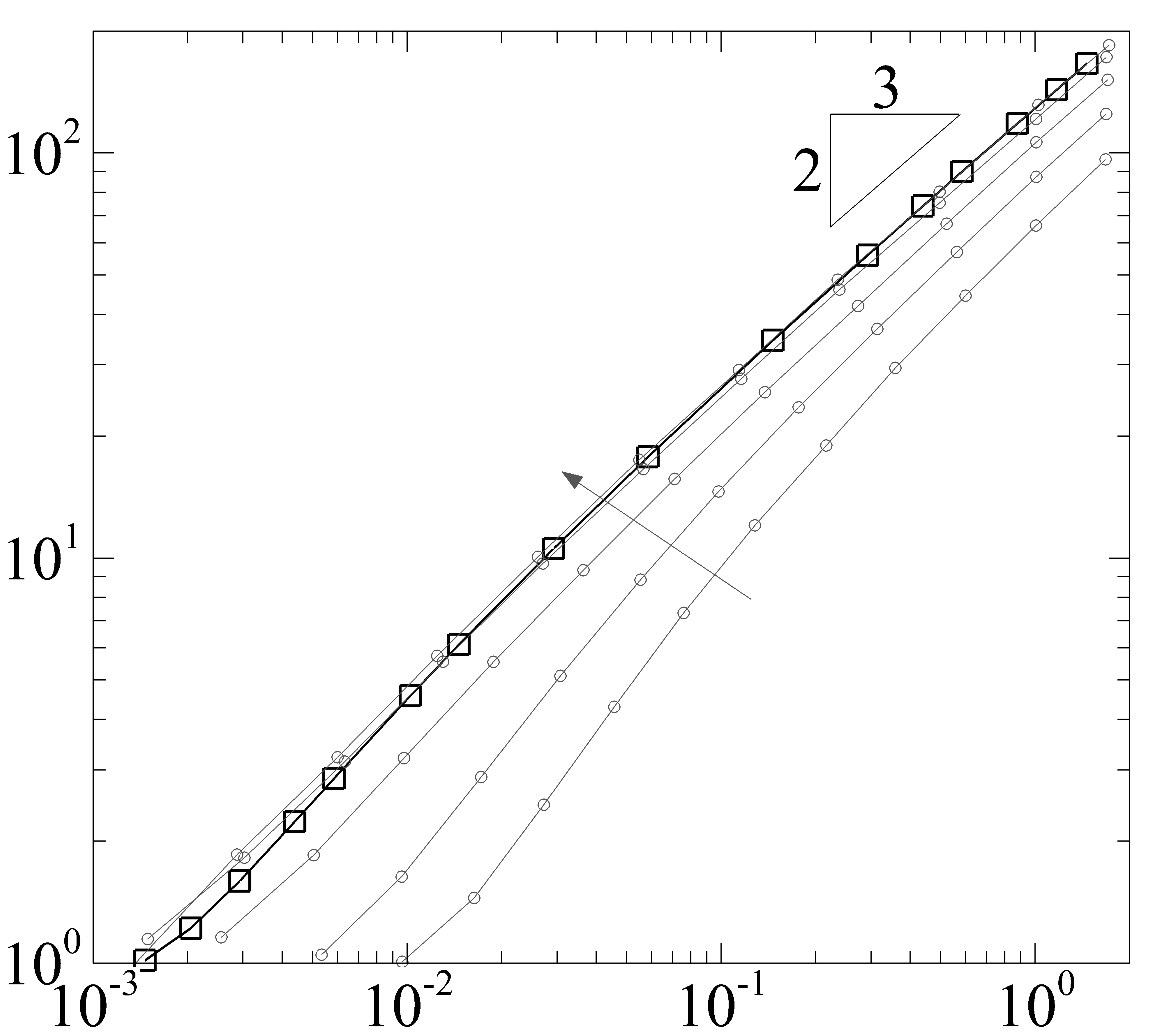}} &
\makecell{\vspace{24mm} \footnotesize{(b)} \\  \vspace{33mm} \rotatebox{90}{\footnotesize{$\alphaOpt$}}}
 & \makecell{\includegraphics[width=0.458\textwidth]{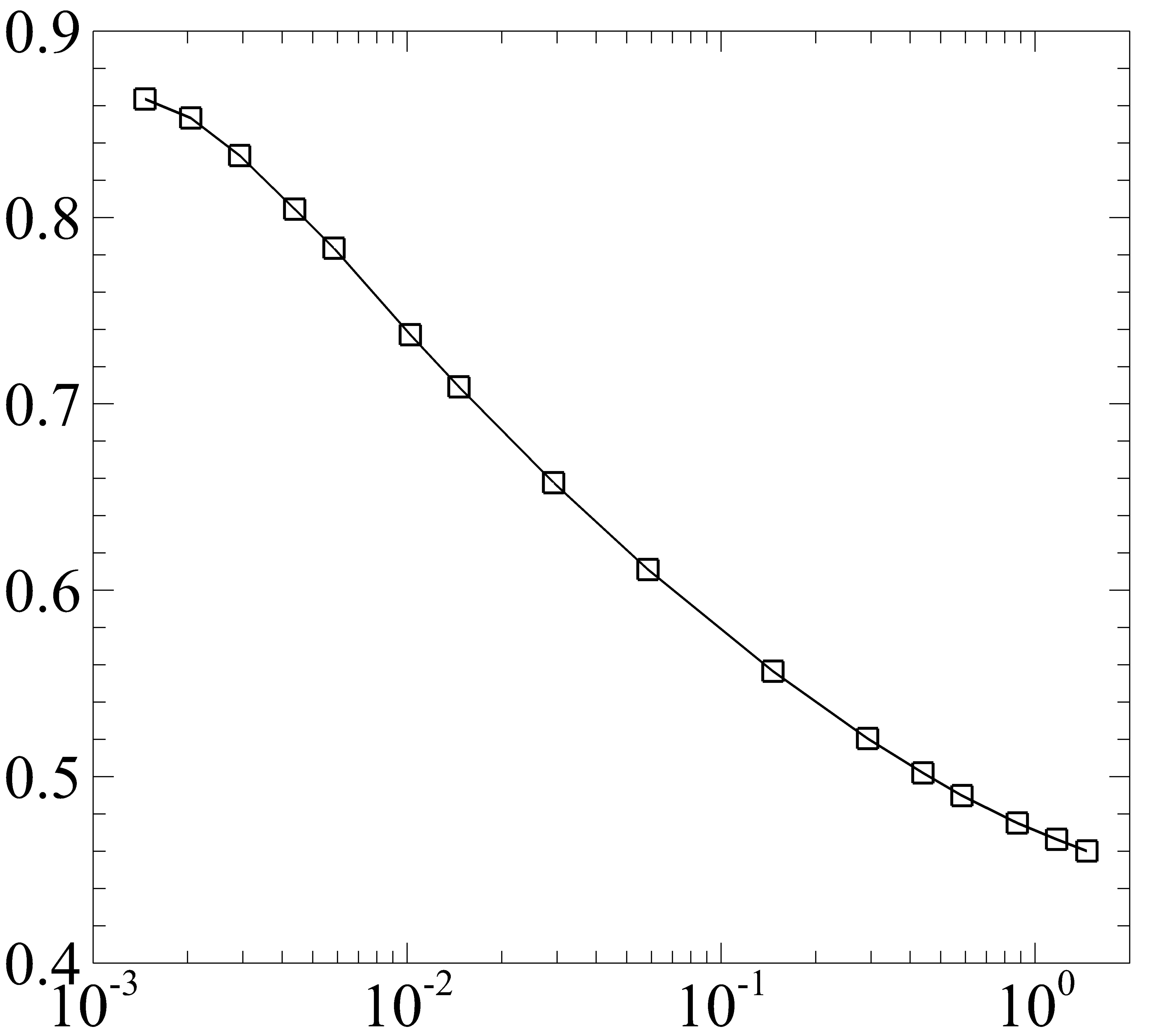}} \\
 & \hspace{36mm} \footnotesize{$\rrc$} & & \hspace{36mm} \footnotesize{$\rrc$} \\
\end{tabular}
\begin{tabular}{ ll }
 \makecell{\vspace{23mm} \footnotesize{(c)} \\  \vspace{33mm} \rotatebox{90}{\footnotesize{$\tauOpt$}}} & \makecell{\includegraphics[width=0.458\textwidth]{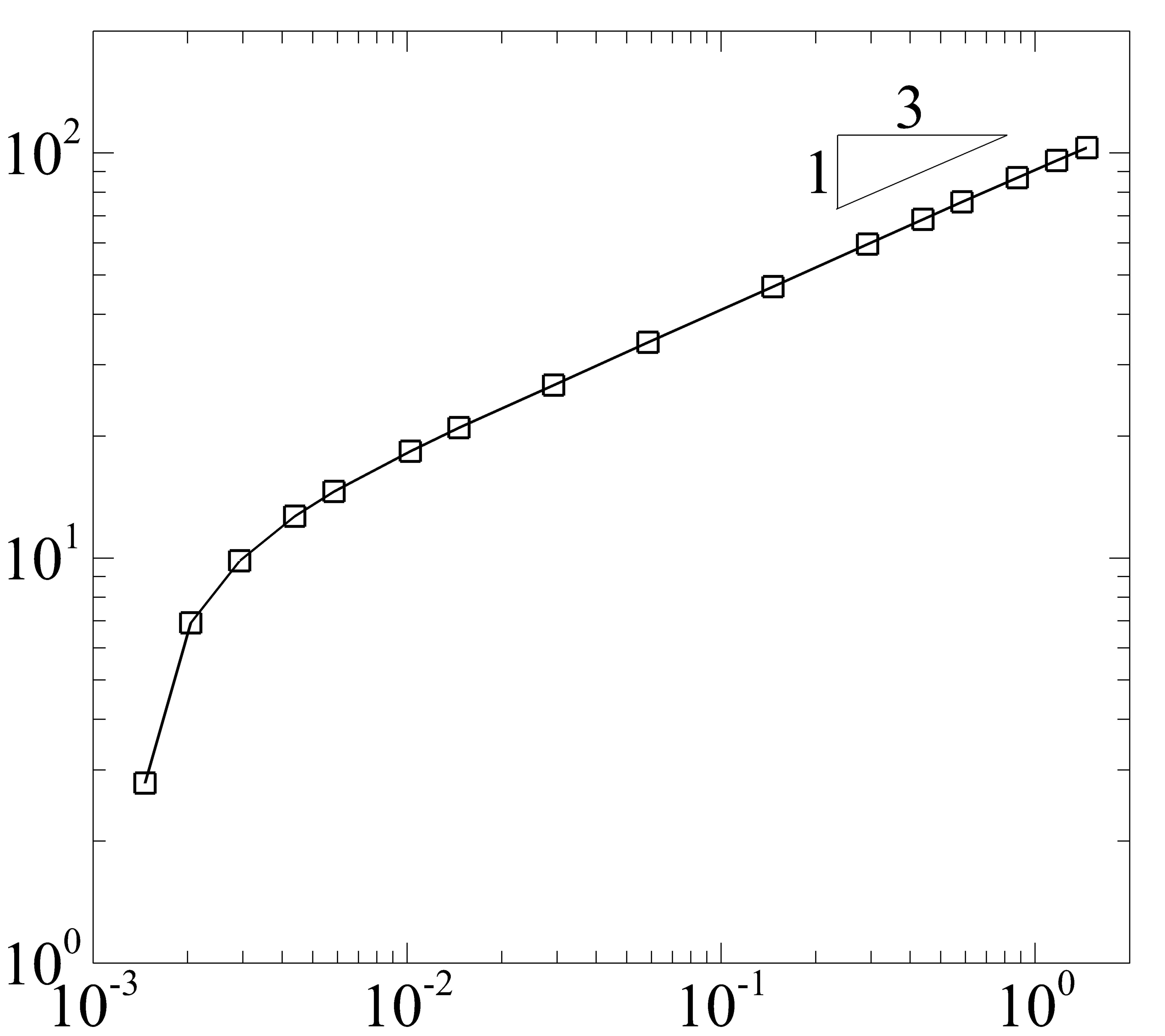}} \\
 & \hspace{36mm} \footnotesize{$\rrc$}  \\
\end{tabular}
\addtolength{\tabcolsep}{+2pt}
\addtolength{\extrarowheight}{+10pt}
\end{center}
    \caption{Linear transient growth of an exponential boundary layer as a function of $\rrc = \ReyS / \ReyCrit$. (a) Growth optimised over initial field, wave number and time interval. Present data (squares) are compared against Q2D duct results from \cite{Potherat2007quasi} (circles). The arrow indicates increasing $H$ through $1$, $3$, $10$, $100$ and $1000$. With increasing $H$, the duct results \cite{Potherat2007quasi} approach the isolated exponential boundary layer results (this work). (b) Optimal wave number. (c) Optimal time interval. }
    \label{fig:lin_trans_opts}
\end{figure}

\begin{figure}
\begin{center}
\addtolength{\extrarowheight}{-10pt}
\addtolength{\tabcolsep}{-2pt}
\begin{tabular}{ ll ll ll ll }
\makecell{\vspace{9mm} \footnotesize{(a)} \\  \vspace{17.5mm} \rotatebox{90}{\footnotesize{$y$}}} & \makecell{\includegraphics[width=0.23\textwidth]{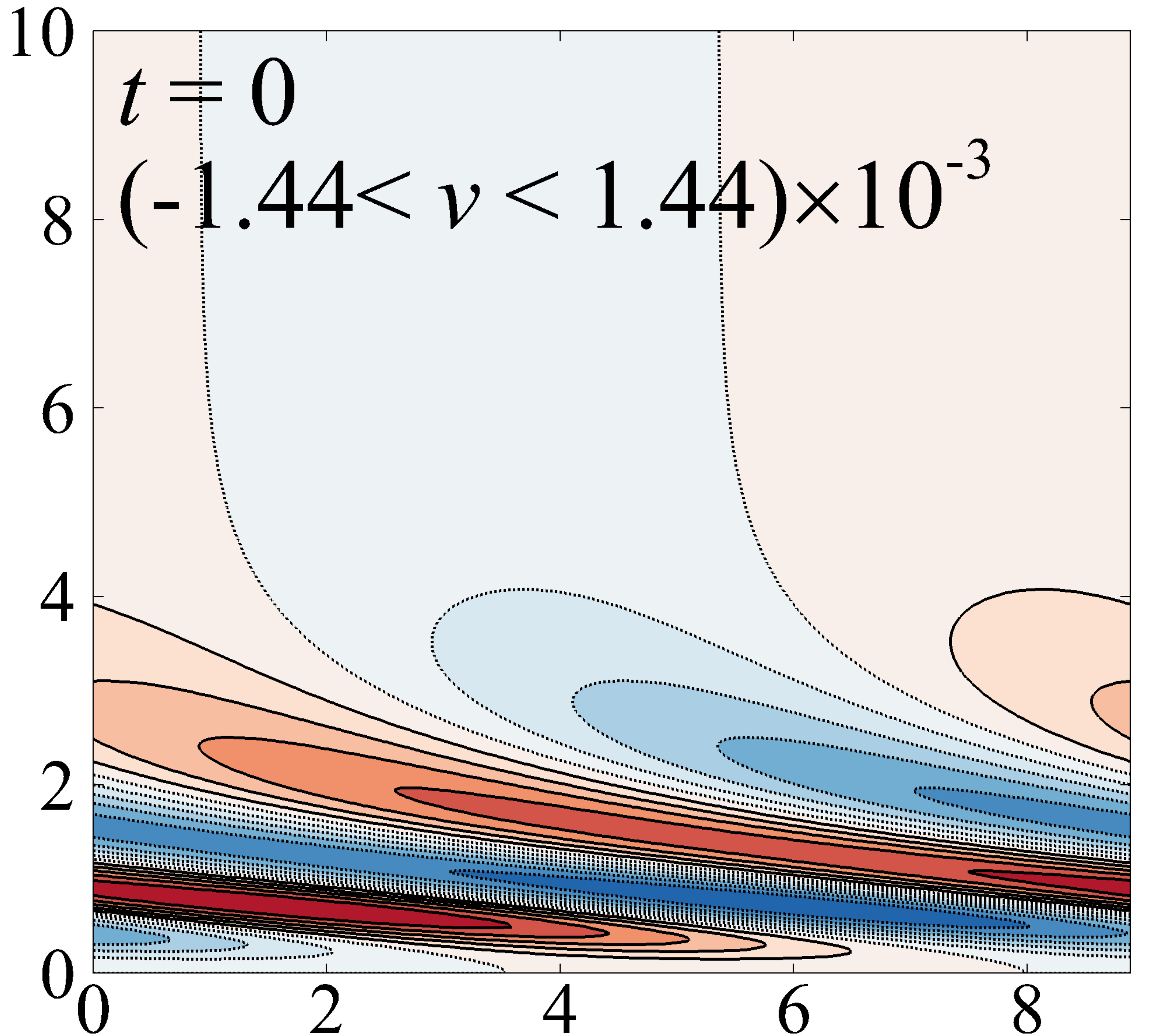}} & 
 & \makecell{\includegraphics[width=0.23\textwidth]{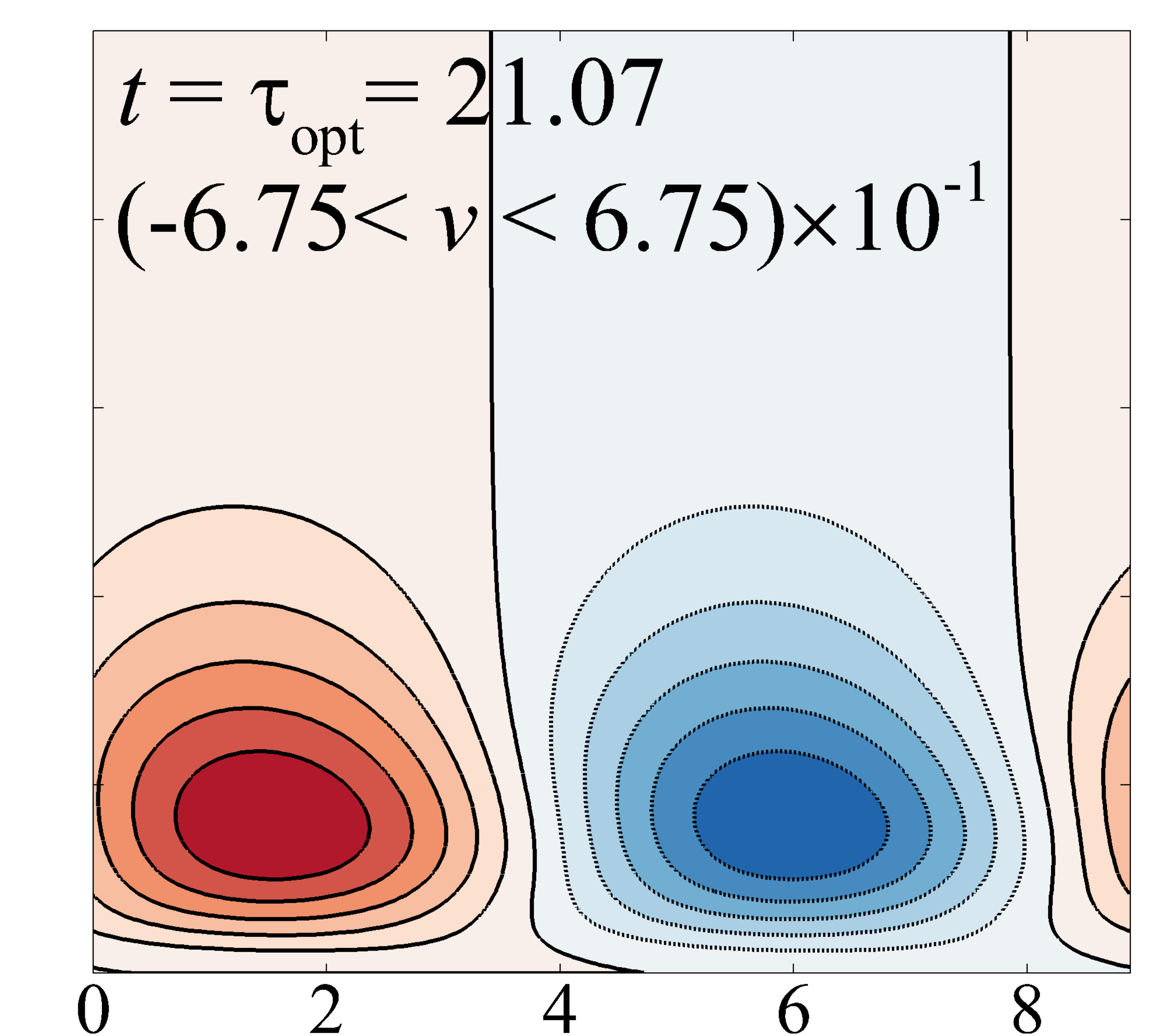}} &
\makecell{\vspace{9mm} \footnotesize{(b)} \\  \vspace{17.8mm} \rotatebox{90}{\footnotesize{$y$}}}  & \makecell{\includegraphics[width=0.23\textwidth]{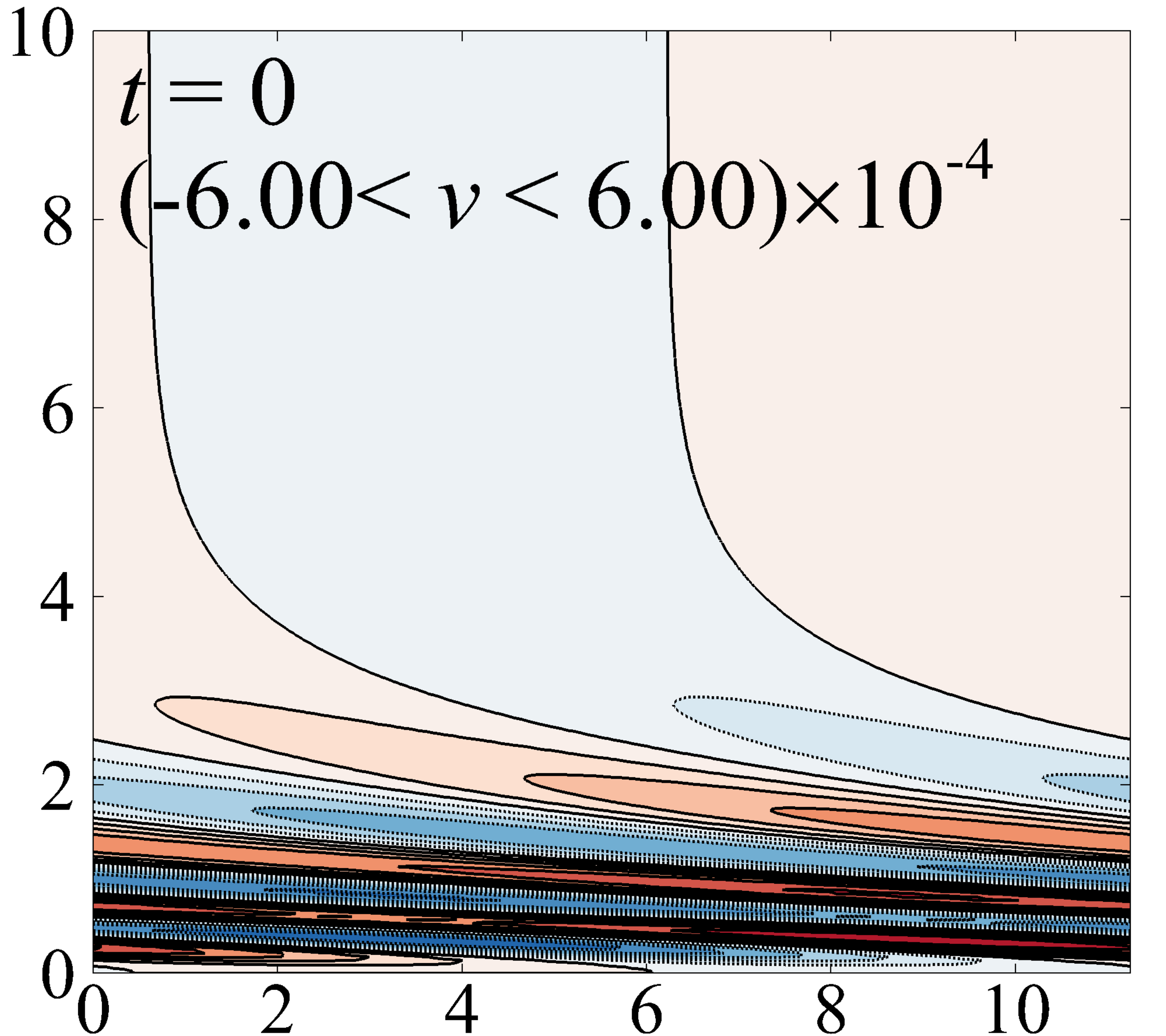}} &
 & \makecell{\includegraphics[width=0.23\textwidth]{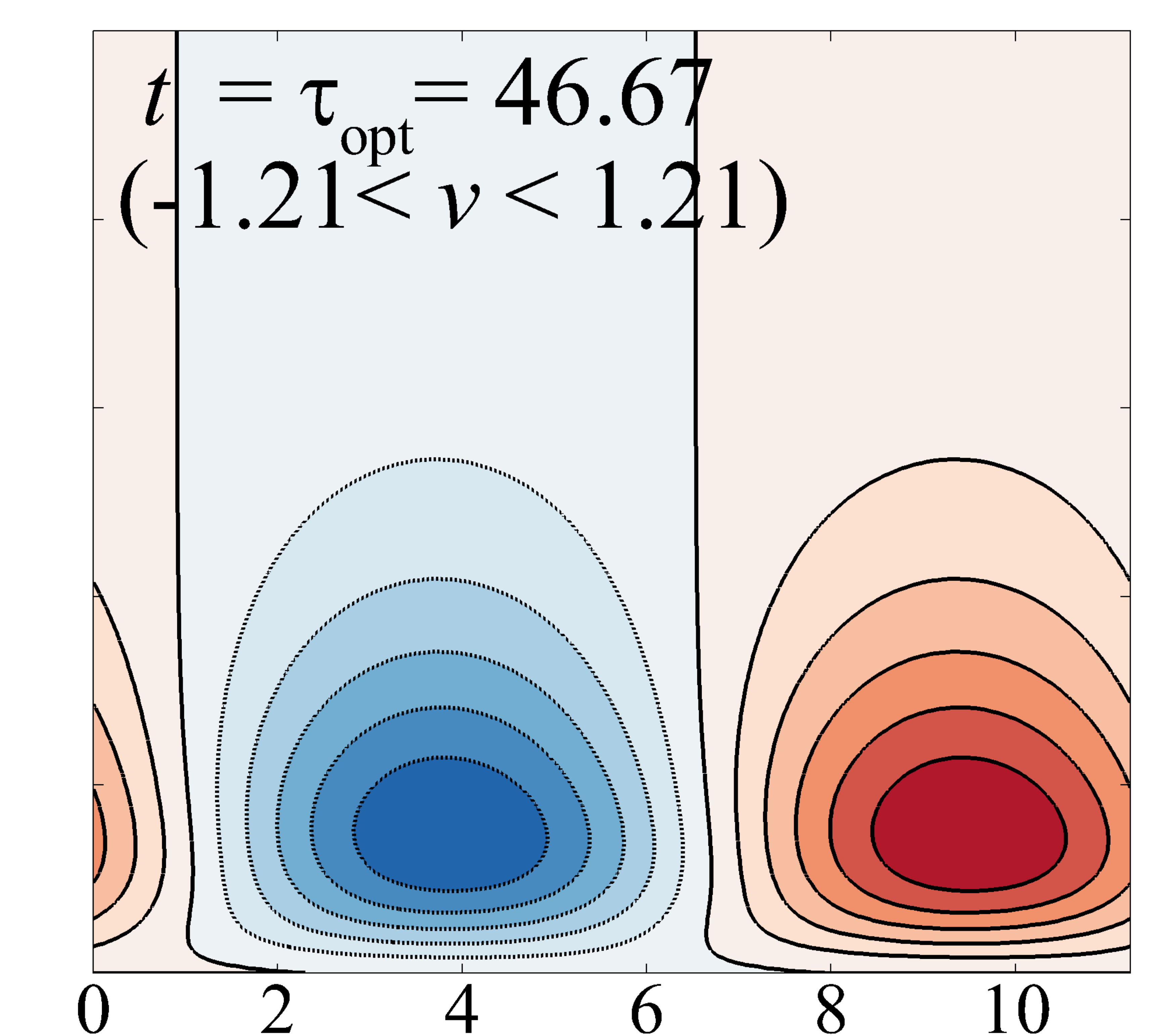}} \\
 & \hspace{16mm} \footnotesize{$x$} &  & \hspace{16mm} \footnotesize{$x$} &  & \hspace{16mm} \footnotesize{$x$} &  & \hspace{16mm} \footnotesize{$x$} \\
\end{tabular}
\addtolength{\tabcolsep}{+2pt}
\addtolength{\extrarowheight}{+10pt}
\end{center}
    \caption{Optimised $\hat{v}$-velocity fields. (a) $\rrc=0.0146$, $\alphaOpt=0.7071$. (b) $\rrc=0.146$, $\alphaOpt=0.5586$. Simulations computed with $\HsubD=28.28$ and images clipped at $y=10$. Solid lines (red flooding) positive; dotted lines (blue flooding) negative.}
    \label{fig:lin_init_and_opt}
\end{figure}

At least one infinitisemal disturbance can achieve exponential growth at Reynolds numbers above the critical Reynolds number $\ReyCrit$. $\ReyCrit$ thereby forms a bound above which transition to turbulence is possible, so long as the domain length has a corresponding wavenumber within the neutral curve. For this problem, $\ReyCrit$ can be determined by rescaling the results of \cite{Potherat2007quasi}; changing length scale from $L$ to $\deltaS$. Thus $\ReyCrit = 4.835\times10^4$ and $\alphaCrit = 0.1615$. The ratio $\rrc = \ReyS / \ReyCrit$ is then defined.

Linear transient growth results are presented in \fig\ \ref{fig:lin_trans_opts}. Duct results from \cite{Potherat2007quasi} at finite $H$ are also included in \fig\ \ref{fig:lin_trans_opts}(a), supporting the argument that the boundary layer at each duct wall is sufficiently isolated at large $H$, and can be modelled separately. At $\rrc=0.00135$, $\Gmax=1$, while by $\rrc=1$, $\Gmax \approx 100$. This modest rise in gain with increasing $\rrc$ may be attributed to two factors. The first is that the base flow is naturally highly stable \citep{Albrecht2006stability}. The second is that two-dimensional systems only permit growth via the Orr mechanism \citep{Butler1992optimal}. This greatly reduces optimal growth, and produces the modest scaling of $\Gmax \sim \ReyS^{2/3}$ for large $\ReyS$. Representative initial and optimal fields are provided in \fig\ \ref{fig:lin_init_and_opt}, which exhibit the classic initial condition of a strongly sheared wave which transiently grows as it is advected upright, until $\tauOpt$. The modes otherwise resemble those of \cite{Potherat2007quasi}, excepting wall confinement effects at low $H$ in the aforementioned work.

\section{Nonlinear transient growth}\label{sec:nltg}
\subsection{Formulation and validation}\label{sec:nlg_form}
%

In this work, nonlinear transient growth is employed solely to assess the similarities between the linear and nonlinear optimals for small target times ($\tau \sim \tauOpt$). Admittedly, nonlinear transient growth routines can identify the initial energy representing separatrix 1, if the target time specified is long enough to allow the minimal seed to reach the turbulent attractor  \citep{Pringle2012minimal, Kerswell2014optimization}. This target time is not known a priori. It is shown in \S~\ref{sec:nonlin} that the turbulent attractor is reached between $t=1.4\times10^3$ and $t=2\times10^3$ at $\rrc=0.585$. As $\tauOpt=75.94$ at $\rrc=0.585$ (figure \ref{fig:lin_trans_opts}) the additional computation cost is proportional to $t/\tauOpt = 18.44-26.34$. In contrast, the hydrodynamic pipe flow work in \citep{Pringle2012minimal} had $\tauOpt \lesssim 30$, while the minimal seed reached the turbulent attractor by $t=75$, so $t/\tauOpt \lesssim 2.5$. Thus, for this problem, it was not amenable to determine separatrix 1 directly from the nonlinear transition growth algorithm.


The scheme to determine the nonlinear growth $G_\mathrm{N} = \left\lVert\hat{\bm u}(\tau)\right\rVert/\left\lVert\hat{\bm u}(0)\right\rVert$, for a specified target time $\tau$, optimised over all initial perturbations, requires maximizing the functional \citep{Pringle2012minimal, Pringle2015fully}
\begin{eqnarray}\label{eq:nl_func}
\mathscr{L} &:=& \langle \frac{1}{2}\hat{\vect{u}}(\tau)^2 \rangle
 -\lambda_0 \left[\langle \frac{1}{2}\hat{\vect{u}}(0)^2 \rangle - E_\mathrm{P} \right]
 -\int_0^\tau \langle \Pi \vect{\nabla}_\perp \bcdot \hat{\vect{u}}  \rangle \mathrm{d}t
 -\int_0^\tau  \Gamma (t)  \langle \hat{\vect{u}}  \bcdot \vect{e_\mathrm{z}}  \rangle \mathrm{d}t \nonumber\\
&& 
 - \int_0^\tau \langle \hat{\vect{u}}^\ddag \bcdot \bigg[\frac{\partial \hat{\vect{u}}}{\partial t} + (\vect{U} \bcdot \vect{\nabla}_\perp) \hat{\vect{u}} + (\hat{\vect{u}} \bcdot \vect{\nabla}_\perp)\vect{U} + (\hat{\vect{u}} \bcdot \vect{\nabla}_\perp) \hat{\vect{u}} + \frac{1}{\rho}[\Lambda(t)\vect{e_\mathrm{z}} + \vect{\nabla}_\perp p'] \nonumber\\
&&  - \frac{1}{\ReyS}\nabla^2_\perp \hat{\vect{u}} + \frac{1}{\ReyS}\hat{\vect{u}}\bigg] \rangle \mathrm{d}t
\end{eqnarray}
where the Lagrange multipliers $\lambda_0$, $\Pi$ and $\Gamma(t)$ are constraints on the specified initial energy of the perturbation $E_\mathrm{P}=(1/2)\int \hat{\vect{u}}(0)^2\mathrm{d}\Omega$, mass conservation and flow rate, respectively. Pressure is  decomposed into a time-varying pressure gradient $\Lambda(t)$, to maintain the flow rate, and fluctuating component $p'$. $\langle \dots \rangle$ represent integrals over the computational domain. The Lagrange multiplier $\hat{\vect{u}}^\ddag$ ensures that the full nonlinear Navier--Stokes equations are enforced over all times $0<t<\tau$ \citep{Pringle2010using}. Each iteration $j$ of the optimisation procedure begins with the forward evolution, from $t=0$ to $t=\tau$, of the nonlinear perturbation equation (within the square brackets of the last term of equation (\ref{eq:nl_func})). If $G_\mathrm{N}$ for iteration $j$ is larger than for iteration $j-1$, the adjoint `initial' field is $\hat{\vect{u}}^\ddag(\tau)=\hat{\vect{u}}(\tau)$ and the iteration continues with backward evolution via the adjoint equations
\begin{eqnarray}\label{eq:nl_adj}
\frac{\partial \hat{\vect{u}}^\ddag}{\partial t} = (\vect{\nabla}_\perp\vect{U})^\mathrm{T} \bcdot \hat{\vect{u}}^\ddag - (\vect{U} \bcdot \vect{\nabla}_\perp)\hat{\vect{u}}^\ddag &+& (\vect{\nabla}_\perp\hat{\vect{u}})^\mathrm{T} \bcdot \hat{\vect{u}}^\ddag - (\hat{\vect{u}} \bcdot \vect{\nabla}_\perp)\hat{\vect{u}}^\ddag \nonumber\\
&&  + \Gamma(t)\vect{e_\mathrm{z}} - \vect{\nabla}_\perp\Pi - \frac{1}{\ReyS}\nabla^2_\perp\hat{\vect{u}}^\ddag - \frac{1}{\ReyS}\hat{\vect{u}}^\ddag
\end{eqnarray}
\begin{equation}\label{eq:nl_adj_con}
\vect{\nabla}_\perp \bcdot \hat{\vect{u}}^\ddag = 0
\end{equation}
%
from time $t=\tau$ to $t=0$. An under-relaxation factor $\epsilon_\mathrm{N}$ is chosen (say, $0.5$) for the first iteration, or adjusted as described in \cite{Pringle2012minimal}. The initial field for the $j+1$ iteration is $\hat{\vect{u}}^{j+1}(0)=\hat{\vect{u}}^{j}(0) + \epsilon_\mathrm{N}(-\lambda_0\hat{\vect{u}}^{j}(0) + \hat{\vect{u}}^{\ddag,j}(0))/\lambda_0$, where $\lambda_0$ is sought such that $\langle \hat{\vect{u}}^{j+1}(0) \bcdot \hat{\vect{u}}^{j+1}(0) \rangle=2E_\mathrm{P}$. However, if $G_\mathrm{N}$ does not increase in iteration $j$, adjoint evolution is not performed, as the updated field (iteration $j$) is further from the optimal than the previous ($j-1$) field. An additional adjustment is then made to the under-relaxation factor, $\epsilon_\mathrm{N} \rightarrow \epsilon_\mathrm{N}/4$. The forward iteration restarts with $\hat{\vect{u}}^{j}(0)=\hat{\vect{u}}^{j-1}(0) + \epsilon_\mathrm{N}(-\lambda_0\hat{\vect{u}}^{j-1}(0) + \hat{\vect{u}}^{\ddag,j-1}(0))/\lambda_0$. This ensures monotonic growth in successive iterations, and avoids contaminating the initial field after iterations with too large an $\epsilon_\mathrm{N}$. Iterations continue until the relative change in $\lambda_0$ and residual $(\delta\mathscr{L}/\delta\hat{\vect{u}}(0))/\lambda_0^2$ are both below a specified tolerance, following \cite{Pringle2012minimal}.

Validation of the nonlinear transient growth is provided in \tbl\ \ref{tab:res_nltg} at $\rrc = 0.293$, considering the polynomial order and time step, for two initial energies. The same mesh for determination of the linear optimals is used, with $\HsubD=28.28$. As the nonlinear transient growth scheme does not evolve the perturbations through turbulent states, the resolution requirements are similar to those of the linear computations, \S~\ref{sec:ltg_form}, rather than the nonlinear forward evolutions, \S~\ref{sec:nln_form}. For consistency, the same time step of $\Delta t=1.25 \times 10^{-3}$ was selected, with $\Np=15$.

\begin{table}
\begin{center}
\begin{tabular}{ cccccc } 
\hline
$\Delta t$  &  $G_\mathrm{N}$; $E_\mathrm{P}=10^{-6}$  & $|$\% Error$|$ & $\Np$ & $G_\mathrm{N}$; $E_\mathrm{P}=10^{-4}$   & $|$\% Error$|$ \\
$5\times10^{-3}$ & 55.9721743040676 & $1.88\times10^{-5}$  & 11 & 54.6714139912327 & $5.24\times10^{-4}$  \\
$2.5\times10^{-3}$ & 55.9721692244256 & $9.69\times10^{-6}$ & 13 & 54.6711233880979 & $7.81\times10^{-6}$ \\
$1.25\times10^{-3}$ & 55.9721654578752 & $2.96\times10^{-6}$ & 15 & 54.6711274190738 & $4.31\times10^{-7}$ \\
$6.25\times10^{-4}$ & 55.9721633006764 & $8.91\times10^{-7}$ & 17 & 54.6711283768056 & $1.32\times10^{-6}$ \\
$3.125\times10^{-4}$ & 55.9721637995307 & 0 & 19 & 54.6711276549269 & 0  \\
\hline
\end{tabular}
\caption{Validation of the time step and polynomial order for the nonlinear transient growth, for initial perturbation energies of $10^{-6}$ and $10^{-4}$, at $\rrc=0.293$, $n=1$. The mesh is based on the $\Nel=154$ case from linear optimisation, except with $\HsubD=28.28$. The tolerance for convergence was $10^{-7}$. Nonlinear computations use the linear $\alphaOpt$ and $\tauOpt$.}
\label{tab:res_nltg}
\end{center}
\end{table}

\subsection{Results}\label{sec:nlg_rslt}

\begin{figure}
\begin{center}
\addtolength{\extrarowheight}{-10pt}
\addtolength{\tabcolsep}{-2pt}
\begin{tabular}{ llll }
\makecell{\vspace{15mm} \footnotesize{(a)} \\  \vspace{20mm} \rotatebox{90}{\footnotesize{$\Gmax-\max_t(G_\mathrm{N})$}}} & \makecell{\includegraphics[width=0.458\textwidth]{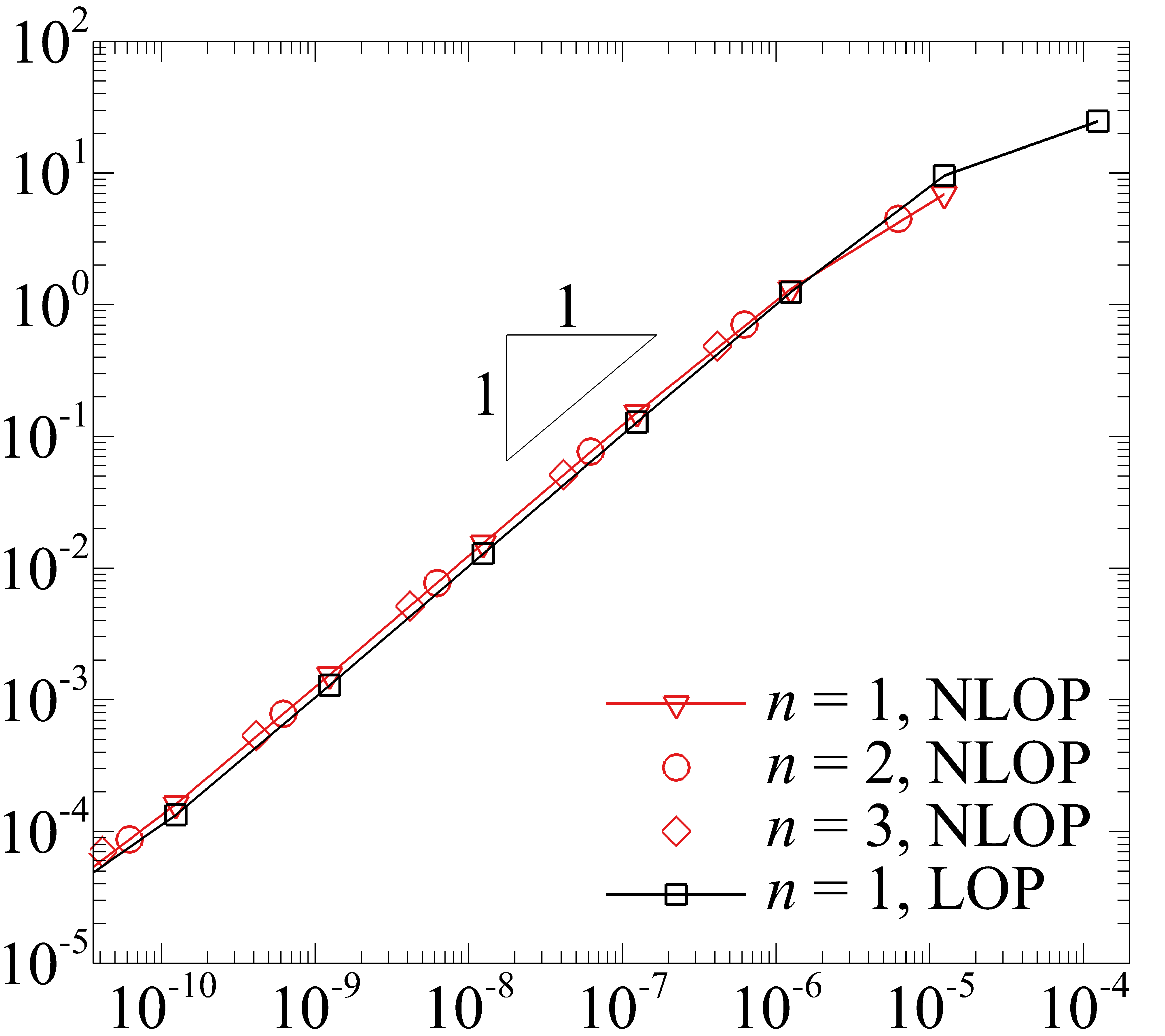}} &
\makecell{\vspace{3mm} \footnotesize{(b)} \\  \vspace{8mm} \rotatebox{90}{\footnotesize{$|\max_t(G_\mathrm{N,LOP})-\max_t(G_\mathrm{N,NLOP})|$}}}
 & \makecell{\includegraphics[width=0.458\textwidth]{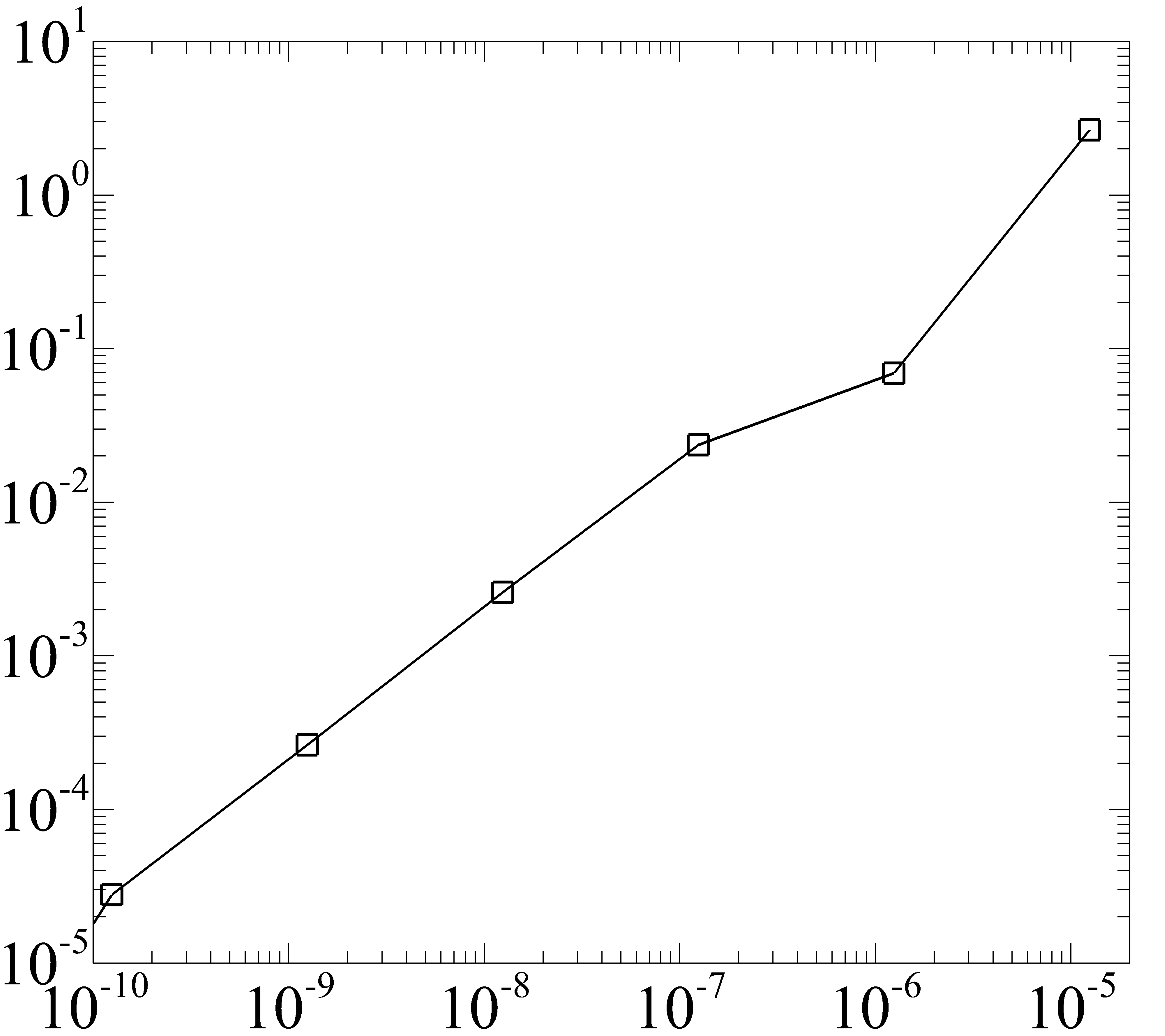}} \\
 & \hspace{36mm} \footnotesize{$\Ezero$} & & \hspace{36mm} \footnotesize{$\Ezero$} \\
\end{tabular}
\addtolength{\tabcolsep}{+2pt}
\addtolength{\extrarowheight}{+10pt}
\end{center}
    \caption{Comparison between linear and nonlinear optimals for various initial energies  $\EzeroR=\int \hat{u}^2+\hat{v}^2\,\dUP\Omega/\int U^2\,\dUP\Omega$ at $\rrc=0.293$. (a) Difference in the maximum linear growth obtained with the linear optimal (LOP) and maximum nonlinear growth with the nonlinear optimal (NLOP), for three domain lengths, and difference in the linear growth of the LOP and the nonlinear growth of the LOP scaled to $\Ezero$ ($n=1$ only). (b) Comparison between the nonlinear growth of the NLOP and nonlinear growth of the LOP scaled to $\Ezero$ ($n=1$). The linear growth of the LOP is $\Gmax=55.9876$.}
    \label{fig:nlop_comp}
\end{figure}

Nonlinear optimals were computed with $\tau = \tauOpt$ and domain lengths based on $n=1$, $n=2$ or $n=3$ repetitions of $\lxopt$, for various initial energies. These results are shown in \fig\ \ref{fig:nlop_comp}(a), which compares the difference between the linear transient growth of the linear optimal and the nonlinear transient growth of the nonlinear optimal (red data points), with the former always greater than the latter (all results are positive valued). As nonlinear collaboration between linear transient growth mechanisms cannot occur, the maximum growth obtained at vanishingly small initial energy is greater than with finite initial energy. \Fig\ \ref{fig:nlop_comp}(a) also shows that for an initial energy defined per unit duct length, the results are not dependent on domain length. Thus, it is the initial energy density that is the important parameter.


Additionaly, \fig\ \ref{fig:nlop_comp}(a) compares the difference in the linear transient growth of the linear optimal and the nonlinear transient growth of the linear optimal scaled to $E_0$ (square symbols). These results are almost coincident with those for the nonlinear growth of the nonlinear optimal (triangle symbols). Thus, the difference between the nonlinear and linear growth is mostly due to the finite energy of the initial field. The mode structure is only very weakly dependent on initial energy (the linear and nonlinear optimals are virtually indistinguishable; not shown). This supports a remark made by \cite{Kerswell2014optimization}, that in two-dimensional systems the nonlinear optimal contains the linear mode trivially. This comparison is further highlighted in \fig\ \ref{fig:nlop_comp}(b), which directly compares the nonlinear growth of the nonlinear optimal to the nonlinear growth of the linear optimal. This difference is very small for initial energies up to $\EzeroR \approx 10^{-6}$, where $\EzeroR=\int \hat{u}^2+\hat{v}^2\,\dUP\Omega/\int U^2\,\dUP\Omega$ is considered to account for the varying domain length. 

For $\EzeroR \gtrsim 10^{-6}$ the nonlinear growth of the nonlinear optimal then slightly exceeds the nonlinear growth of the rescaled linear optimal.  However, the differences are still small at $\EzeroR=10^{-5}$, which is an initial energy more than sufficient to generate large amounts of nonlinear second-stage growth, as is discussed in detail in \S~\ref{sec:nonlin}. Thus, there is little `error' in estimating the minimal seed energy with the linear optimal, for the initial energies of interest. 


\section{Nonlinear evolution at specified initial energies}\label{sec:nonlin}
\subsection{Validation}\label{sec:nln_form}
%
\begin{figure}
\begin{center}
\addtolength{\extrarowheight}{-10pt}
\addtolength{\tabcolsep}{-2pt}
\begin{tabular}{ llll }
\makecell{\vspace{26mm} \footnotesize{(a)} \\  \vspace{33mm} \rotatebox{90}{\footnotesize{$\Ev$}}} & \makecell{\includegraphics[width=0.458\textwidth]{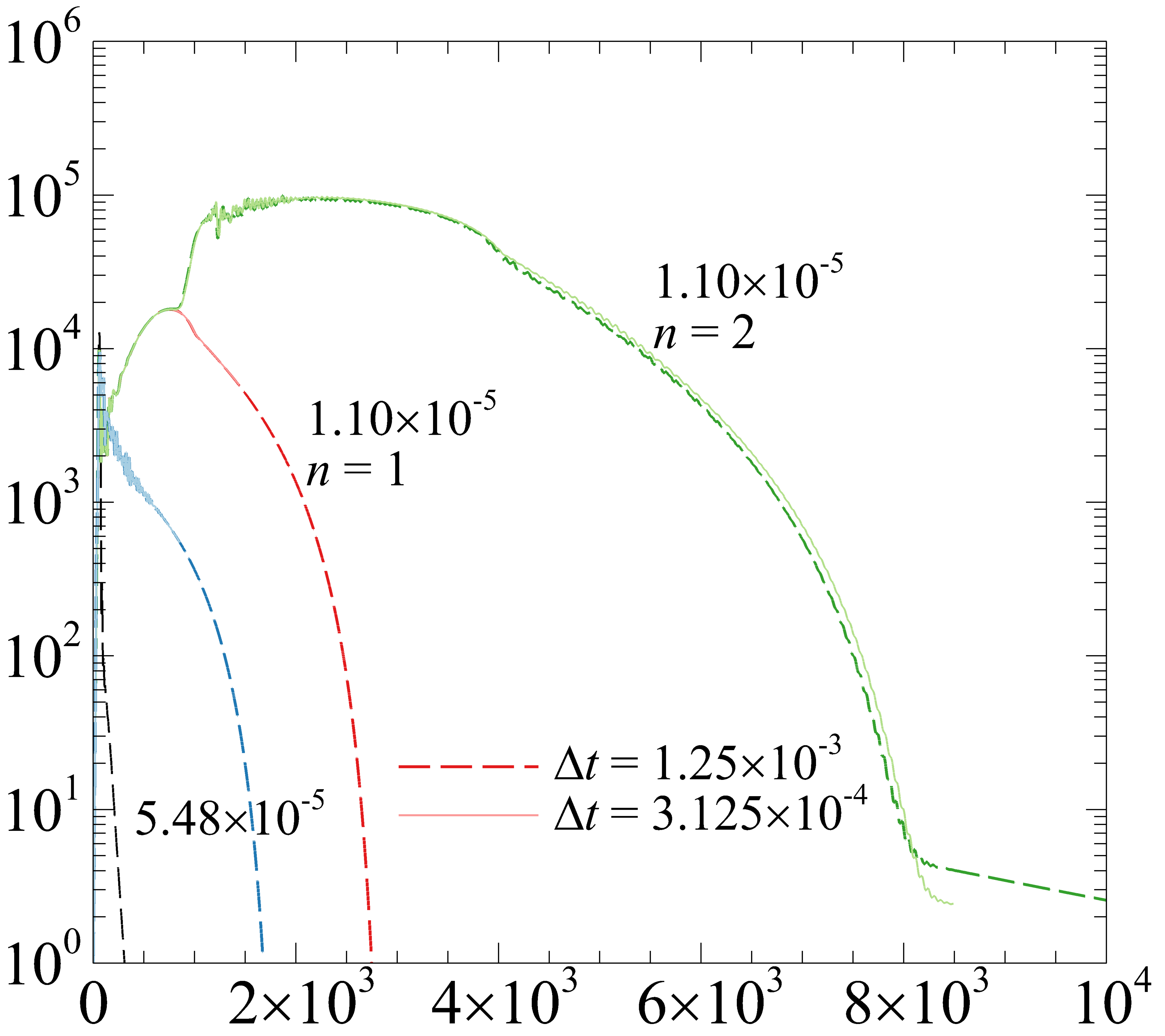}} &
\makecell{\vspace{26mm} \footnotesize{(b)} \\  \vspace{33mm} \rotatebox{90}{\footnotesize{$\Ev$}}}
 & \makecell{\includegraphics[width=0.458\textwidth]{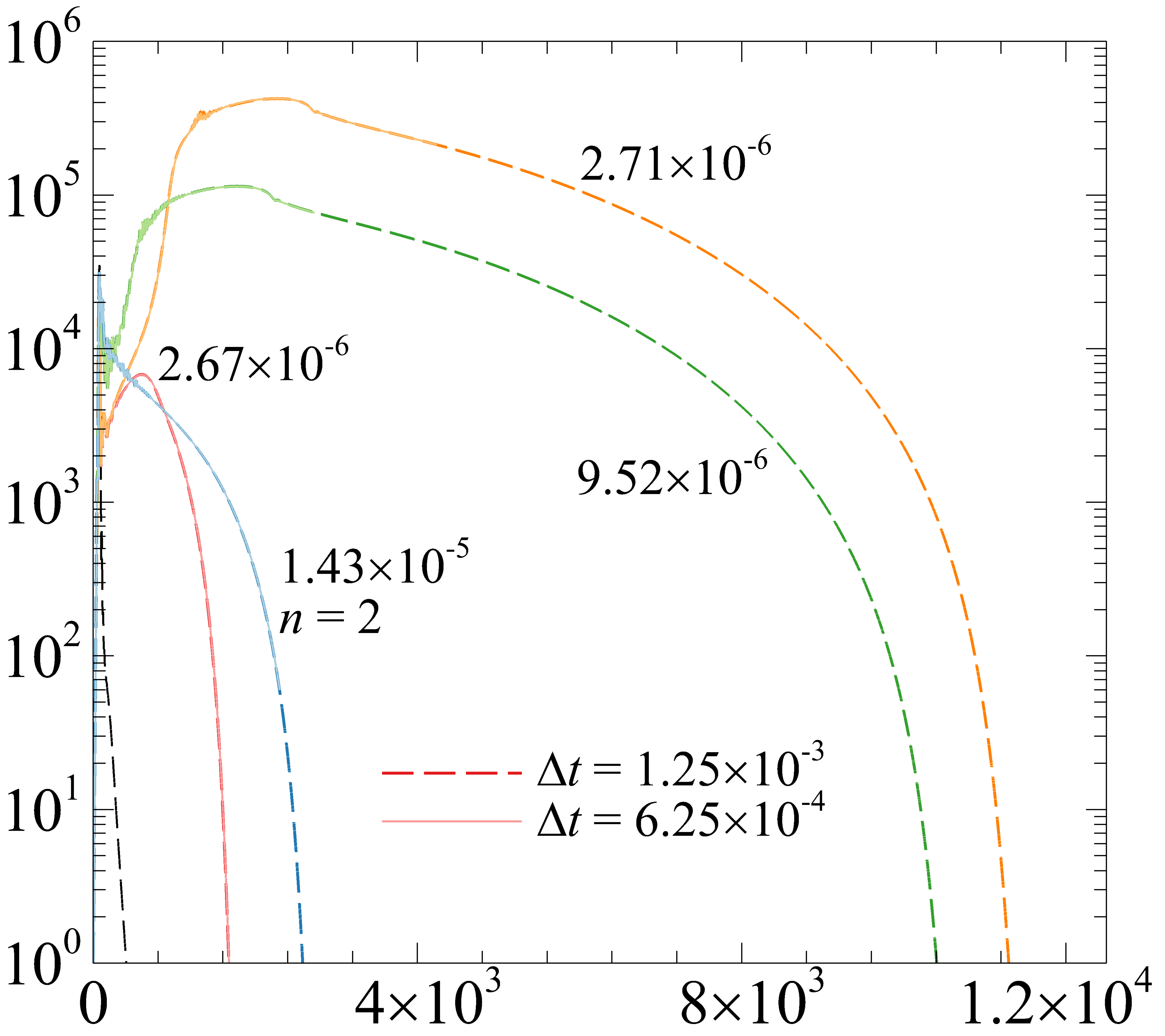}} \\
 & \hspace{36mm} \footnotesize{$t$} & & \hspace{36mm} \footnotesize{$t$} \\
\makecell{\vspace{26mm} \footnotesize{(c)} \\  \vspace{33mm} \rotatebox{90}{\footnotesize{$\Ev$}}} & \makecell{\includegraphics[width=0.458\textwidth]{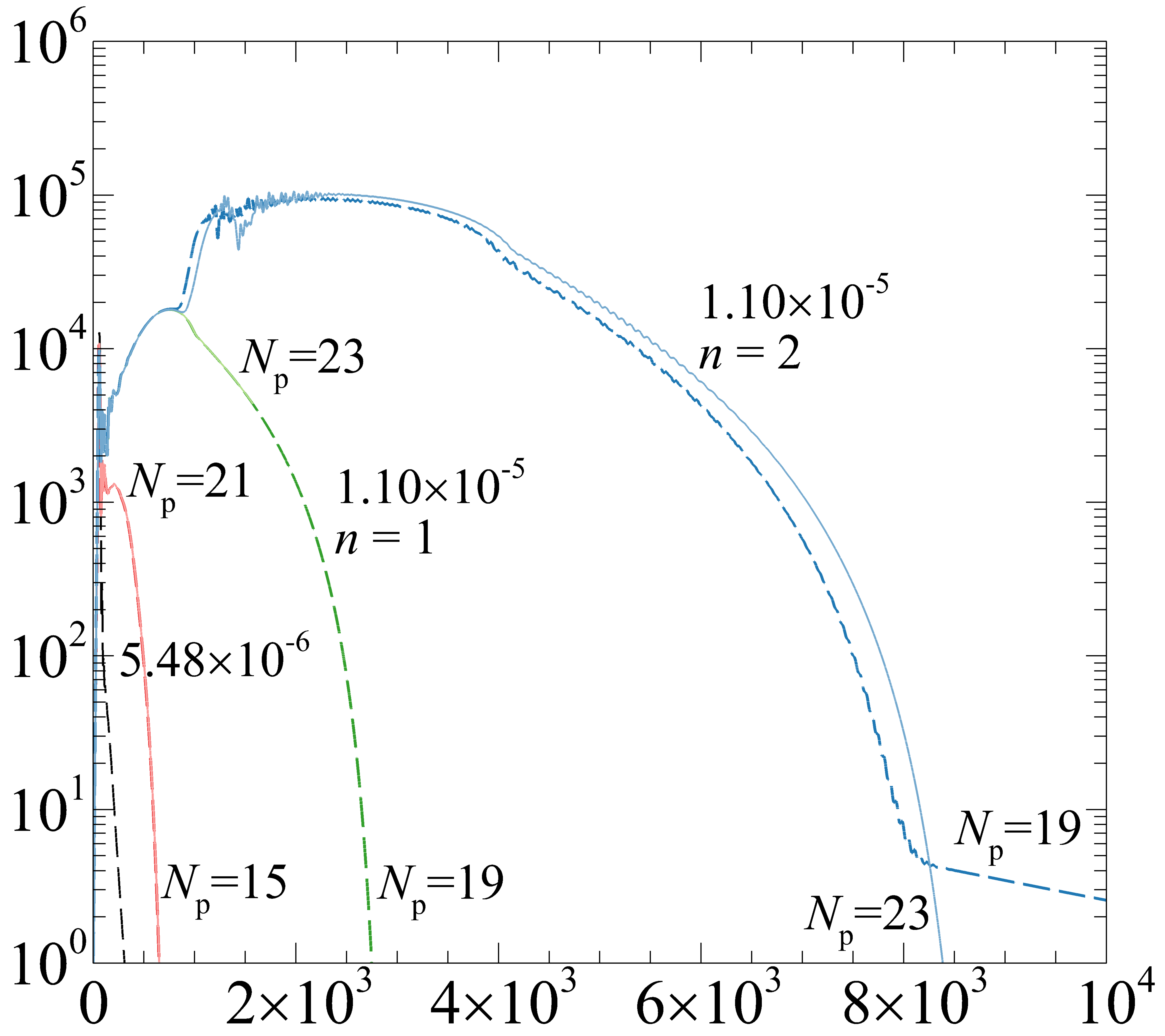}} &
\makecell{\vspace{26mm} \footnotesize{(d)} \\  \vspace{33mm} \rotatebox{90}{\footnotesize{$\Ev$}}}
 & \makecell{\includegraphics[width=0.458\textwidth]{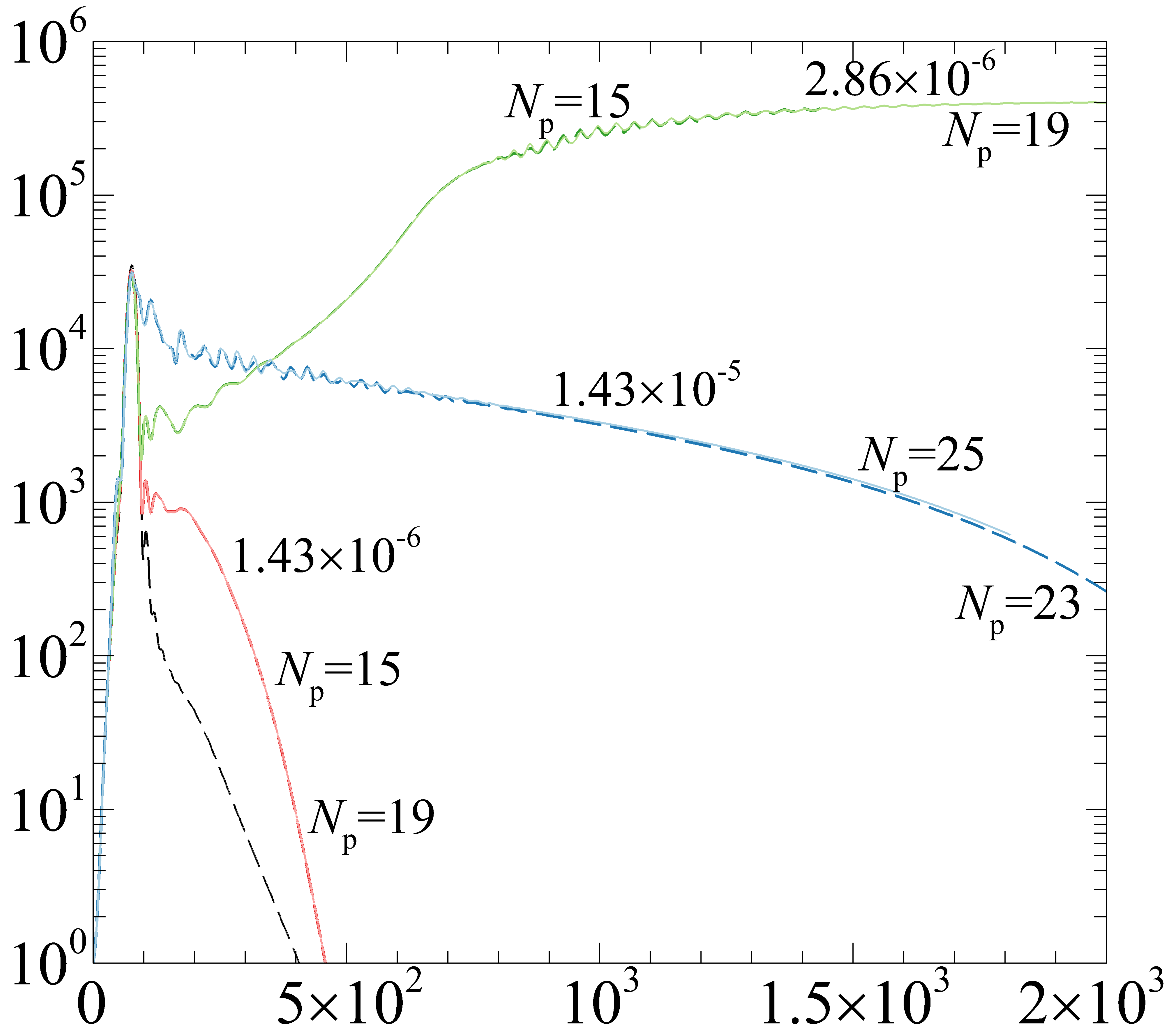}} \\
 & \hspace{36mm} \footnotesize{$t$} & & \hspace{36mm} \footnotesize{$t$} \\
\end{tabular}
\addtolength{\tabcolsep}{+2pt}
\addtolength{\extrarowheight}{+10pt}
\end{center}
    \caption{(a--b) temporal and (c--d) spatial resolution testing of the nonlinear evolution of linear optimals, for various initial energies $\Ezero$. (a \& c) $\rrc=0.293$. (b \& d) at $\rrc=0.585$. The smaller polynomial order (value annotated for each curve), or larger time step (see legend), is denoted by a long dashed line for each $\Ezero$. $n=1$ unless otherwise stated. A black long dashed line represents the linear evolution.}
    \label{fig:res_dt}
\end{figure}

The initial energy of each linear optimal is scaled to $\EzeroR$ when seeded onto the base flow. Forward evolution of the full nonlinear equations (\ref{eq:non_dim_m}) and (\ref{eq:non_dim_c}) then commences. The measures $\Ev = (1/2)\int \hat{v}^2\,\dUP\Omega$ and $\Euv = (1/2)\int \hat{u}^2 + \hat{v}^2 \,\dUP\Omega$ are defined. These separate the growth of the perturbation, captured by $\Ev$, and the effective modulation of the base flow, via a streamwise-independent structure, captured by $\Euv$. 

The effect of time step variation is depicted in \fig\ \ref{fig:res_dt}(a), \ref{fig:res_dt}(b). These show negligible differences between $\Delta t=1.25\times10^{-3}$ and significantly smaller time step sizes. $\Delta t=1.25\times10^{-3}$ was therefore deemed satisfactory. The polynomial order has to be more carefully selected, as the spatial accuracy is strongly dependent on $\ReyS$ and $\EzeroR$, as shown in \fig\ \ref{fig:res_dt}(c), \ref{fig:res_dt}(d). Discrepancies within chaotic regions cannot reasonably be avoided, although the trajectories thereafter match well. A polynomial order of $\Np=19$ is sufficient for smaller initial energies (all $\rrc$), and either $\Np=23$ ($\rrc=0.293$ or $0.585$) or $\Np=29$ ($\rrc=1.463$) for larger initial energies, based on resolution testing approximately 40 different $\ReyS-\EzeroR$ combinations. 


\subsection{Delineation energy}\label{sec:nln_rsts}

\begin{figure}
\begin{center}
\addtolength{\extrarowheight}{-10pt}
\addtolength{\tabcolsep}{-2pt}
\begin{tabular}{ llll }
\makecell{\vspace{26mm} \footnotesize{(a)} \\  \vspace{33mm} \rotatebox{90}{\footnotesize{$\ELD$}}} & \makecell{\includegraphics[width=0.458\textwidth]{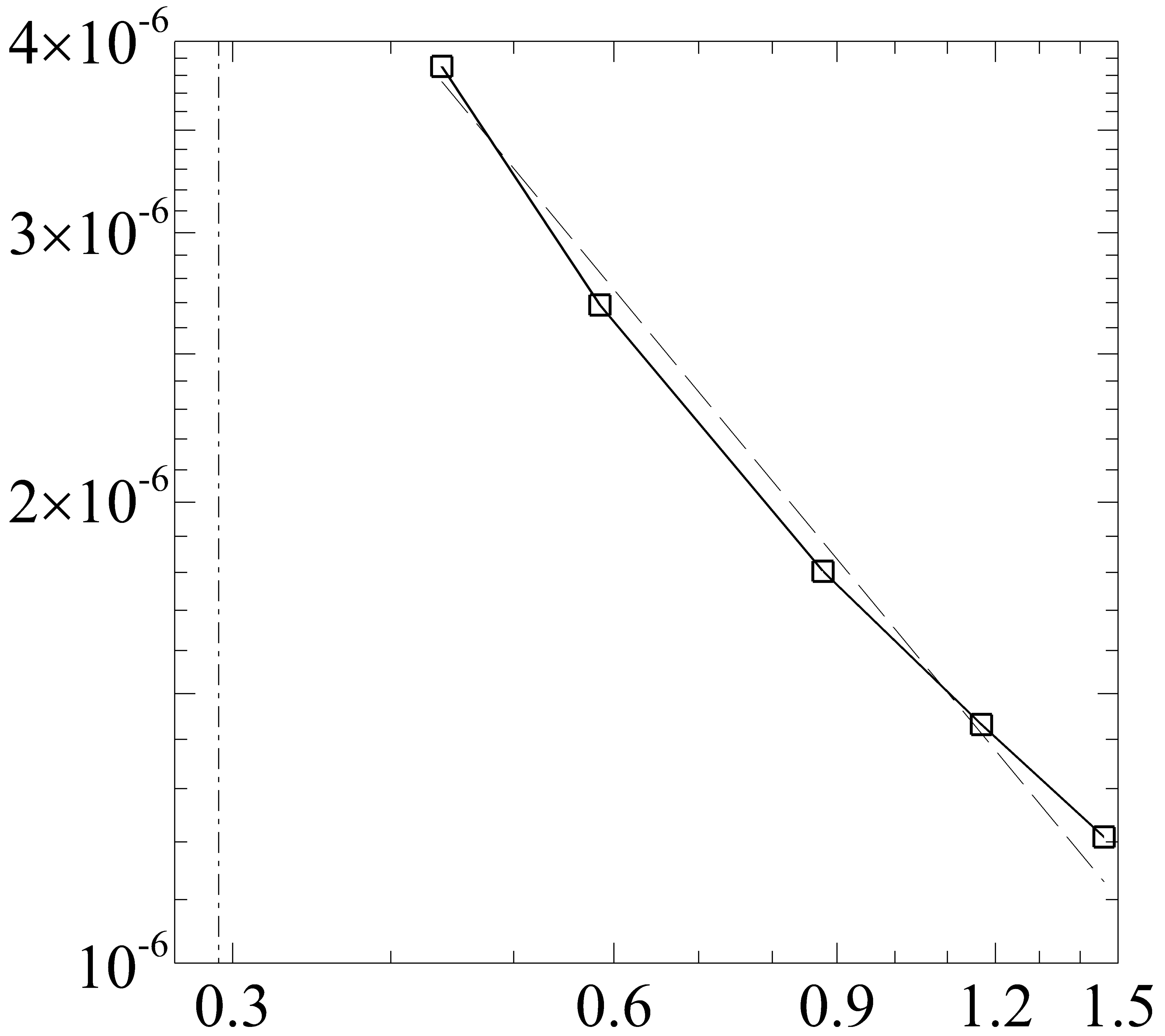}} &
\makecell{\vspace{26mm} \footnotesize{(b)} \\  \vspace{33mm} \rotatebox{90}{\footnotesize{$E$}}}
 & \makecell{\includegraphics[width=0.458\textwidth]{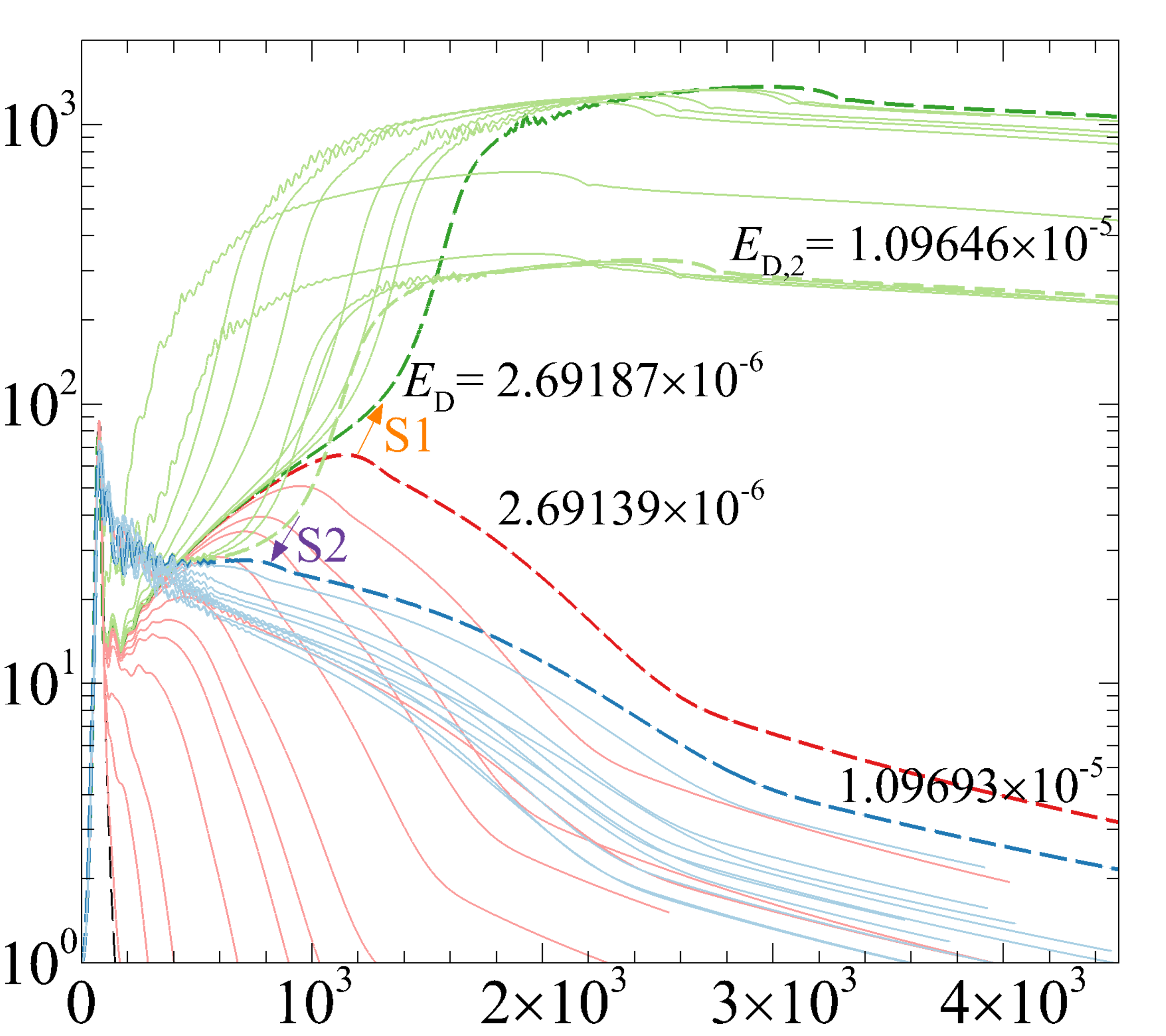}} \\
 & \hspace{42mm} \footnotesize{$\rrc$} & & \hspace{36mm} \footnotesize{$t$} \\
\end{tabular}
\begin{tabular}{ ll }
\makecell{\vspace{26mm} \footnotesize{(c)} \\  \vspace{33mm}}
 & \makecell{\includegraphics[width=0.458\textwidth]{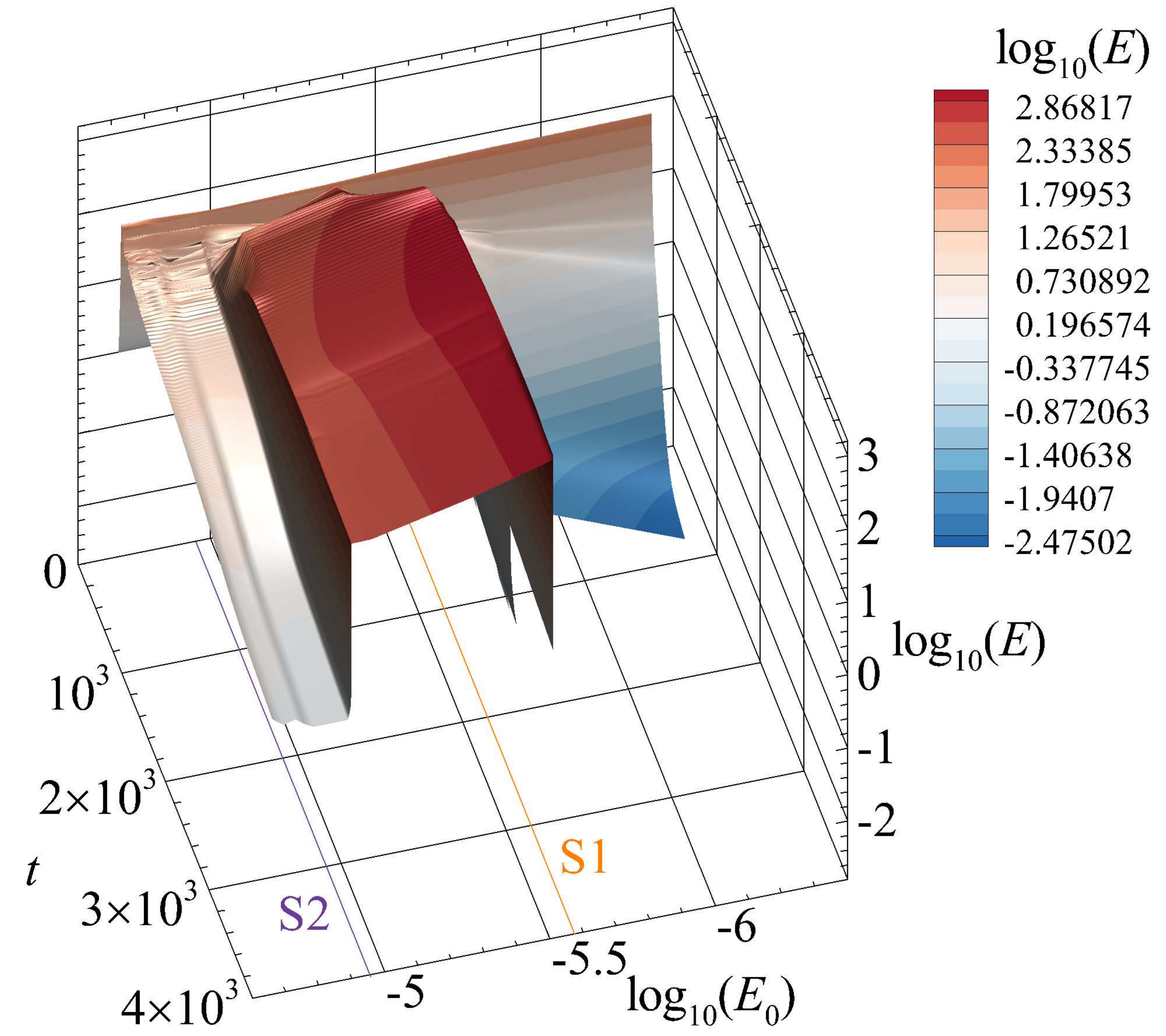}} \\
\end{tabular}
\addtolength{\tabcolsep}{+2pt}
\addtolength{\extrarowheight}{+10pt}
\end{center}
    \caption{
    (a) The lower delineation energy as a function of $\rrc = \ReyS / \ReyCrit$ ($n=1$ domain). The dot-dashed line roughly approximates the maximum $\rrc$ for which the delineation energy is undefined. 
    (b) Energy time histories at $\rrc=0.585$, varying $E_0$. Light red curves with $\Ezero<\ELD$ have a secondary local maximum at best. The orange arrow indicates the switch from local maximum to inflection point, and the lowest initial energy (dashed dark green curve; $\ELD$)  sufficient to cross separatrix 1. All green curves transition to turbulence. The largest initial energy that avoids crossing separatrix 2 ($\EUD$) is also dashed. Light blue curves with $\Ezero>\EUD$, which are briefly chaotic, all cross separatrix 2, with the purple arrow indicating the switch back from an inflection point to a local maximum. All curves are rescaled to start at unity to aid visualization, and the linear curve is denoted with a black long dashed line. At $\rrc=0.585$, $\Gmax = 89.9630$, while the maximum gain at $\Ezero=\ELD$ exceeds $10^3$. (c) Same results as (b), except depicted as a 3D surface, to accentuate the discontinuous changes at the separatrices.}
    \label{fig:amp_delin}
\end{figure}

The nonlinear evolution of linear optimal perturbations in domains with lengths based on $n=1$ repetitions of $\lxopt$ are considered first. The lower delineation energy $\ELD$, representing separatrix 1, is shown in \fig\ \ref{fig:amp_delin}(a) as a function of Reynolds number. \Figs\ \ref{fig:amp_delin}(b), (c) demonstrate how the delineation energy is determined at $\rrc=0.585$ ($\ELD=2.69187\times10^{-6}$). $\ELD$ is determined with a bisection method \cite{Duguet2009localized,Beneitez2019edge,Vavaliaris2020optimal}. However, the bisection method is modified as when $E_0 = \ELD$ the energy-time history does not hover about a mean value \cite{Beneitez2019edge}, as the solution is not on the edge of a stable manifold. Furthermore, all turbulent flows eventually relaminarize. Thus, the flow is deemed returning to a laminar state if its energy reaches a secondary local maximum, and is deemed to be turbulent if its energy exhibits a secondary local inflection point. An initial energy between the largest initial energy that remains laminar, and smallest that incurs transition to turbulence, is then tested, and defined as either the new laminar or new turbulent bound. This process is repeated until $\ELD$ is determined to 4 significant figures.

For the $\rrc$ simulated, \fig\ \ref{fig:amp_delin}(a), there is no clear trend in $\ELD$ with $\ReyS$ (the dashed guideline has an $\rrc^{-1}$ trend). A dot-dashed line at $\rrc=0.293$ provides a rough lower estimate for the $\ReyS$ at which no perturbation is capable of reaching the turbulent attractor, with any initial energy (in an $n=1$ domain). At $\rrc=0.293$ nonlinear second-stage growth yielded a maximum in $E$ greater than the initial linear maximum, at best. For $\rrc \leq 0.146$ the linear growth provided the global maximum in $E$. 

A second delineation energy $\EUD=1.09646\times10^{-5}$ could also be defined for $\rrc = 0.585$, denoting seperatrix 2. The bisection method is unchanged, except that now it is the larger initial energy that is considered laminar, and the smaller initial energy that transitions to tuburbulence. Thus, there is only a finite band of initial energies $\ELD\leq\Ezero\leq\EUD$ able to attain a temporary turbulent state. Only perturbations which resemble conventional, linearly grown \TS\ waves were able take advantage of the nonlinear second-stage growth, which appears to be the only subcritical route to high energy turbulent states. This process is disrupted at larger $\EzeroR$, which noticeably distort the perturbation, inducing rapid decay after the linear growth, similar to the discussion in \cite{Budanur2020upper}. These arguments are also supported by additional nonlinear simulations, at $\rrc=0.585$ and $\rrc=1.463$. The initial seeds tested for comparison were the eigenvector field which generates the second largest linear growth in $\tauOpt$, and random noise, in the same size domains and over a wide range of initial energies. In none of these simulations was a \TS\ wave structure observed akin to that necessary to obtain the nonlinear second-stage growth observed in \fig\ \ref{fig:amp_delin}(b). The eigenvector generating the second largest linear growth managed to achieve only very small amounts of nonlinear second-stage growth. Random noise seeds monotonically decayed. Overall, only the eigenvector which generates the largest linear growth was able to transition to turbulence, by virtue of at least an additional order of magnitude of nonlinear growth. It will be shown later that $\ELD$ does not vary with $n$ (for $\rrc \geq 0.439$) but that $\EUD$ does.



\subsection{Temporal evolution of optimals}\label{sec:nln_topo}

\begin{figure}
\begin{center}
\addtolength{\extrarowheight}{-10pt}
\addtolength{\tabcolsep}{-2pt}
\begin{tabular}{ lllll}
\makecell{\\  \vspace{8mm} \rotatebox{90}{\footnotesize{$y$}}} & 
\makecell{\includegraphics[width=0.244\textwidth]{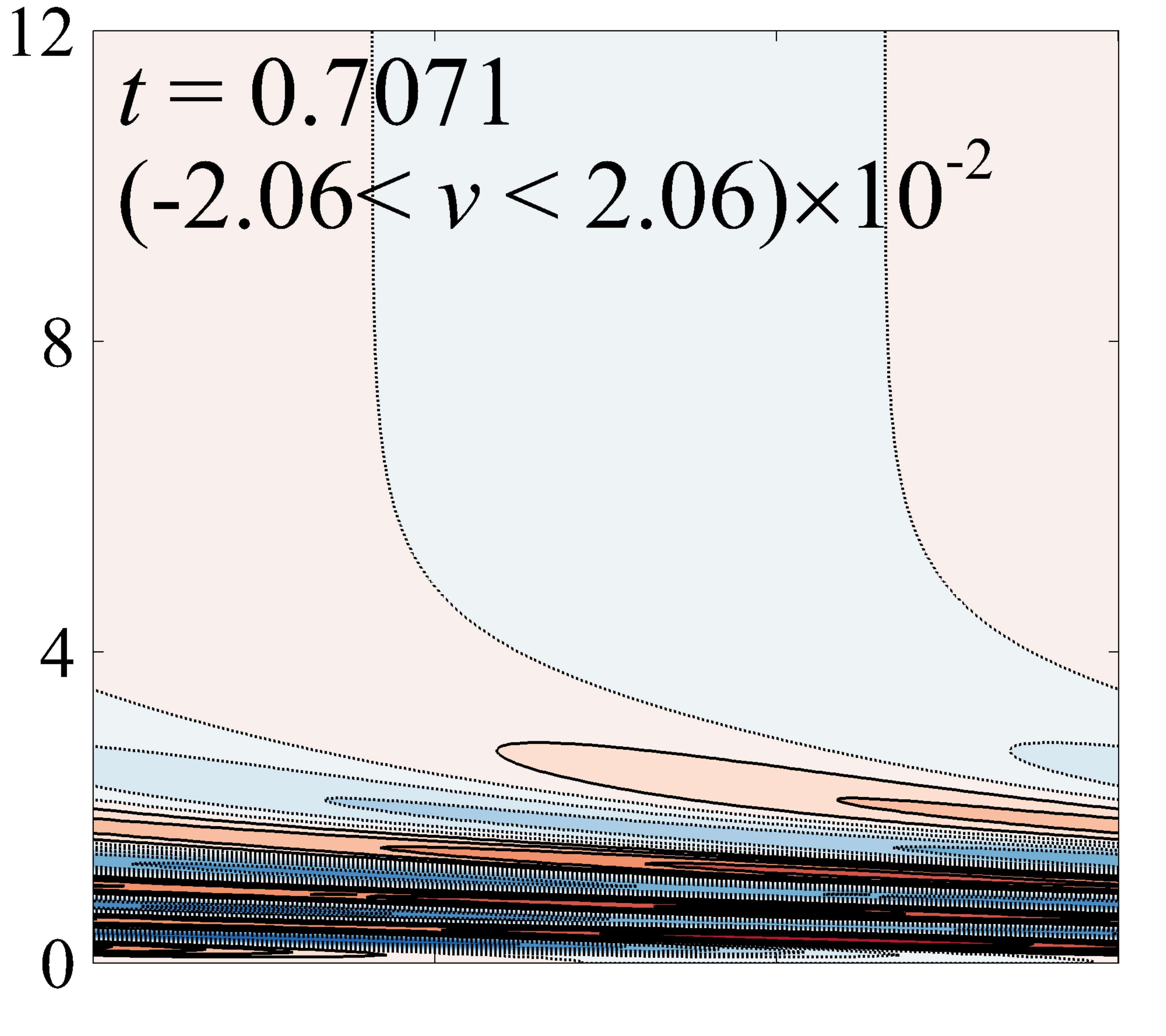}} & 
\makecell{\includegraphics[width=0.244\textwidth]{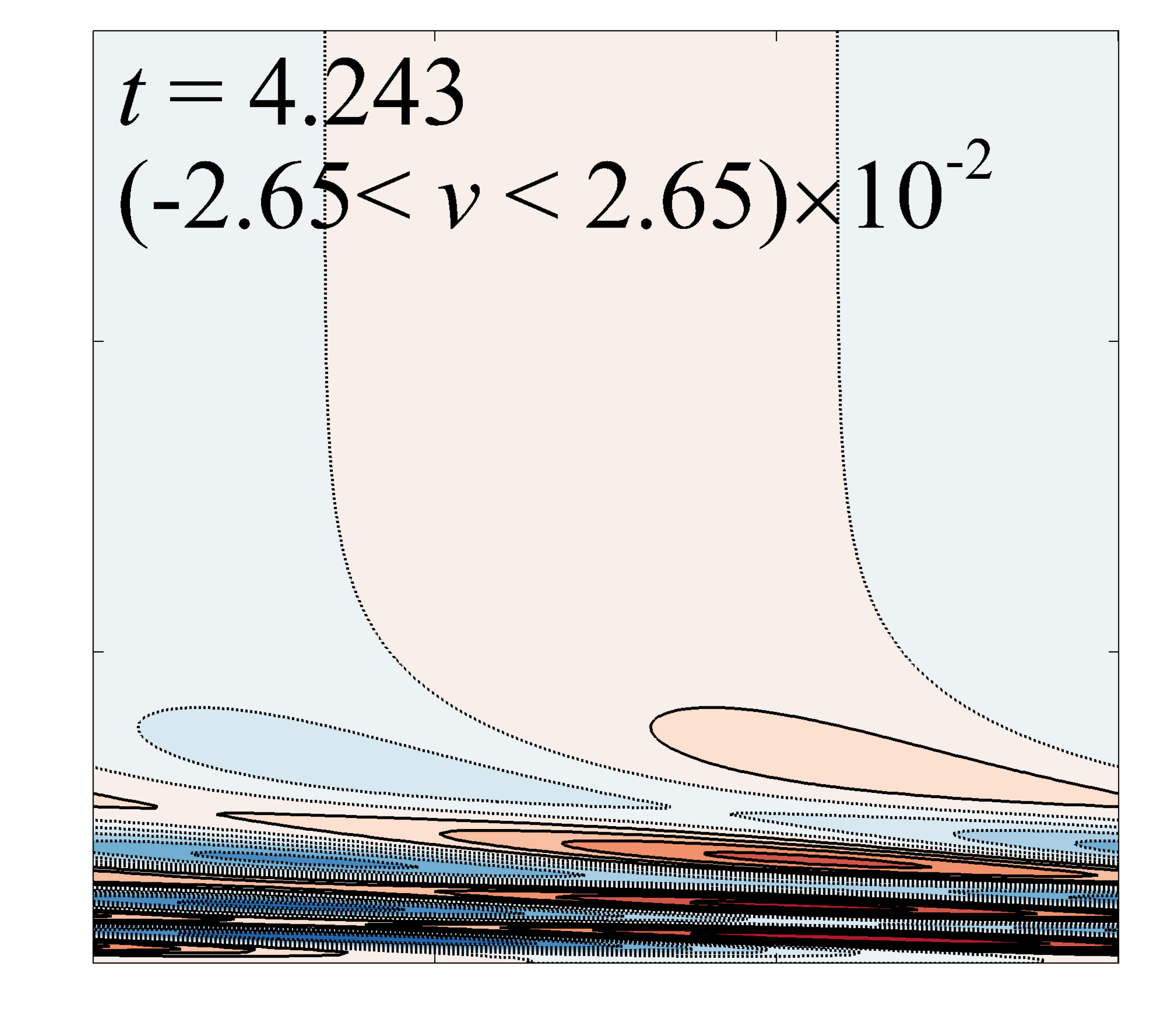}} &
\makecell{\includegraphics[width=0.244\textwidth]{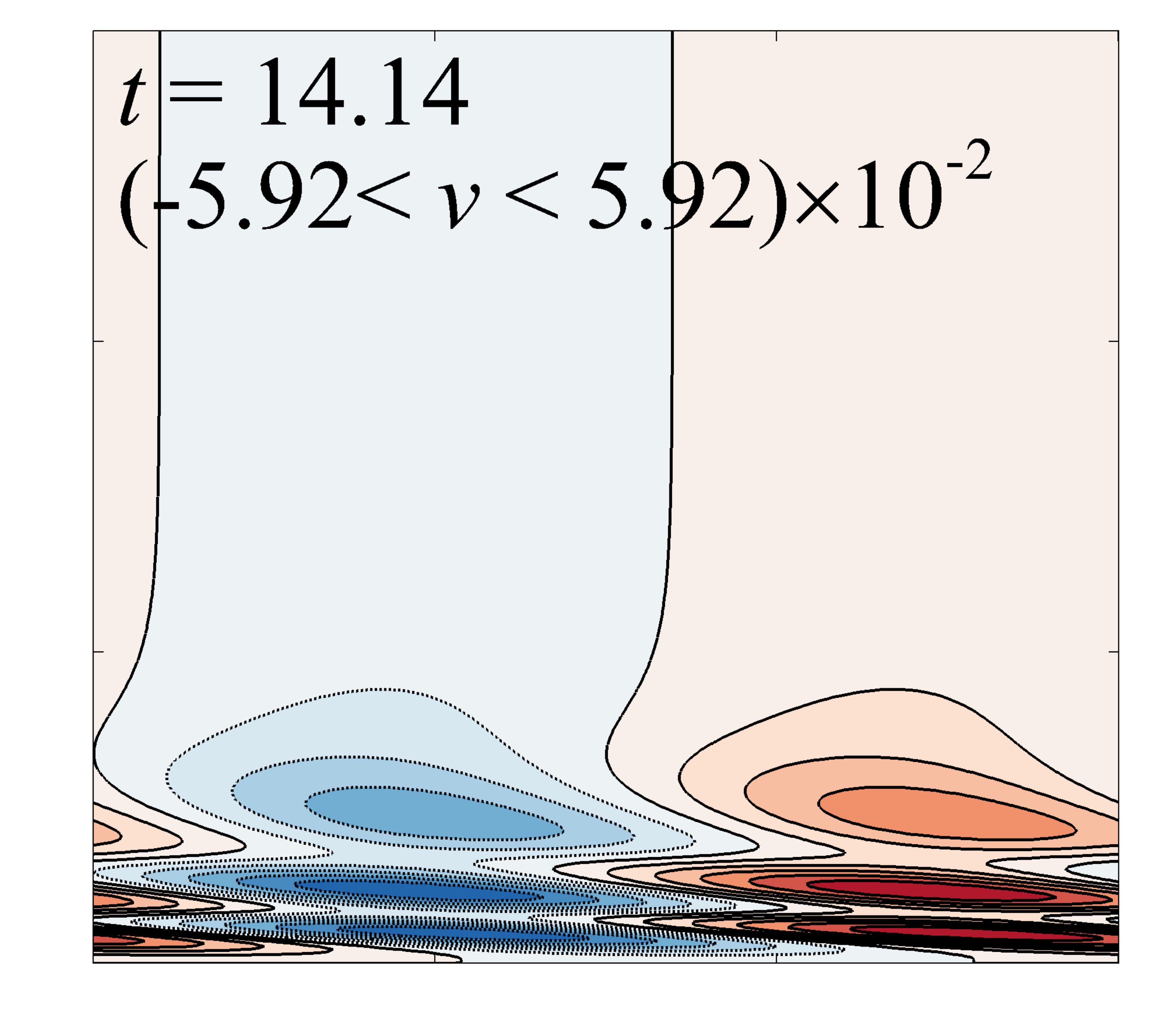}} & 
\makecell{\includegraphics[width=0.244\textwidth]{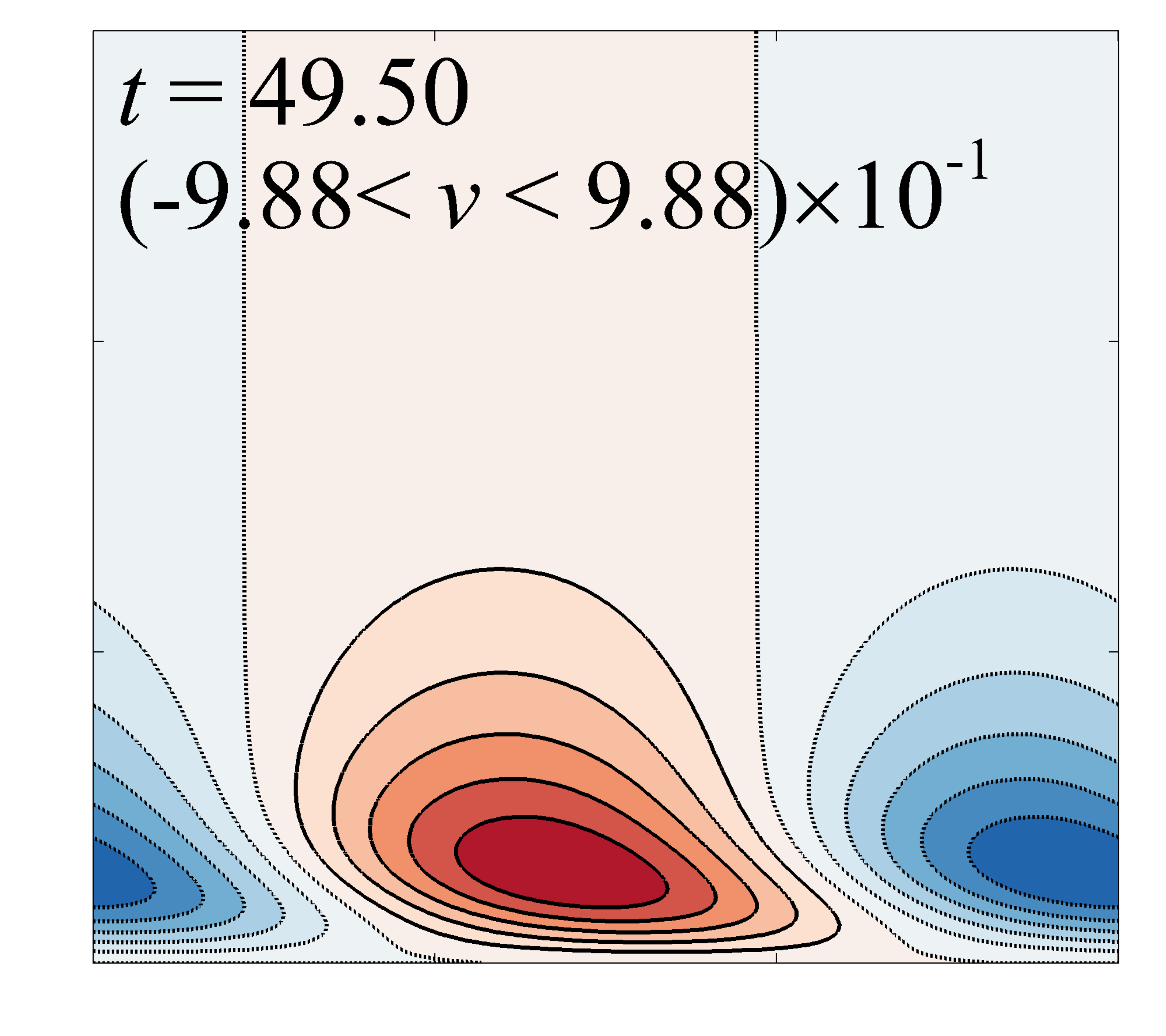}} \\
\makecell{\\  \vspace{8mm} \rotatebox{90}{\footnotesize{$y$}}} & 
\makecell{\includegraphics[width=0.244\textwidth]{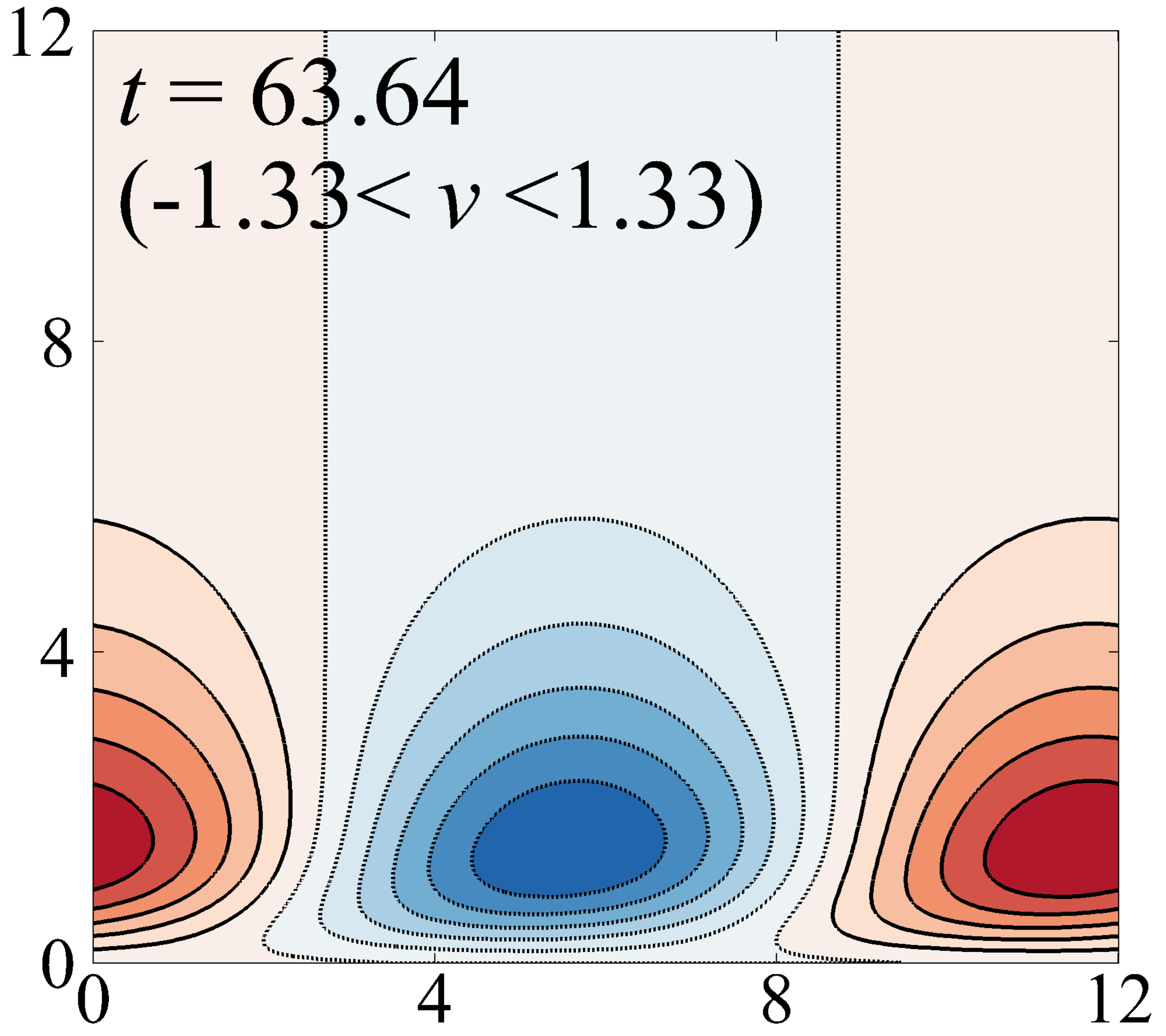}} & 
\makecell{\includegraphics[width=0.244\textwidth]{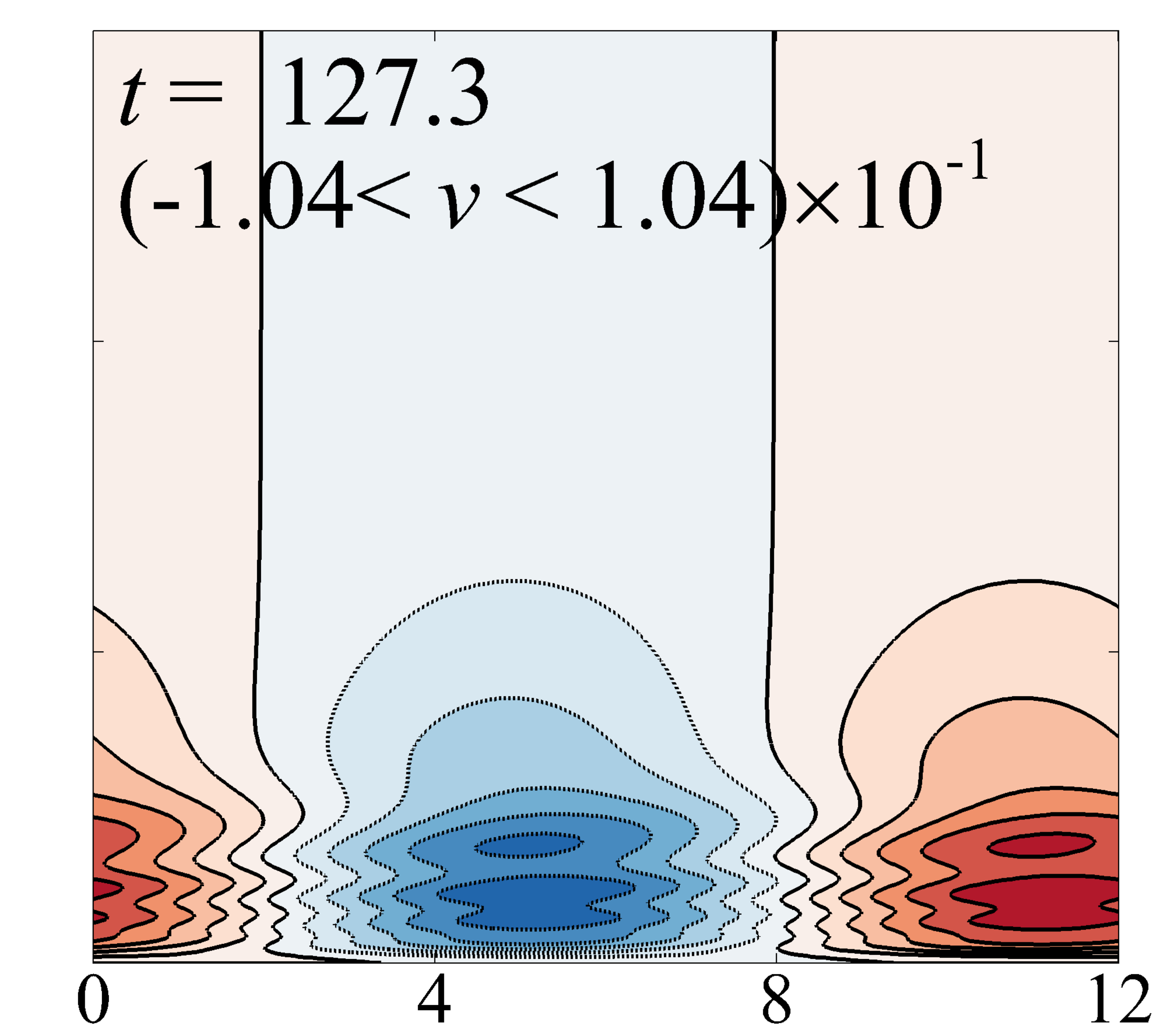}} &
\makecell{\includegraphics[width=0.244\textwidth]{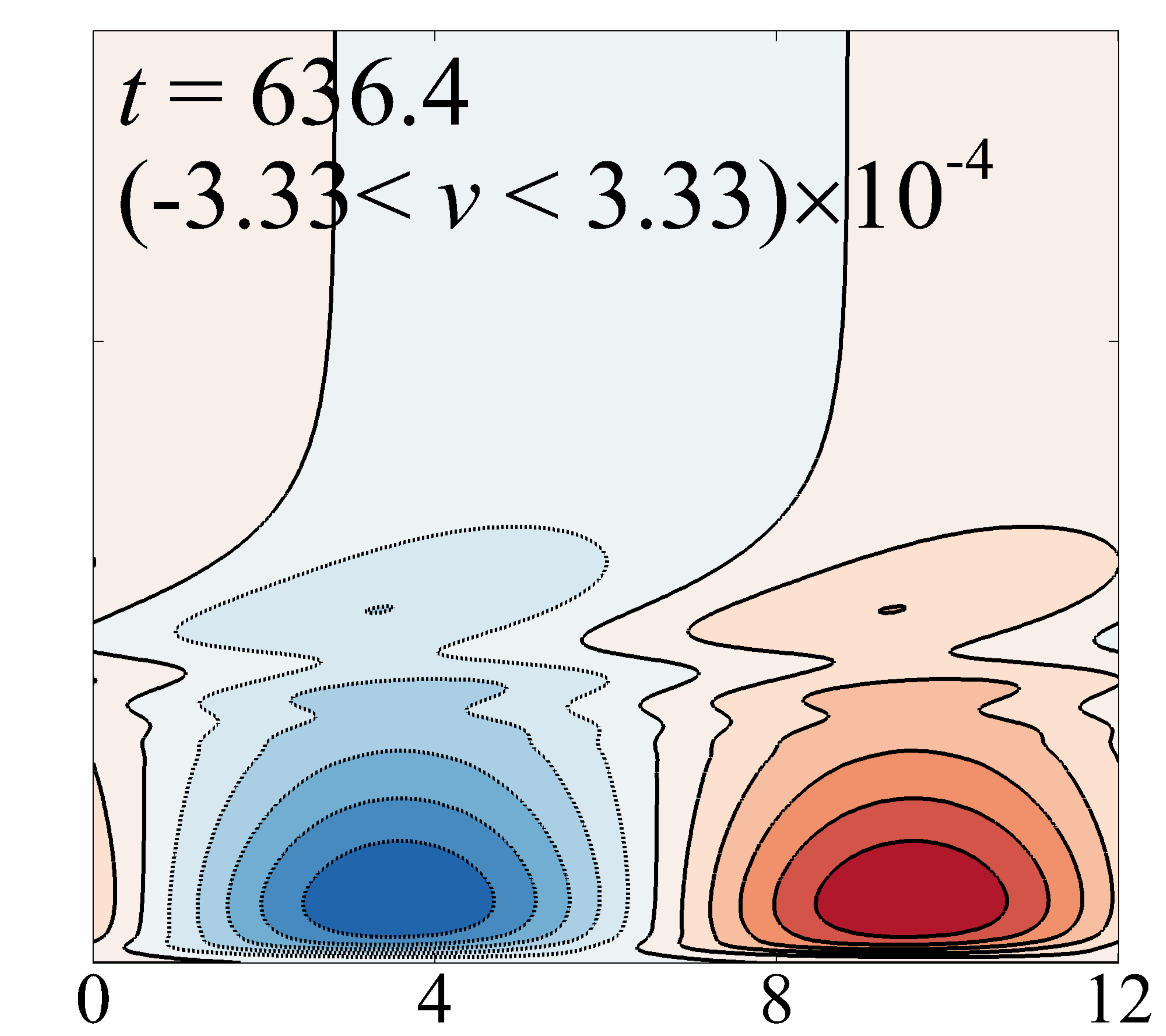}} & 
\makecell{\includegraphics[width=0.244\textwidth]{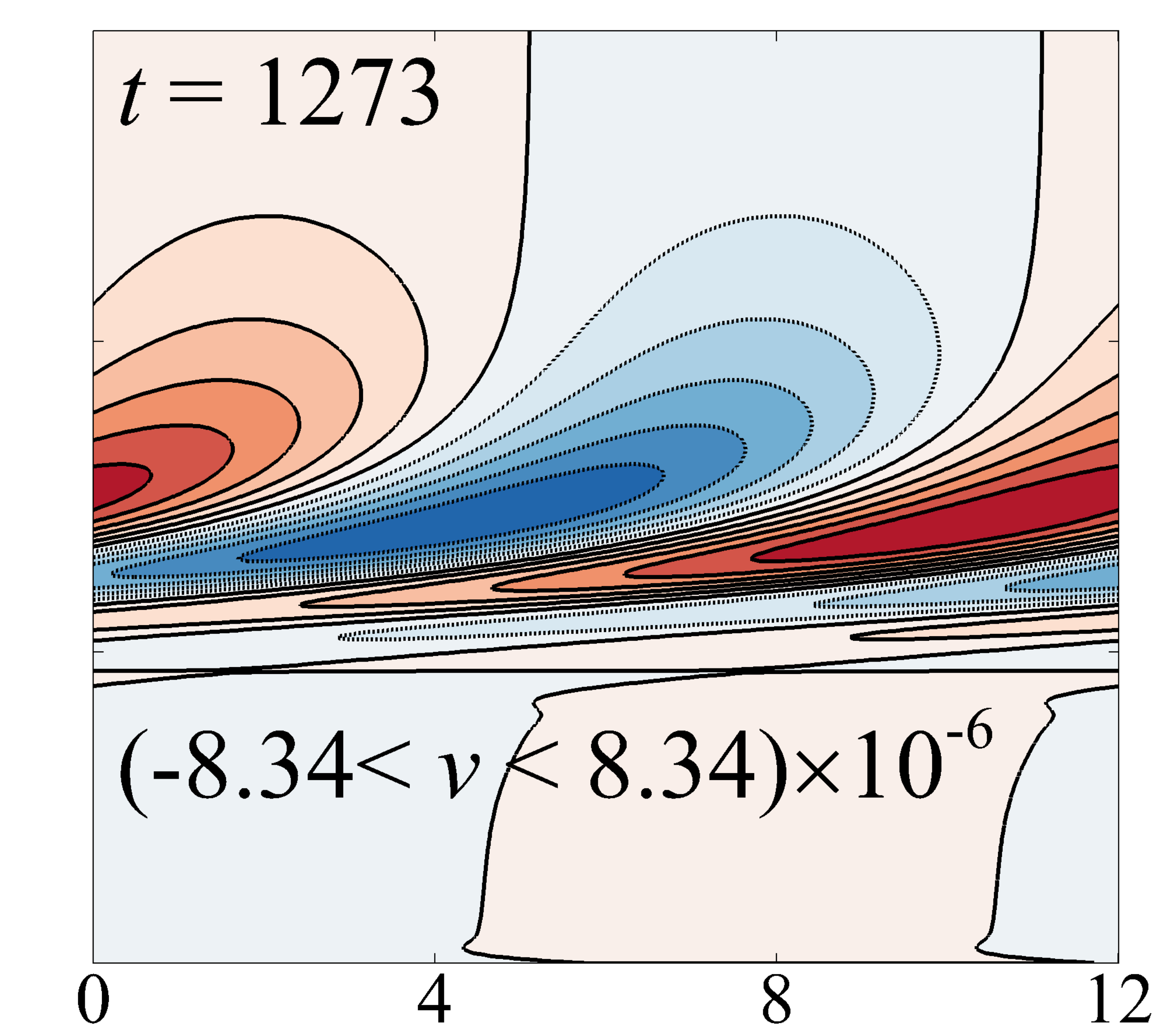}} \\
 & \hspace{20mm} \footnotesize{$x$} & \hspace{20mm} \footnotesize{$x$} & \hspace{20mm} \footnotesize{$x$} & \hspace{20mm} \footnotesize{$x$} \\
\end{tabular}
\addtolength{\tabcolsep}{+2pt}
\addtolength{\extrarowheight}{+10pt}
\end{center}
    \caption{Linearised evolution at $\rrc=0.293$, $\HsubD=28.28$; $\hat{v}$-velocity contours. Solid lines (red flooding) positive; dotted lines (blue flooding) negative.}
    \label{fig:4e4_lin}
\end{figure}

\begin{figure}
\begin{center}
\addtolength{\extrarowheight}{-10pt}
\addtolength{\tabcolsep}{-2pt}
\begin{tabular}{ lllll}
\makecell{\\  \vspace{8mm} \rotatebox{90}{\footnotesize{$y$}}} & 
\makecell{\includegraphics[width=0.244\textwidth]{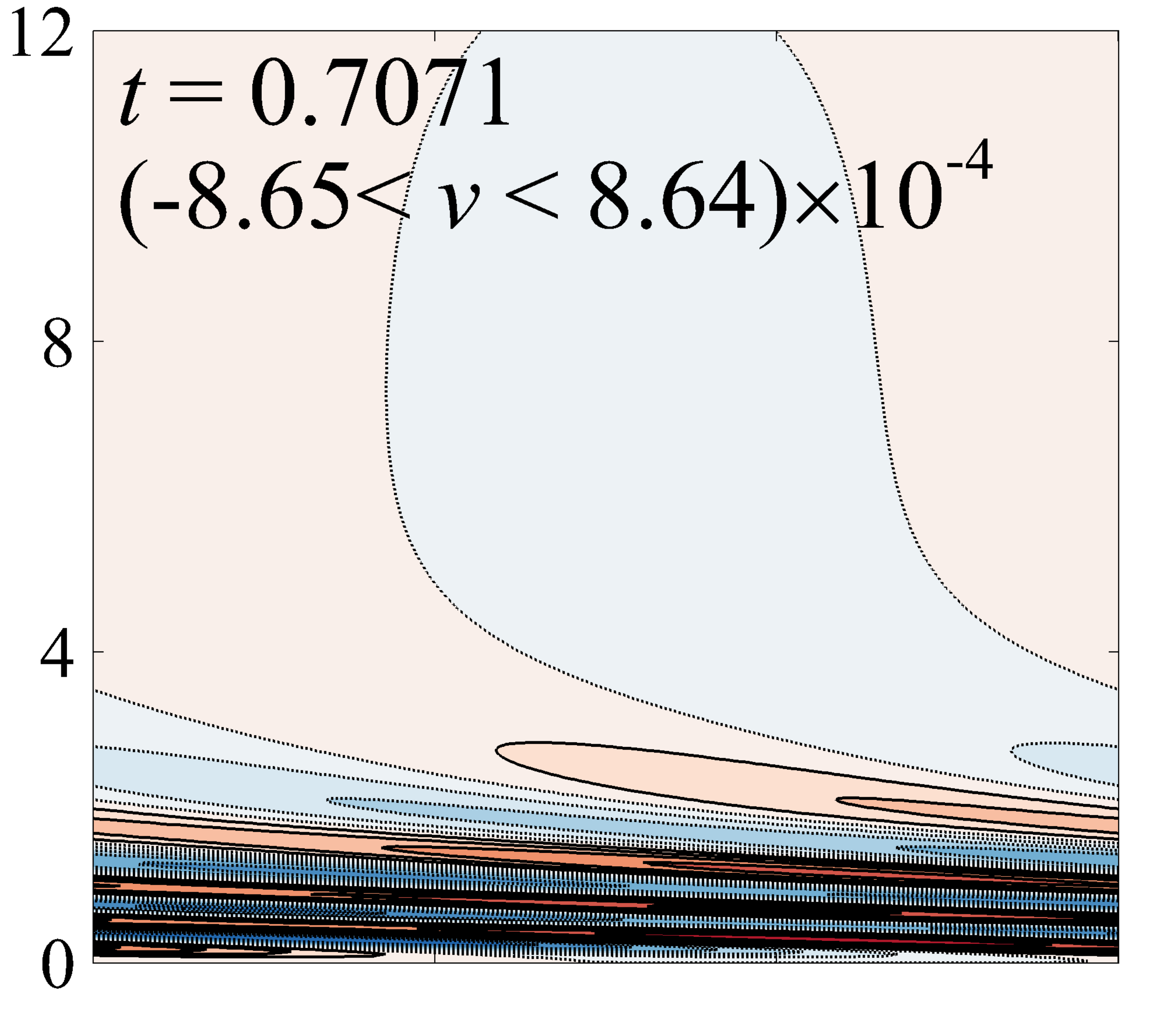}} & 
\makecell{\includegraphics[width=0.244\textwidth]{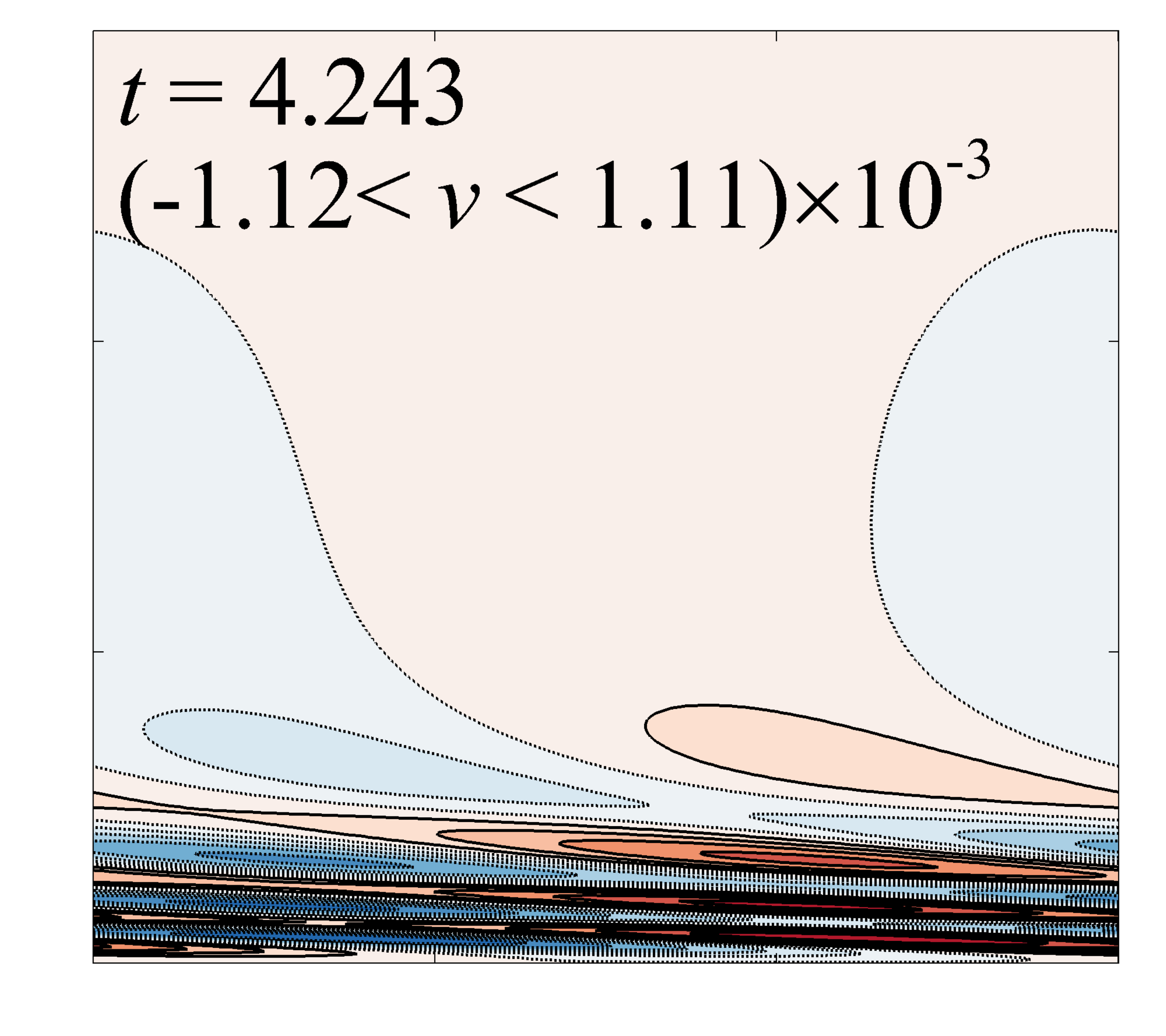}} &
\makecell{\includegraphics[width=0.244\textwidth]{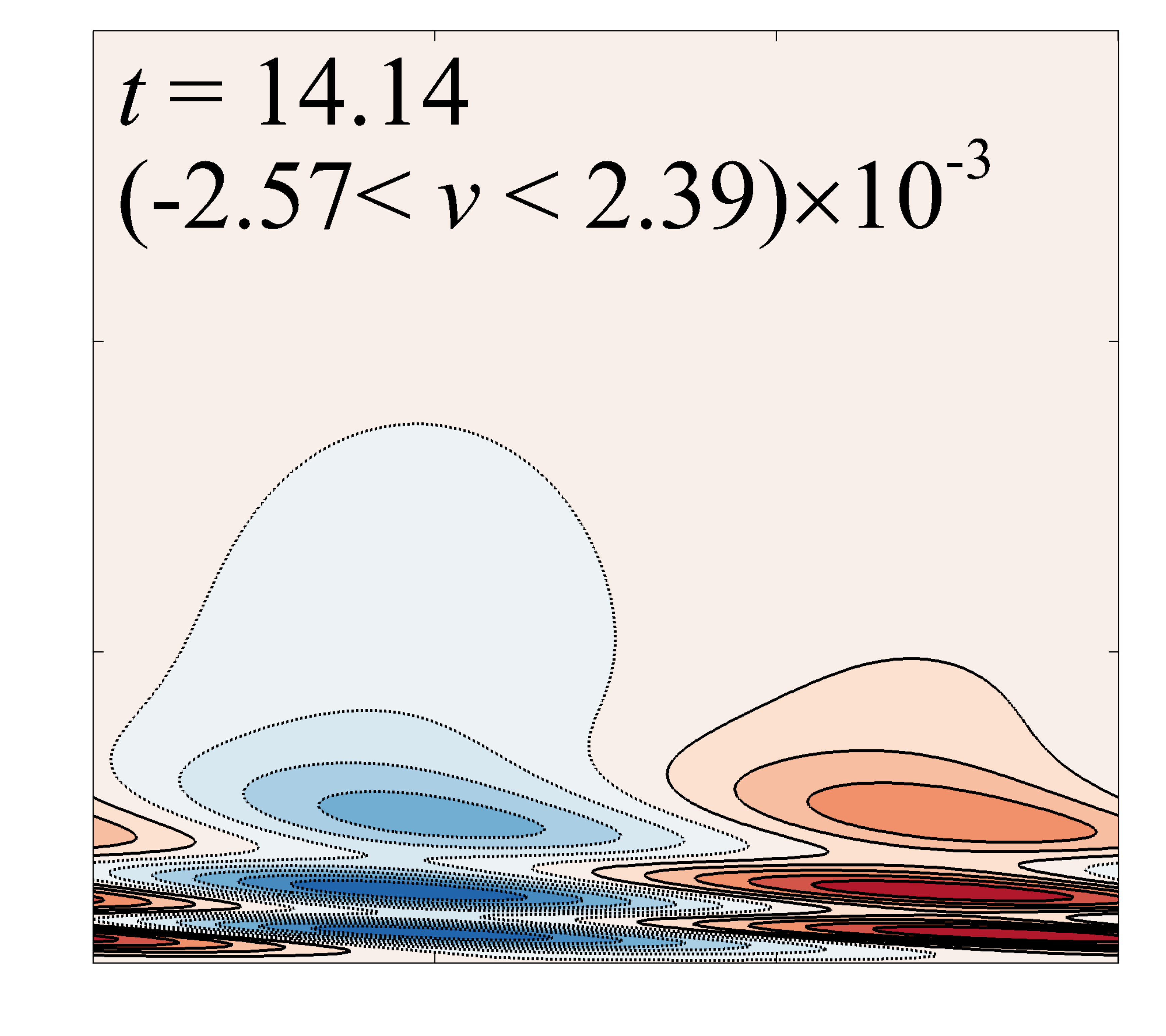}} & 
\makecell{\includegraphics[width=0.244\textwidth]{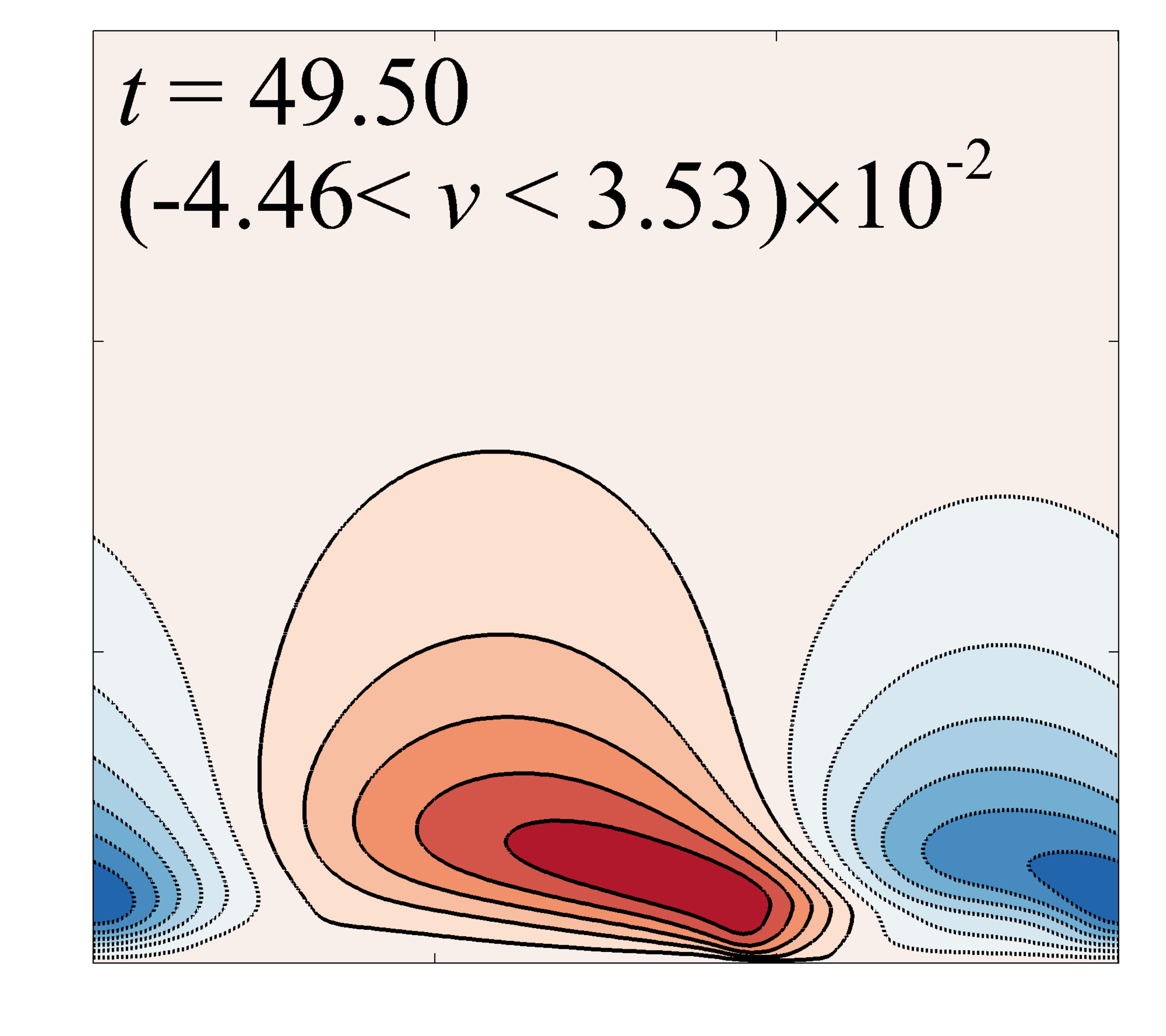}} \\
\makecell{\\  \vspace{8mm} \rotatebox{90}{\footnotesize{$y$}}} & 
\makecell{\includegraphics[width=0.244\textwidth]{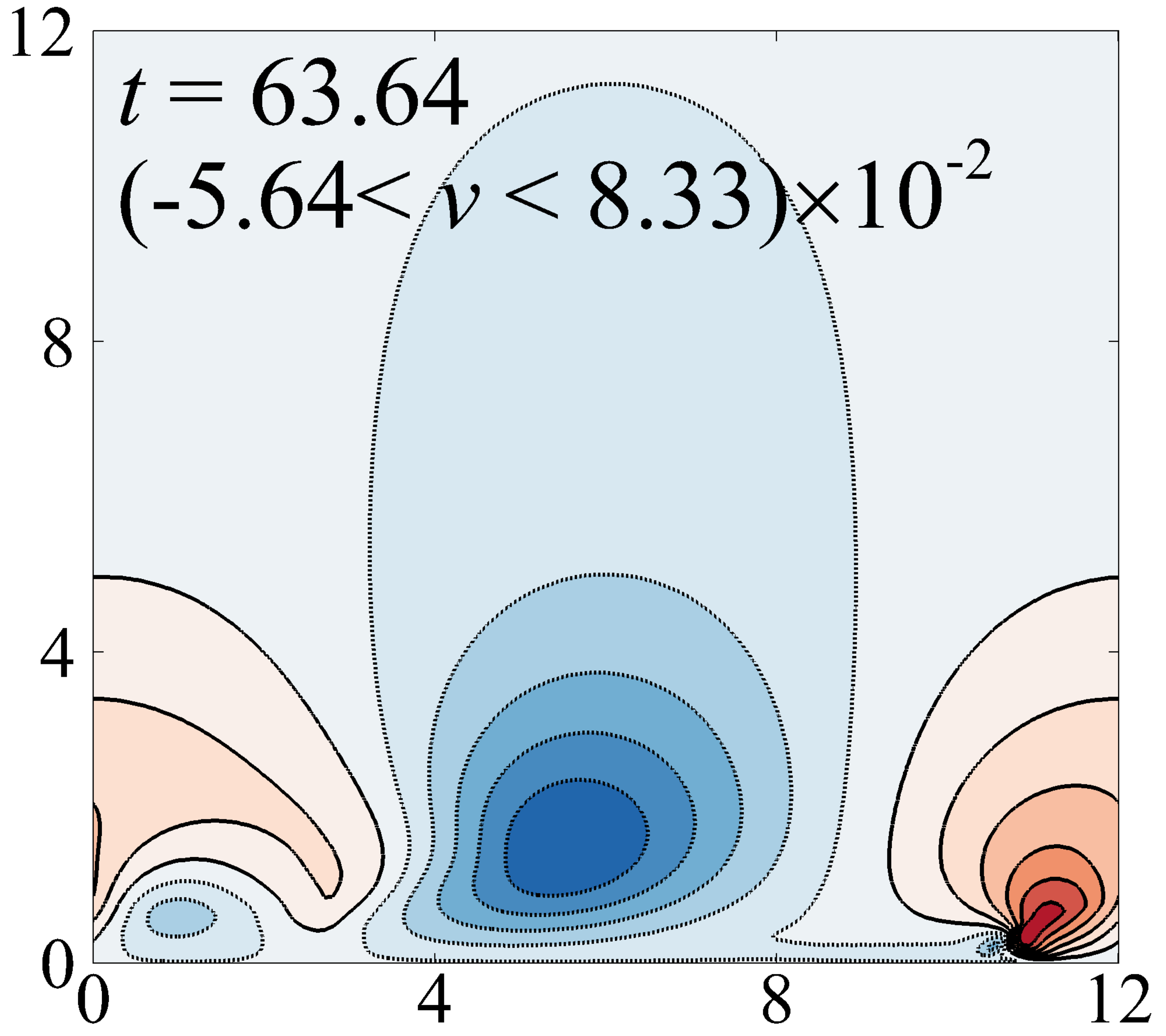}} & 
\makecell{\includegraphics[width=0.244\textwidth]{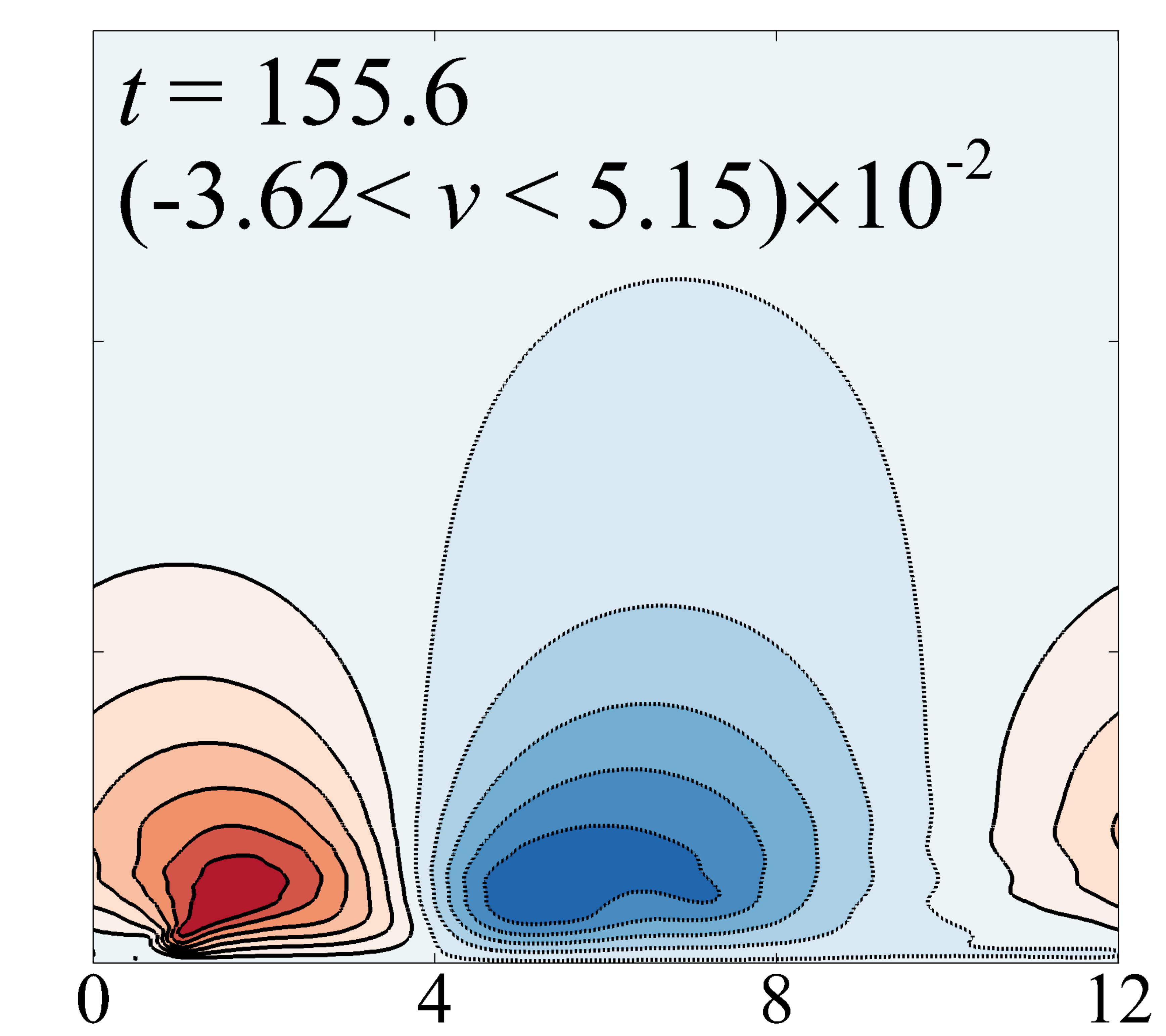}} &
\makecell{\includegraphics[width=0.244\textwidth]{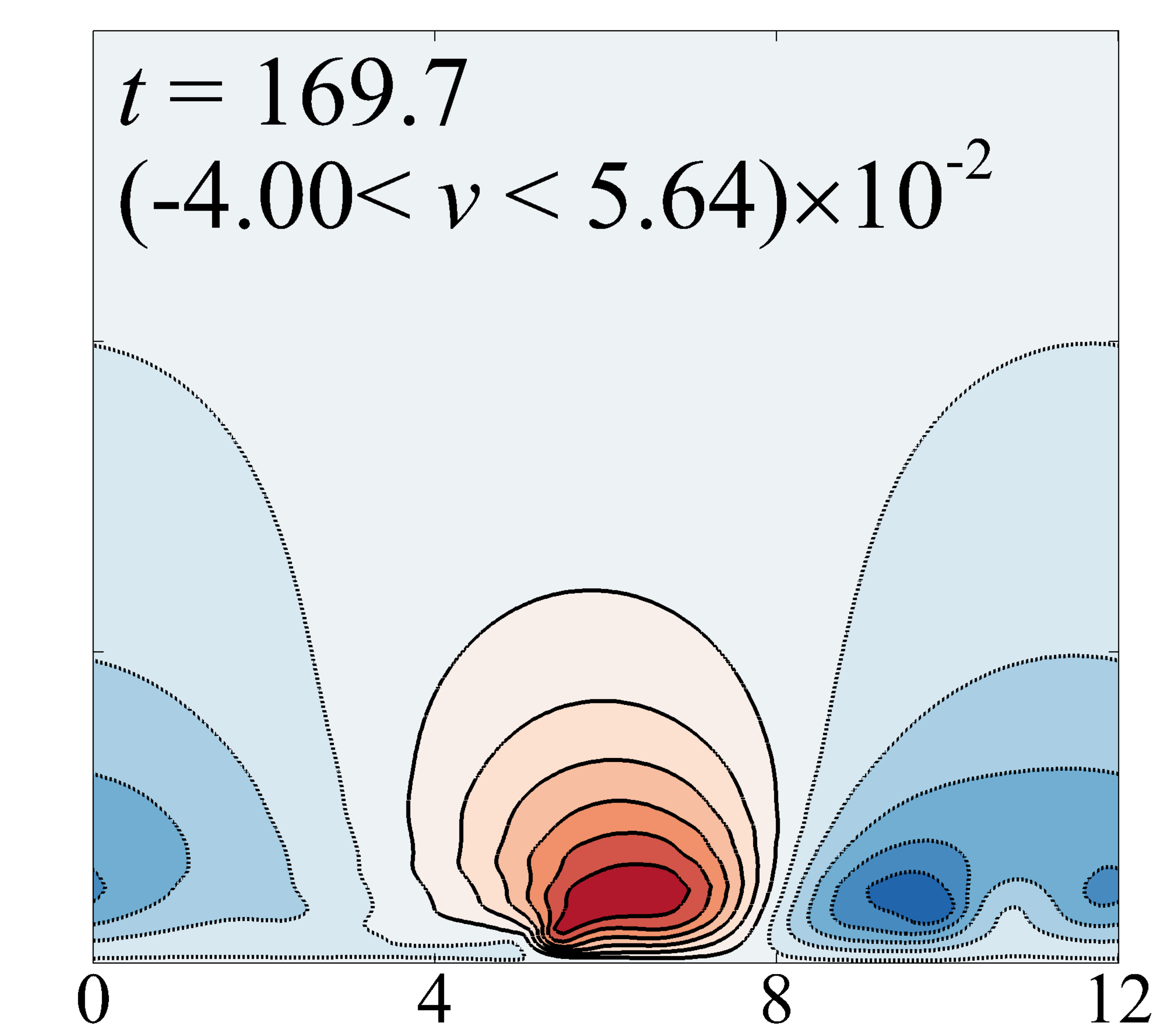}} & 
\makecell{\includegraphics[width=0.244\textwidth]{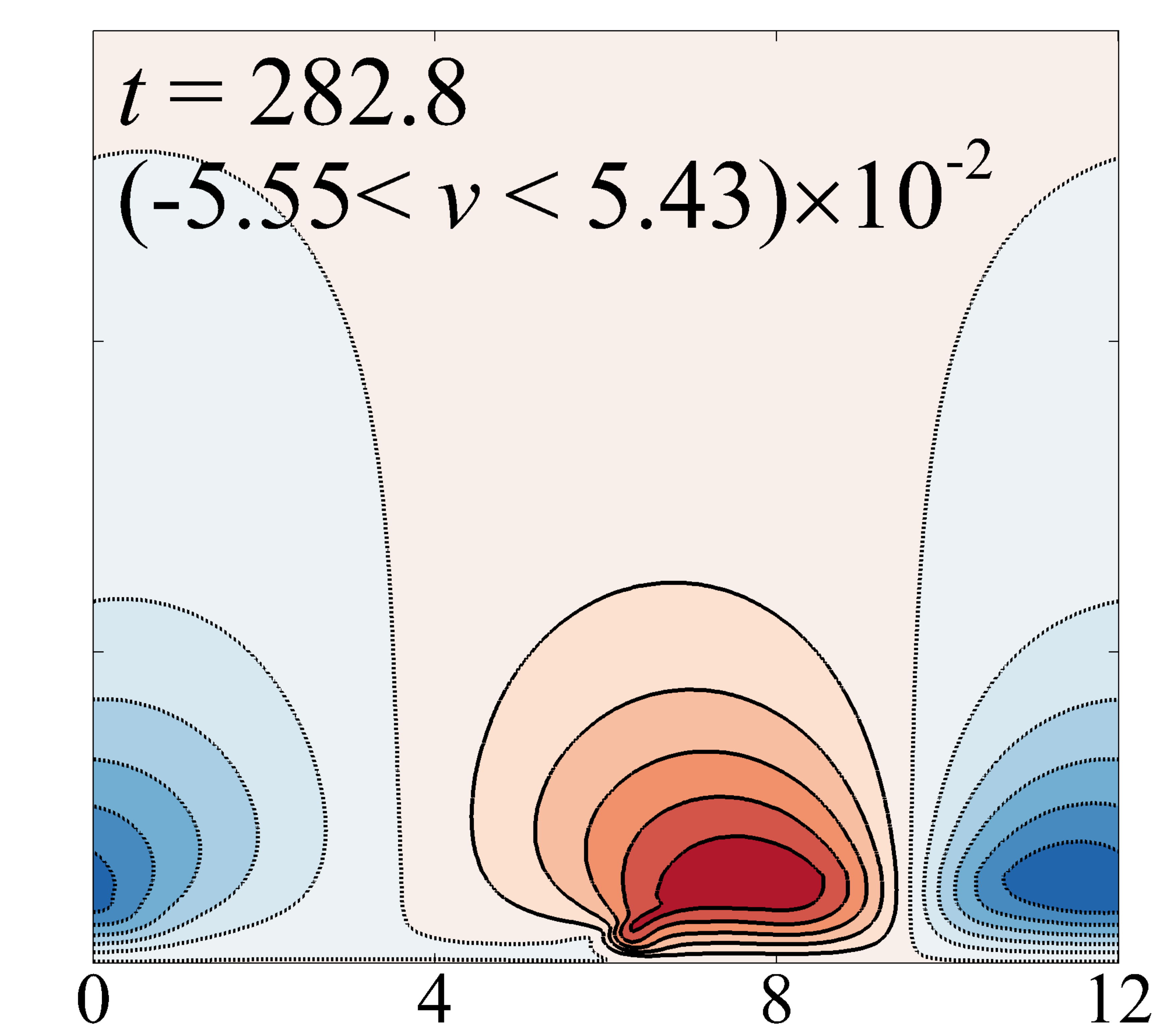}} \\
 & \hspace{20mm} \footnotesize{$x$} & \hspace{20mm} \footnotesize{$x$} & \hspace{20mm} \footnotesize{$x$} & \hspace{20mm} \footnotesize{$x$} \\
\end{tabular}
\addtolength{\tabcolsep}{+2pt}
\addtolength{\extrarowheight}{+10pt}
\end{center}
    \caption{Nonlinear evolution at $\rrc=0.293$, $\HsubD=28.28$, $\EzeroR = 1.10\times10^{-5}$; $\hat{v}$-velocity contours. Solid lines (red flooding) positive; dotted lines (blue flooding) negative.}
    \label{fig:4e4_e1}
\end{figure}

The observable effects of nonlinearity are similar so long as nonlinear second-stage growth occurs and regardless whether $\Ezero>\ELD$, $\Ezero<\ELD$ or if $\ELD$ is even defined ($\rrc = 0.293$). As such, a linearised evolution at $\rrc = 0.293$ is depicted in \fig\ \ref{fig:4e4_lin}, and compared to the corresponding nonlinear evolution at $\EzeroR=1.10\times10^{-5}$ in \fig\ \ref{fig:4e4_e1}. Animations comparing the linear and nonlinear evolutions are also provided as supplementary material \cite{Supvideos2020}. The first relevant differences are discerned at $t=49.50$. The nonlinear evolution shows a mode which appears pinched at the wall, while the linear structure remains flat-bottomed. Following the nonlinear case, as time progresses, the structure rolls over this more slowly moving pinch point. At $t=63.64$, additional localised circulation has appeared near the wall, with a very small region of negative velocity immediately upstream of the pinch point (at $x\sim10.5$). Nonlinear second-stage growth then occurs, as the structure alternates between an arched \TS\ wave ($t=155.6$) and structures which break apart ($t=169.7$) and coalesce into an arched \TS\ wave again ($t=282.8$). After this occurs a few times, the arched \TS\ wave structure retains the form seen at $t=282.8$ for over a thousand times units (see \fig\ \ref{fig:amp_large_time}(b) for the corresponding energy time history), unlike the rapidly decaying linear counterpart. The advecting arched \TS\ wave structure is eventually smoothed out near the wall (online animation only), and finally decays in the same manner as the linear counterpart. The linearised evolution monotonically decays as the structure leans into the mean shear ($t=63.64$). This decay is more rapid for the near wall structure, leaving teardrop-shaped remnants outside the boundary layer as shown at $t=1273$.

\begin{figure}
\begin{center}
\addtolength{\extrarowheight}{-10pt}
\addtolength{\tabcolsep}{-2pt}
\begin{tabular}{ llll }
\makecell{\vspace{26mm} \footnotesize{(a)}  \\  \vspace{33mm} \rotatebox{90}{\footnotesize{$y$}}} & \makecell{\includegraphics[width=0.458\textwidth]{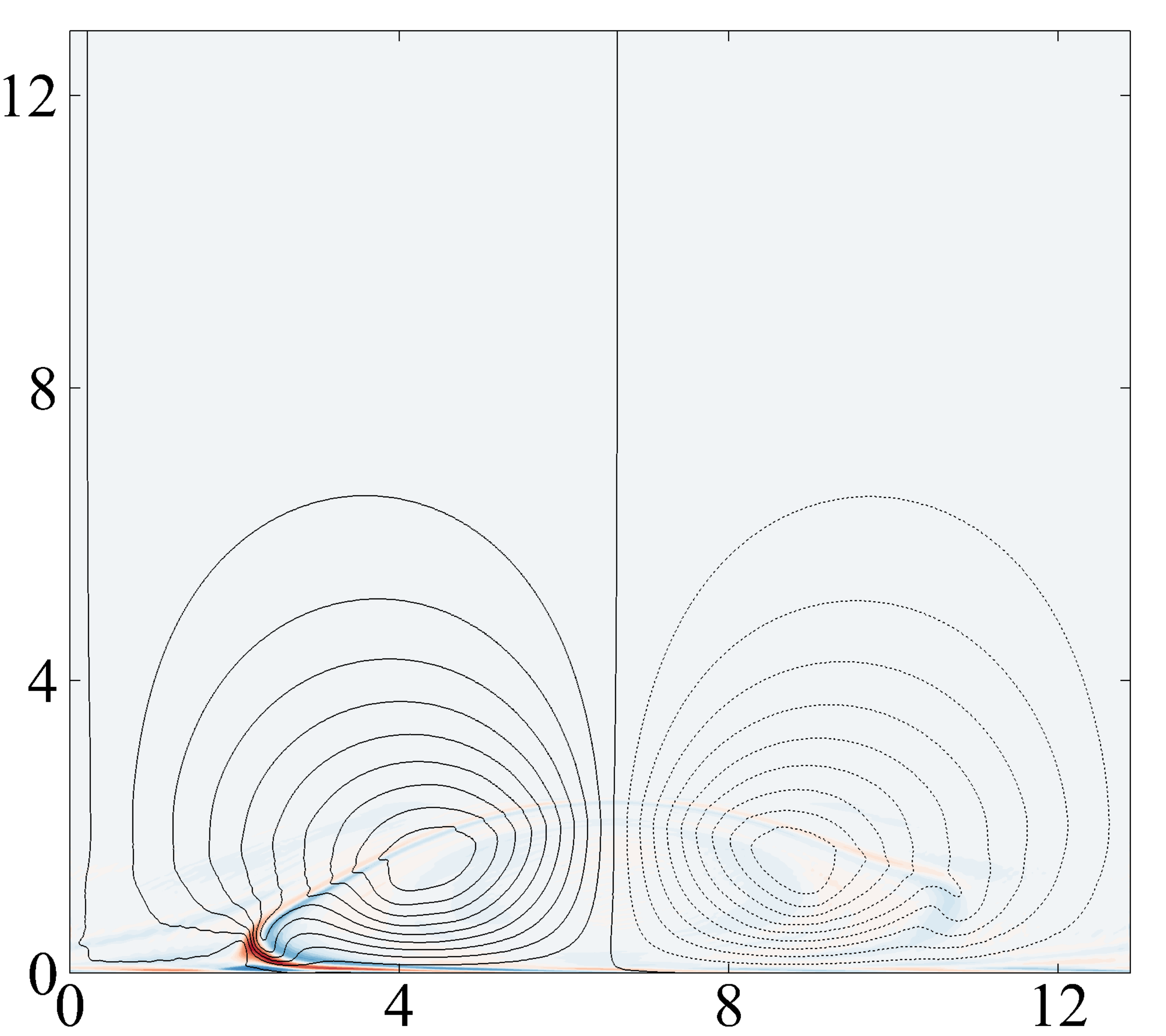}}  & 
\makecell{\vspace{26mm} \footnotesize{(b)}  \\  \vspace{33mm} \rotatebox{90}{\footnotesize{$y$}}} & \makecell{\includegraphics[width=0.458\textwidth]{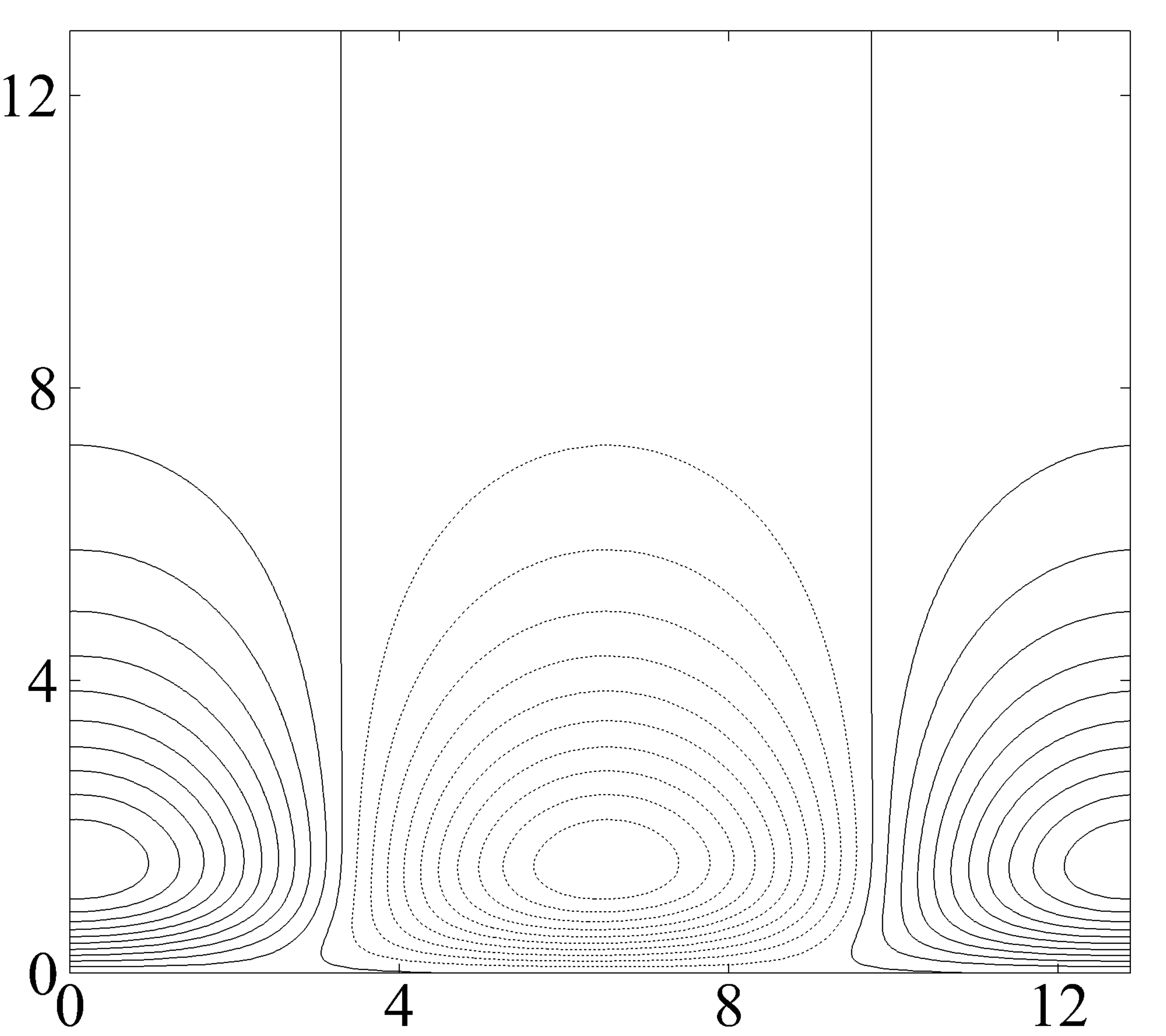}} \\
 & \hspace{36mm} \footnotesize{$x$} & & \hspace{36mm} \footnotesize{$x$} \\
\end{tabular}
\addtolength{\tabcolsep}{+2pt}
\addtolength{\extrarowheight}{+10pt}
\end{center}
    \caption{
    (a) An example of the arched \TS\ wave depicted by the $\hat{v}$-velocity contour lines (solid positive; dotted negative), at $\rrc=0.585$, $E_0 = 2.69187\times10^{-6} > \ELD$, $t=2.121\times10^3$. The underlying backbone of the arch is highlighted by overlaying the high-pass-filtered vorticity $\hat{\omega}_z$, where streamwise Fourier coefficients of modes $\kappa \leq 3$ have been removed. (b) An example of the conventional \TS\ wave from the linear transient growth analysis, at $\rrc=0.585$, $t=77.78$.     
    }
    \label{fig:arch_high}
\end{figure}

The arching of the \TS\ wave appears paramount to the second-stage growth, as flatter \TS\ waves only decay, if outside the neutral curve. An enlarged arched \TS\ wave is shown in \fig\ \ref{fig:arch_high}(a). A high-pass-filtered in-plane vorticity  $\hat{\omega}_z=\partial\hat{v}/\partial x - \partial\hat{u}/\partial y$ is overlaid (streamwise Fourier coefficients of modes $\kappa \leq 3$ have been removed) to help guide the eye along the backbone of the arch, which is a thin, highly sheared layer. The largest vorticity magnitudes are still near the pinch point. To highlight the differences, a conventional \TS\ wave is provided in \fig\ \ref{fig:arch_high}(b), in its upright position, from the linear simulation. The arch is distinctly nonlinear, as the high-pass-filtered vorticity is zero for the conventional, linear \TS\ wave. With increasing time, the conventional \TS\ wave will tilt into the mean shear, whereas the arched \TS\ wave remains upright, and will continue advecting through the domain relatively unchanged.




\subsection{Roles of streamwise and wall-normal velocity components}\label{sec:nln_comp}

\begin{figure}
\begin{center}
\addtolength{\extrarowheight}{-10pt}
\addtolength{\tabcolsep}{-2pt}
\begin{tabular}{ llll }
\makecell{\vspace{26mm} \footnotesize{(a)} \\  \vspace{33mm} \rotatebox{90}{\footnotesize{$\Euv$}}} & \makecell{\includegraphics[width=0.458\textwidth]{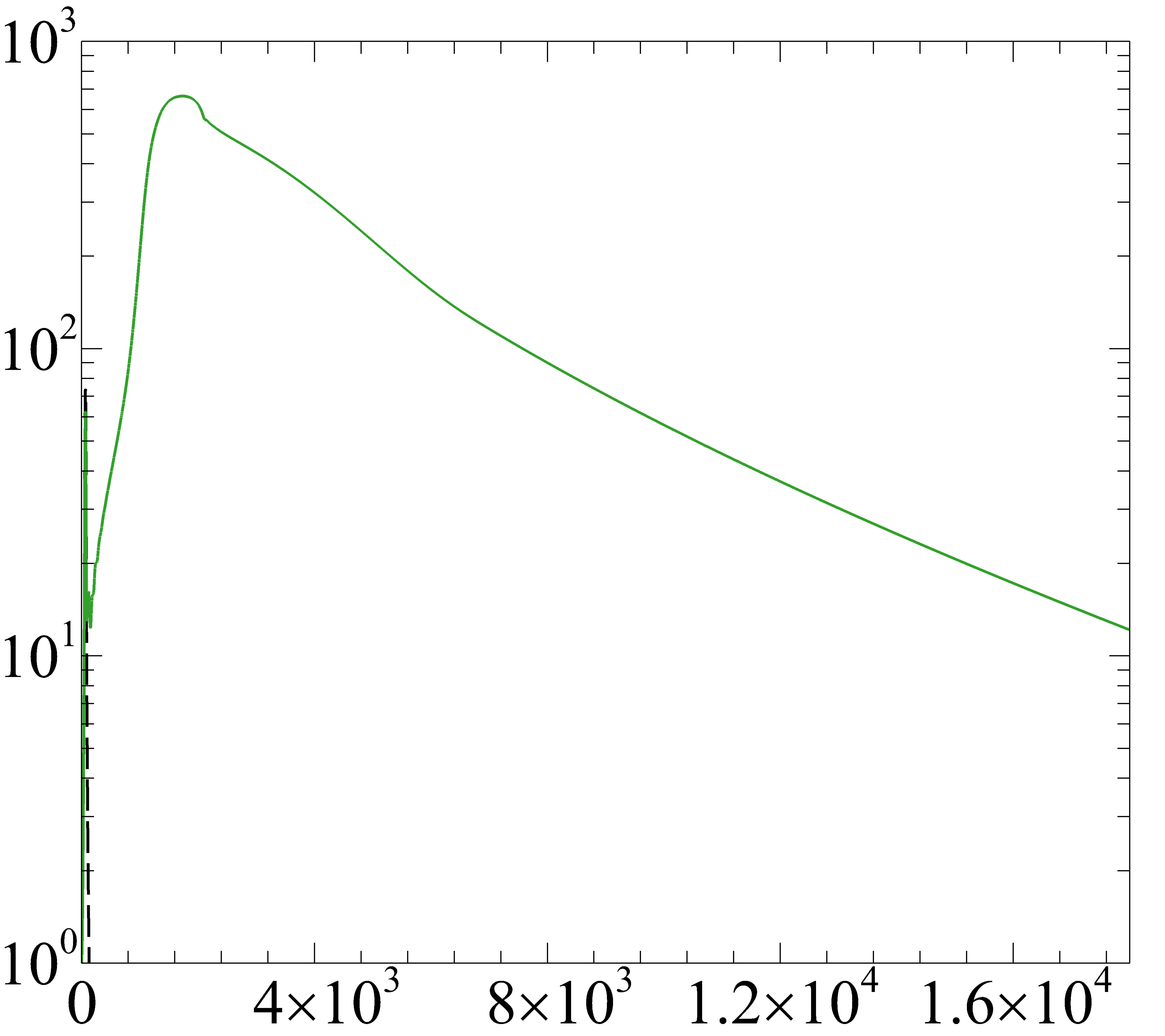}} &
\makecell{\vspace{26mm} \footnotesize{(b)} \\  \vspace{33mm} \rotatebox{90}{\footnotesize{$\Ev$}}}
 & \makecell{\includegraphics[width=0.458\textwidth]{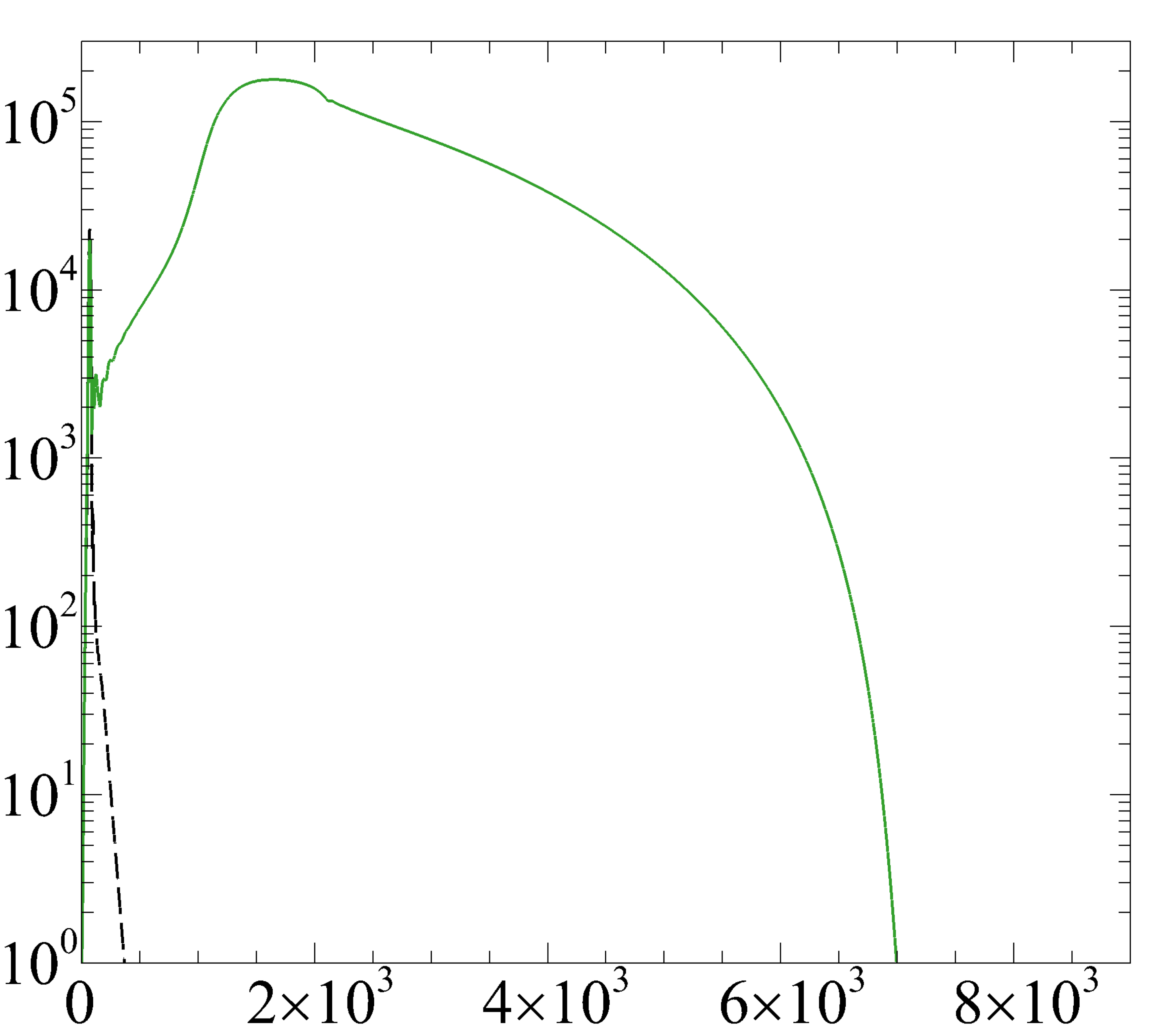}} \\
 & \hspace{36mm} \footnotesize{$t$} & & \hspace{36mm} \footnotesize{$t$} \\
\end{tabular}
\addtolength{\tabcolsep}{+2pt}
\addtolength{\extrarowheight}{+10pt}
\end{center}
    \caption{Energy growth at $\rrc = 0.439$, $\EzeroR = 3.869\times10^{-6} > \ELD$, $n=1$. (a) $\Euv = (1/2)\int \hat{u}^2 + \hat{v}^2 \,\dUP\Omega$. (b) $\Ev =(1/2)\int \hat{v}^2 \,\dUP\Omega$. At $\rrc = 0.439$, $\Gmax = 73.9706$ and $\ELD=3.853\times10^{-6}$. All curves are rescaled to unit initial energy. The linear evolution is shown as a black long dashed line.}
    \label{fig:FF_vs_PG}
\end{figure}

The disturbance is now considered in more detail by separating growth solely in $\Euv$, \fig\ \ref{fig:FF_vs_PG}(a), and $\Ev$, \fig\ \ref{fig:FF_vs_PG}(b), for $\Ezero$ just greater than $\ELD$. Growth appears larger in the latter measure as the wall-normal velocity makes up a smaller fraction of the energy in the initial field. Both $\hat{u}^2$ and $\hat{v}^2$ show noticeable second-stage growth. However, the $\hat{v}$-velocity magnitudes rapidly decrease after the second-stage growth, while the $\hat{u}$-velocity magnitudes, and thus $\Euv$, decrease slowly. 

\begin{figure}
\begin{center}
\addtolength{\extrarowheight}{-10pt}
\addtolength{\tabcolsep}{-2pt}
\begin{tabular}{ llllll }
\makecell{\vspace{15mm} \footnotesize{(a)} \\  \vspace{24mm} \rotatebox{90}{\footnotesize{$y$}}} & \makecell{\includegraphics[width=0.322\textwidth]{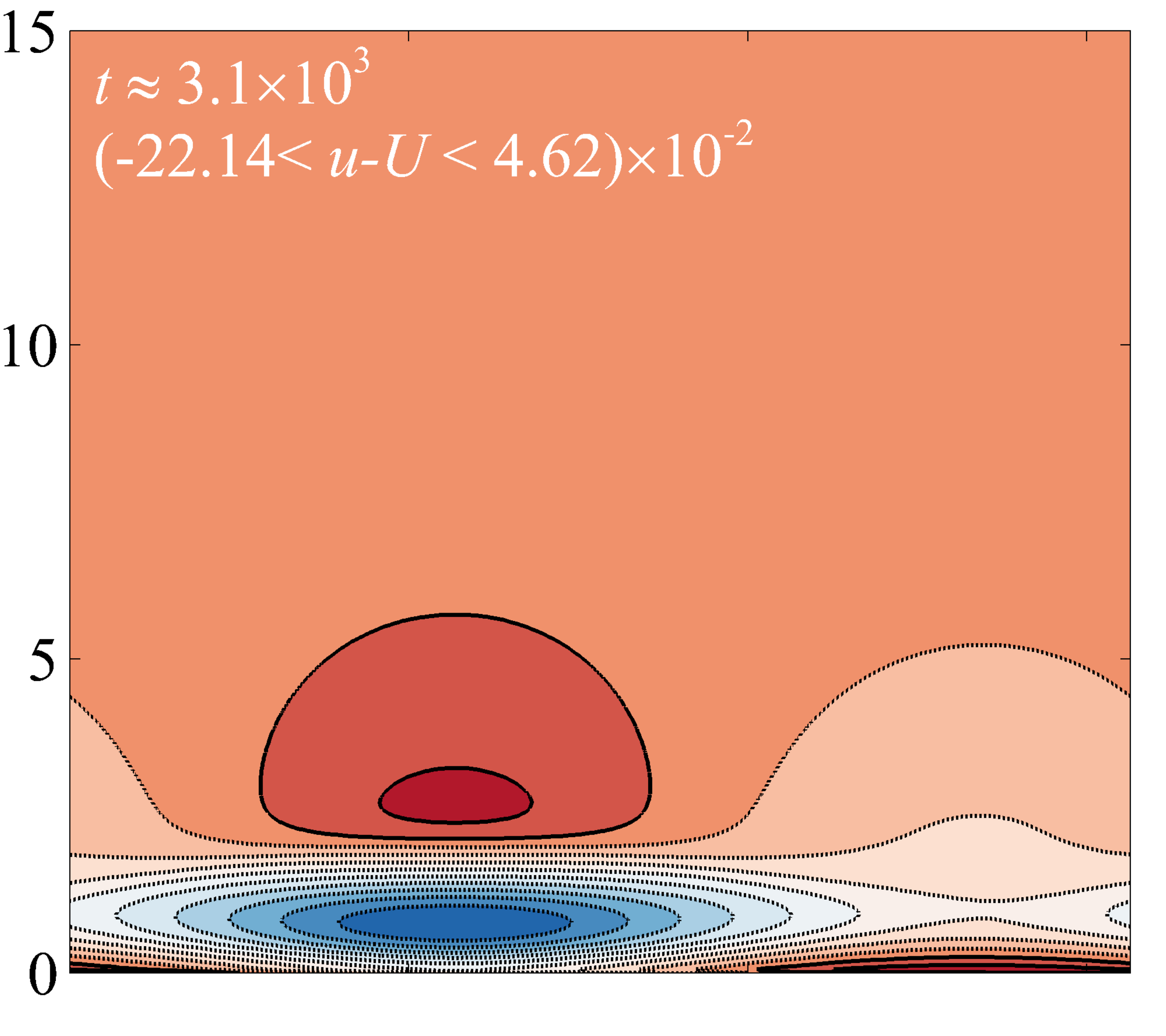}} &
 & \makecell{\includegraphics[width=0.322\textwidth]{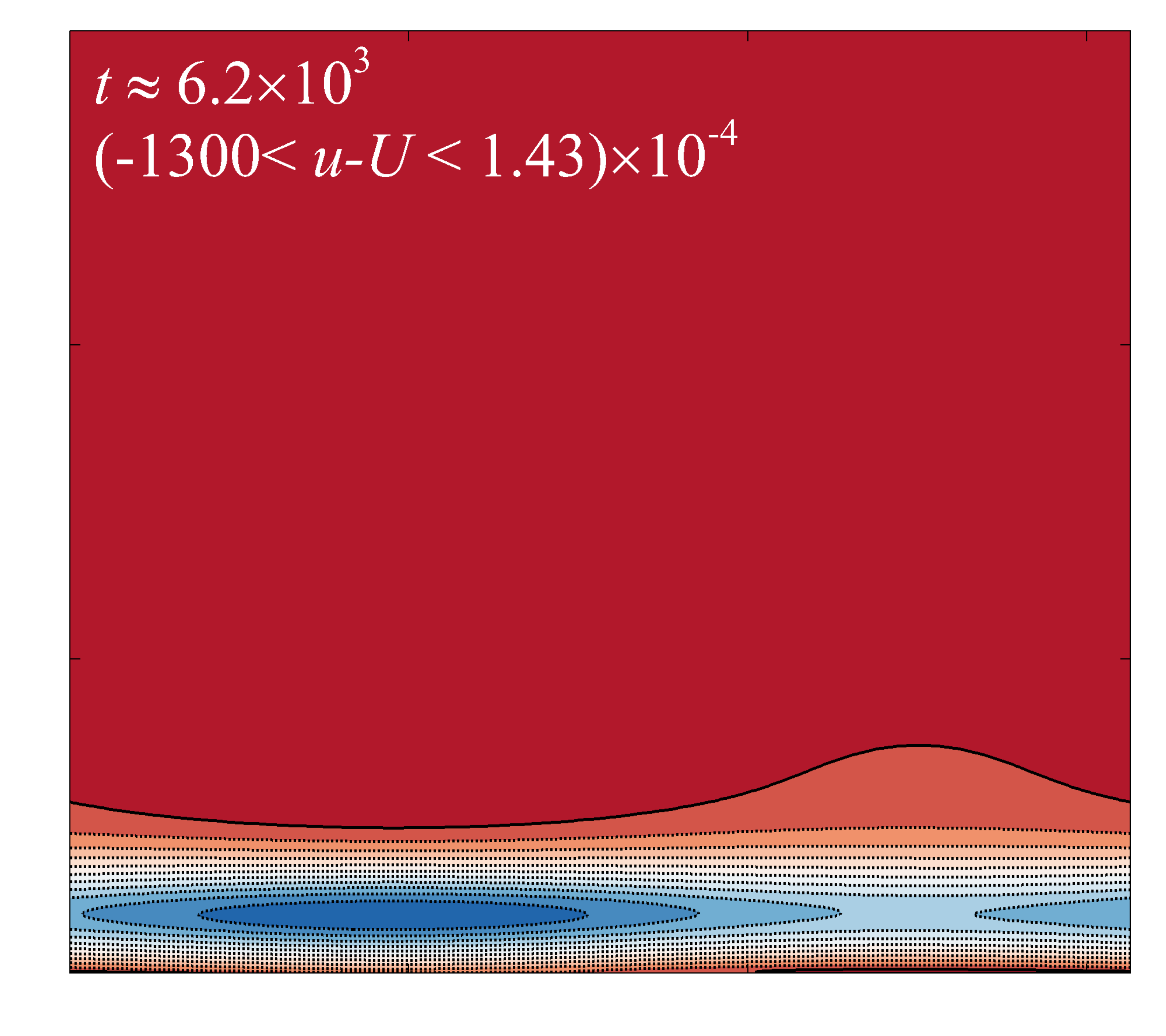}} &
 & \makecell{\includegraphics[width=0.322\textwidth]{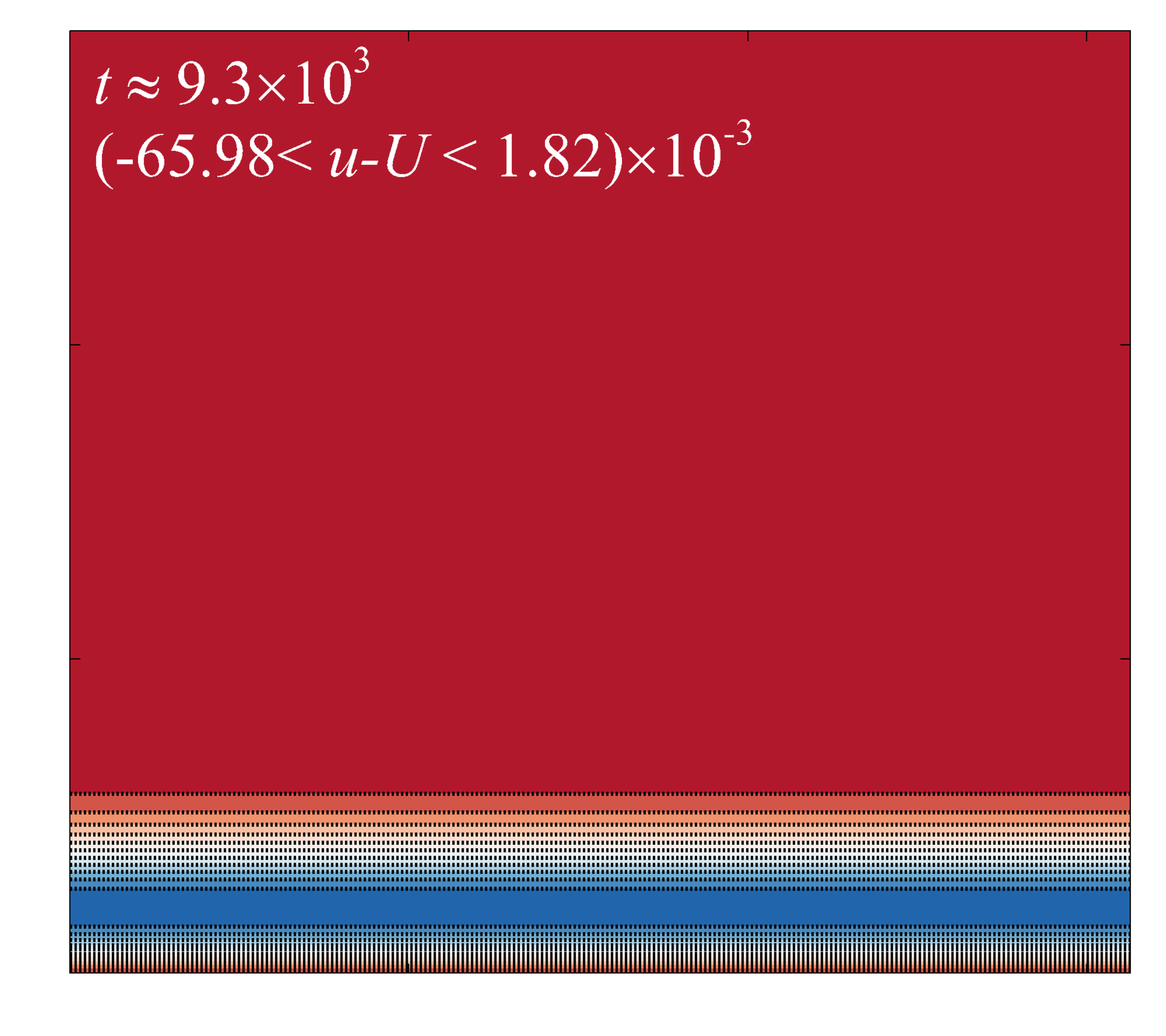}} \\
 \makecell{\vspace{15mm} \footnotesize{(b)} \\  \vspace{24mm} \rotatebox{90}{\footnotesize{$y$}}} & \makecell{\includegraphics[width=0.322\textwidth]{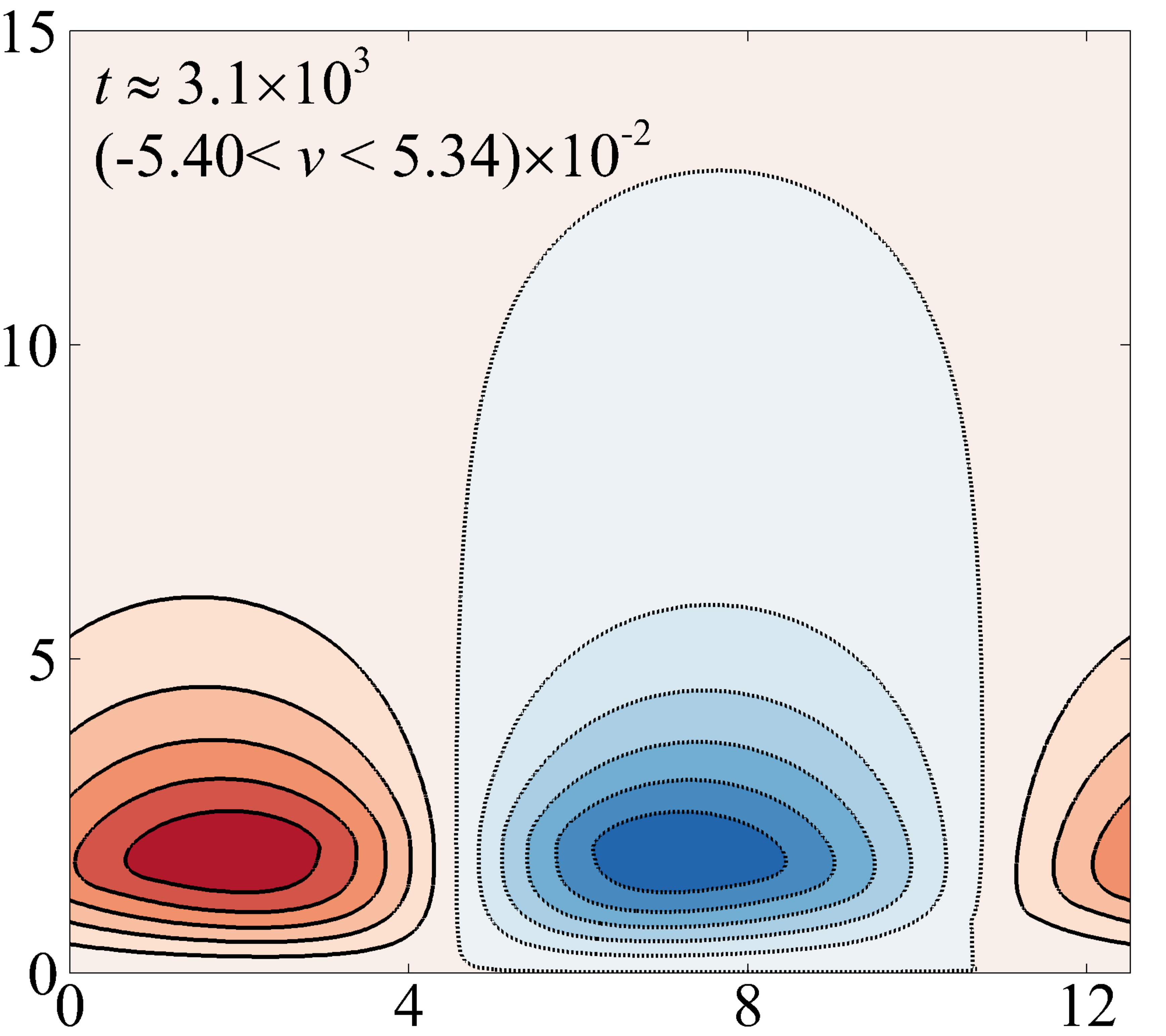}} &
 & \makecell{\includegraphics[width=0.322\textwidth]{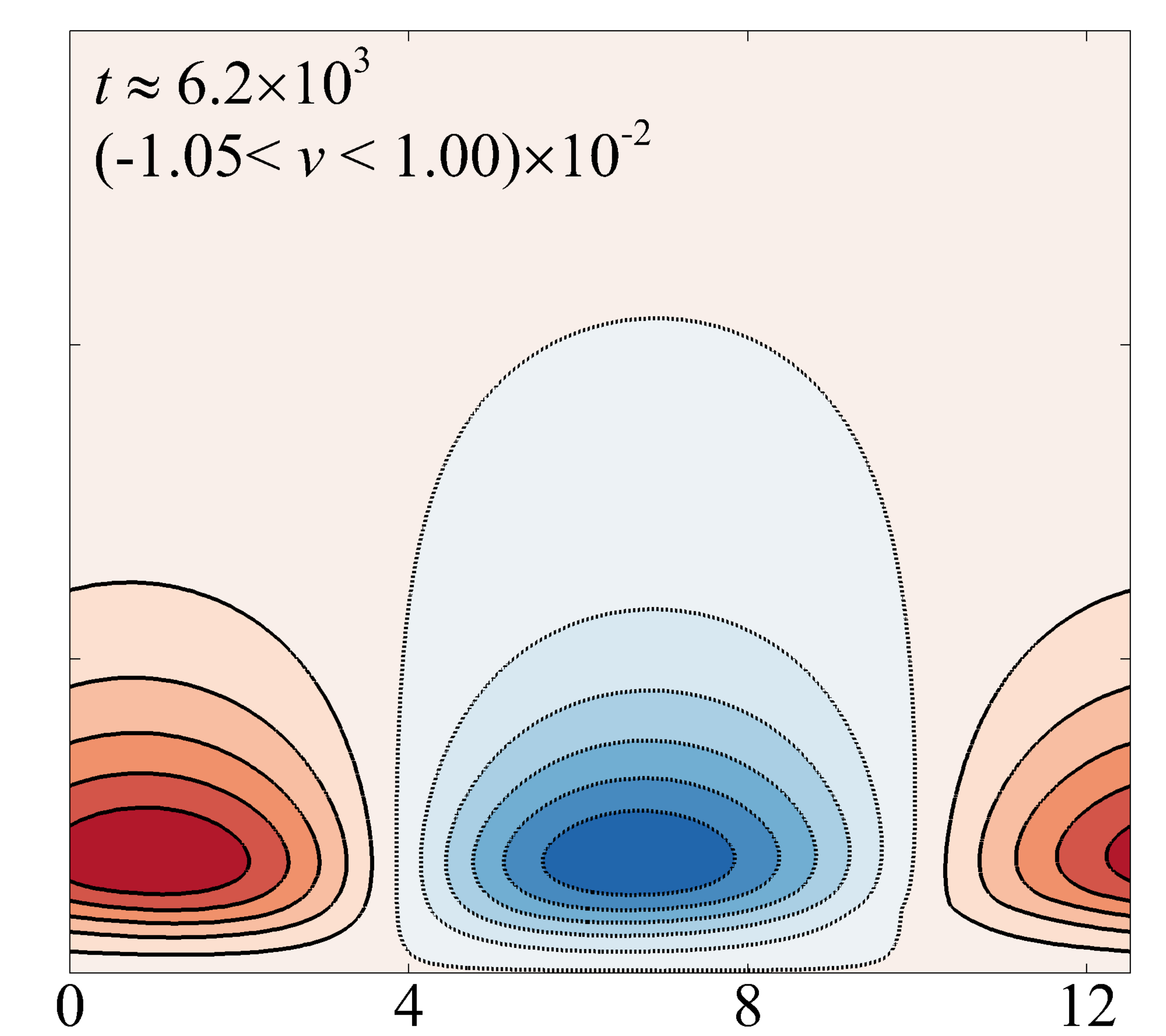}} &
 & \makecell{\includegraphics[width=0.322\textwidth]{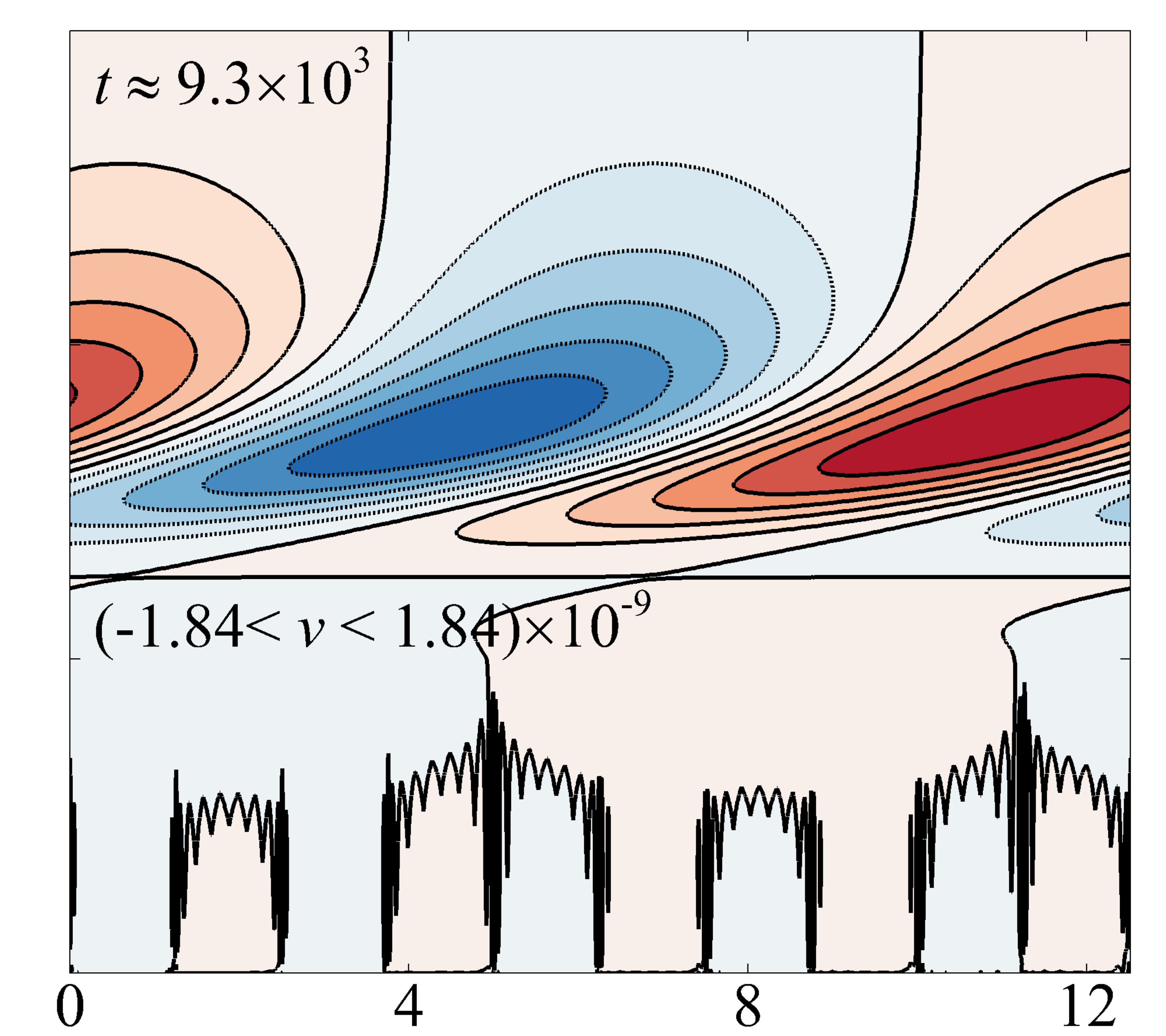}} \\
 & \hspace{24mm} \footnotesize{$x$} & & \hspace{24mm} \footnotesize{$x$} & & \hspace{24mm} \footnotesize{$x$} \\
\end{tabular}
\addtolength{\tabcolsep}{+2pt}
\addtolength{\extrarowheight}{+10pt}
\end{center}
    \caption{Temporal evolution at $\rrc = 0.439$, $\HsubD=28.28$, $n=1$, with $\EzeroR = 3.869\times10^{-6} > \ELD$. (a) Streamwise perturbation $\hat{u}=u-U$. (b). Wall-normal perturbation $\hat{v}=v$. Solid lines (red flooding) positive; dotted lines (blue flooding) negative.}
    \label{fig:umU_vs_v}
\end{figure}

The flow structures throughout this evolution are depicted in \figs\ \ref{fig:umU_vs_v}(a) for $\hat{u}$ and \figs\ \ref{fig:umU_vs_v}(b) for $\hat{v}$. While the maximum and minimum $\hat{v}$-velocities have similar magnitude, the $\hat{u}$ structures have a much larger magnitude minimum velocity (compared to the positive maximum). The $\hat{u}$ structures elongate until they eventually become uniform in the streamwise direction. Thus, as $\hat{v}$ decays, rather than reducing the magnitude of $\hat{u}$, continuity (equation~\ref{eq:non_dim_c}) is instead satisfied by reducing $\partial \hat{u}/\partial x$. This stores perturbation energy, recalling the slow decay of $E$ in \fig\ \ref{fig:FF_vs_PG}(a). The streamwise-independent structure forms regardless if $\Ezero>\ELD$ or $\Ezero<\ELD$. However, there is more perturbation energy to store if the flow transitions to turbulence, when $\Ezero>\ELD$. Lastly, it is worth noting that in this configuration, any non-sinusoidal streamwise variation indicates nonlinearity. Thus, the formation of the streamwise-independent structure is distinctly nonlinear. Streamwise-independent structures are also commonly observed in the final form of 3D simulations, \eg\ \cite{Krasnov2004numerical}. By comparison, the $\hat{v}$ structures maintain similar size until they rapidly decay to a structure resembling the long time state of the linear optimal.

\subsection{Influence of domain length}\label{sec:nln_doml}

\begin{figure}
\begin{center}
\addtolength{\extrarowheight}{-10pt}
\addtolength{\tabcolsep}{-2pt}
\begin{tabular}{ llll }
\makecell{\vspace{26mm} \footnotesize{(a)} \\  \vspace{33mm} \rotatebox{90}{\footnotesize{$\Euv$}}} & \makecell{\includegraphics[width=0.458\textwidth]{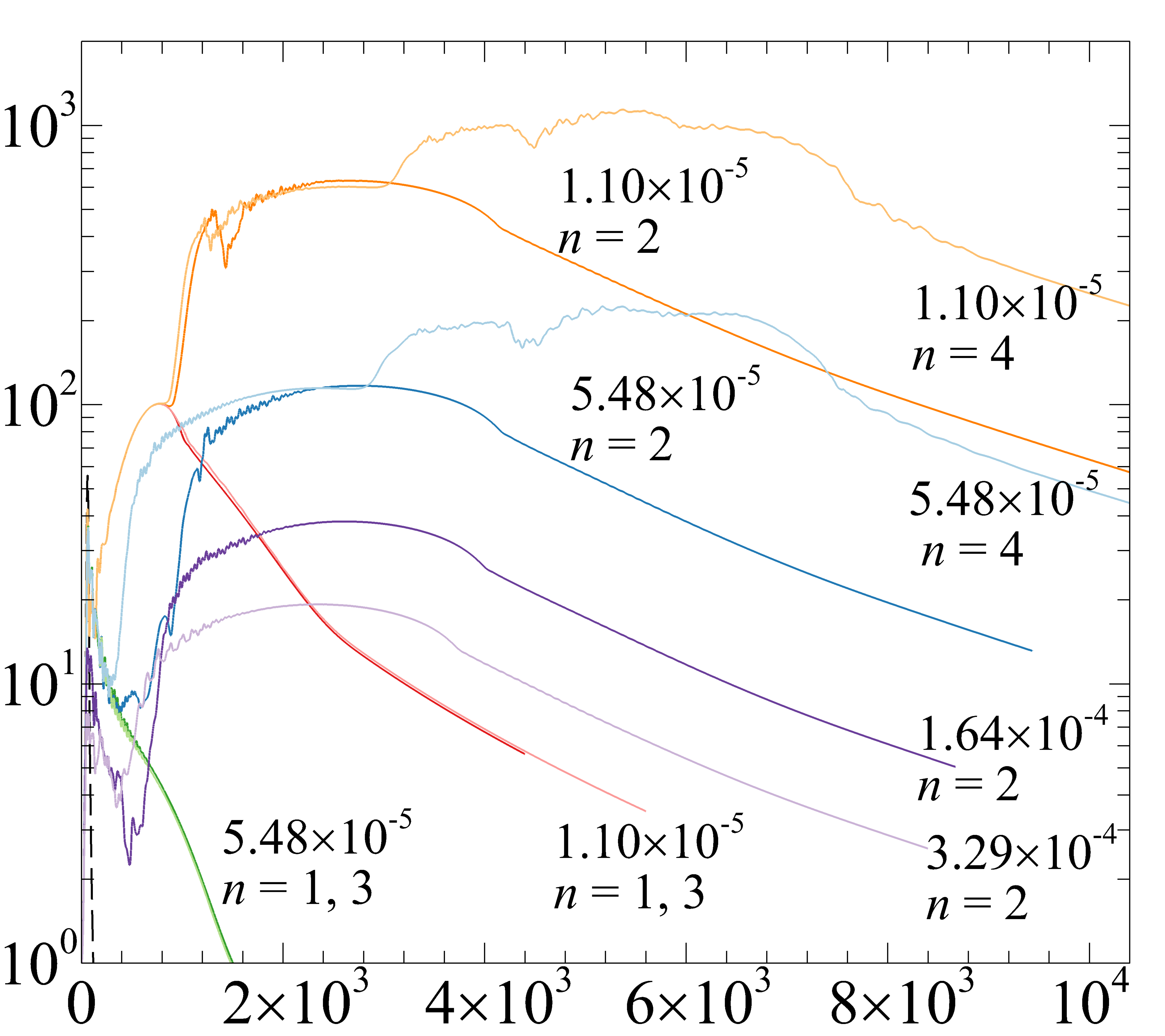}} &
\makecell{\vspace{26mm} \footnotesize{(b)} \\  \vspace{33mm} \rotatebox{90}{\footnotesize{$\Ev$}}}
 & \makecell{\includegraphics[width=0.458\textwidth]{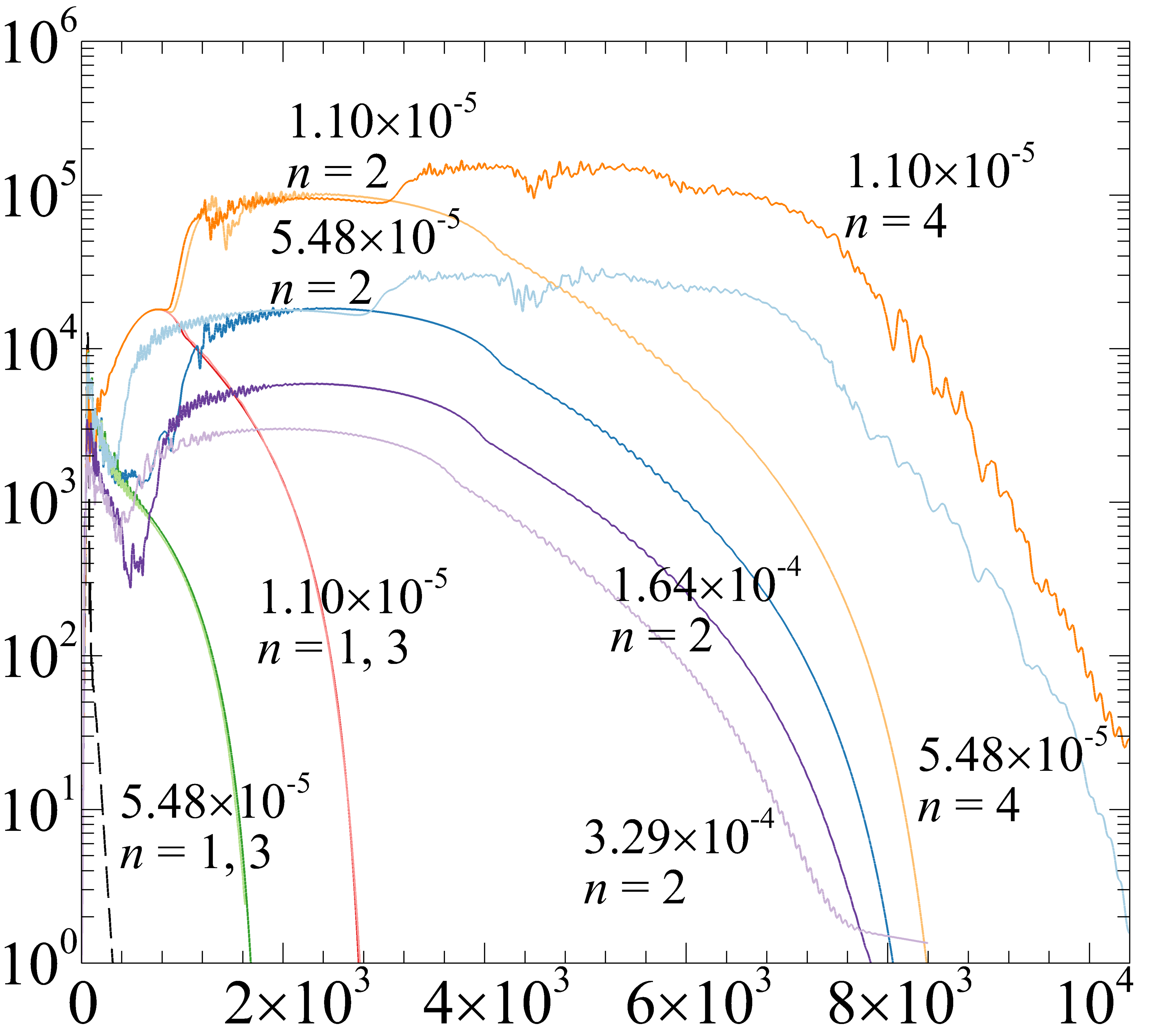}} \\
 & \hspace{36mm} \footnotesize{$t$} & & \hspace{36mm} \footnotesize{$t$} \\
\end{tabular}
\addtolength{\tabcolsep}{+2pt}
\addtolength{\extrarowheight}{+10pt}
\end{center}
    \caption{Energy time histories at $\rrc=0.293$, varying the initial energy and domain length via repetitions $n$ of $\lxopt$. (a) $\Euv = (1/2) \int \hat{u}^2 + \hat{v}^2 \,\dUP\Omega$. (b) $\Ev = (1/2) \int \hat{v}^2 \,\dUP\Omega$. Additional nonlinear growth is provided for even multiples of $n$, for all initial energies tested at $\rrc=0.293$, via pairwise coalescence of \TS\ wave repetitions. All curves are rescaled to unit initial energy. The linear curves are presented with black long dashed lines. At $\rrc=0.293$, $\Gmax=55.9876$.}
    \label{fig:amp_large_time}
\end{figure}


In \S\ \ref{sec:nln_rsts}, $\ELD$ and $\EUD$ were considered in $n=1$ domains. The effect of increasing the domain length on $\ELD$ and $\EUD$ is now discussed, for integer repetitions up to $n=4$ ($\WlenD = n \lxopt$). Growth measures $\Euv$ and $\Ev$ are shown in \fig\ \ref{fig:amp_large_time} for $\rrc=0.293$, with four distinct influences of domain length discussed. Recall that in the $n=1$ domain at $\rrc=0.293$ some $\Ezero$ can attain growth to a secondary local maximum (\eg\ $\EzeroR=1.10\times10^{-5}$) but no $\Ezero$ transition to turbulence (cross separatrix 1). The first influence of domain length is that if two instances of the same perturbation evolve in an $n=2$ domain, an inflection point appears in the energy-time history, indicating a crossing of separatrix 1. This occurs as the two individual repetitions of the \TS\ wave structure coalesce into a single wave structure, with a rapid jump in energy at the secondary maximum from the $n=1$ case. Secondly, at $\EzeroR=1.10\times10^{-5}$, but with an $n=3$ domain, this extra jump in energy does not occur ($n=3$ follows $n=1$). There would be a mismatch in wavelengths if only one pair of structures coalesced, prohibiting the interaction of all three repetitions. Thirdly, again at $\EzeroR=1.10\times10^{-5}$, the $n=4$ case can experience both the $n=2$ pairwise coalescence ($4\rightarrow2$ repetitions), and then another coalescence ($2\rightarrow1$ repetition), which allows for an additional, albeit smaller, jump in energy. In the $\EzeroR=1.10\times10^{-5}$ case, the $n=4$ curve closely follows the $n=2$ curve early on, indicating the time it takes for the lower energy case to sense the full domain length. However, fourthly, the $\EzeroR=5.48\times10^{-5}$ case differs between $n=2$ and $n=4$, with the structure able to increase in size more rapidly in the latter case when reforming to an arched \TS\ wave structure. This is inhibited in smaller ($n=1$) domains, in which the structure decays because it is distorted by too large an initial energy. The same is true of even larger initial energies, $\EzeroR=1.64\times10^{-4}$ and $3.29\times10^{-4}$, which undergo second-stage growth in  the $n=2$ domain, while the $n=1$ cases only decay after the linear maximum.  

\begin{figure}
\begin{center}
\addtolength{\extrarowheight}{-10pt}
\addtolength{\tabcolsep}{-2pt}
\begin{tabular}{ ll ll ll ll }
\makecell{\vspace{10mm} \footnotesize{(a)}  \\  \vspace{15mm} \rotatebox{90}{\footnotesize{$y$}}} & \makecell{\includegraphics[width=0.208\textwidth]{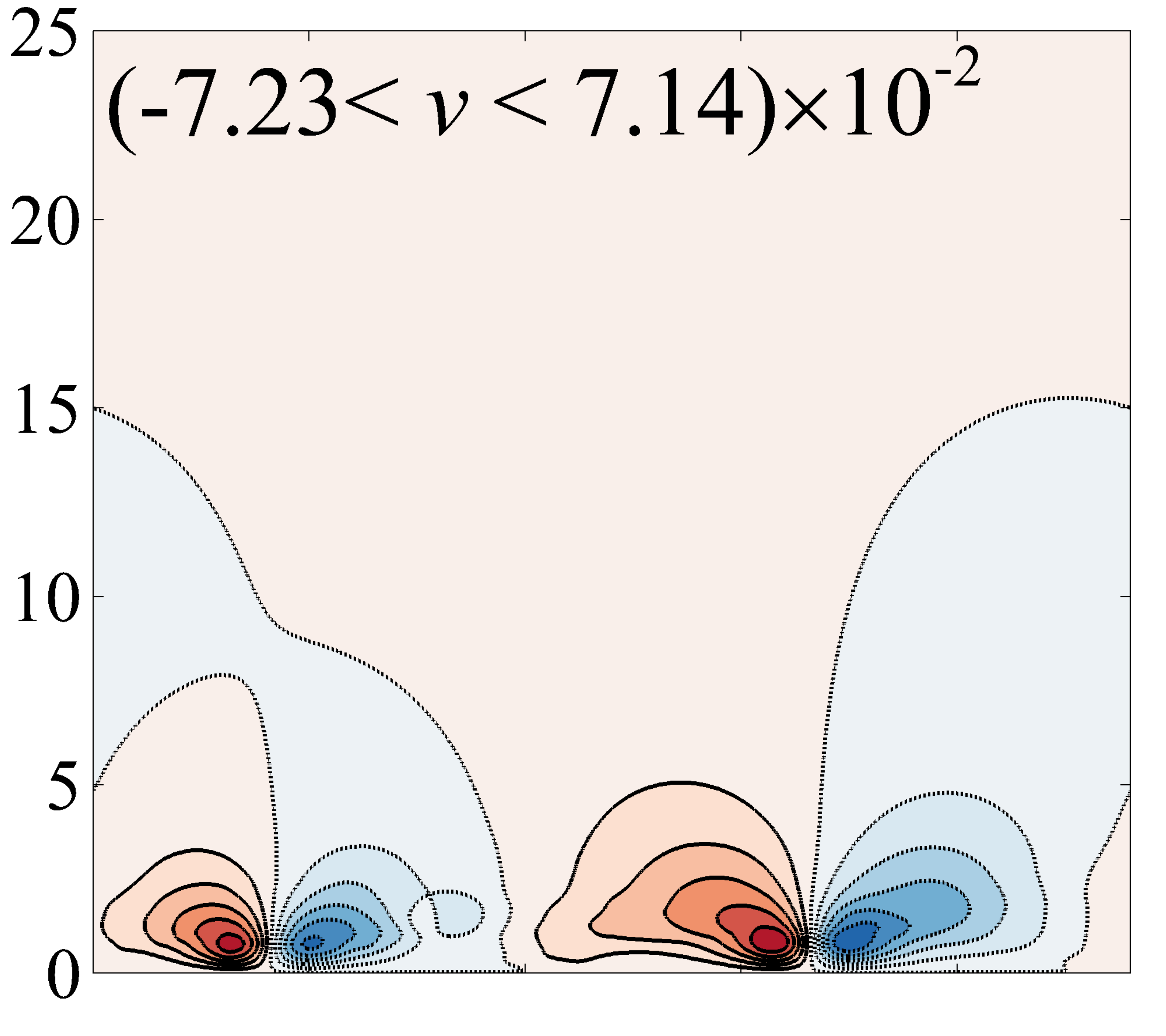}} & 
\makecell{\vspace{10mm} \footnotesize{(b)}  \\  \vspace{15mm} } & \makecell{\includegraphics[width=0.208\textwidth]{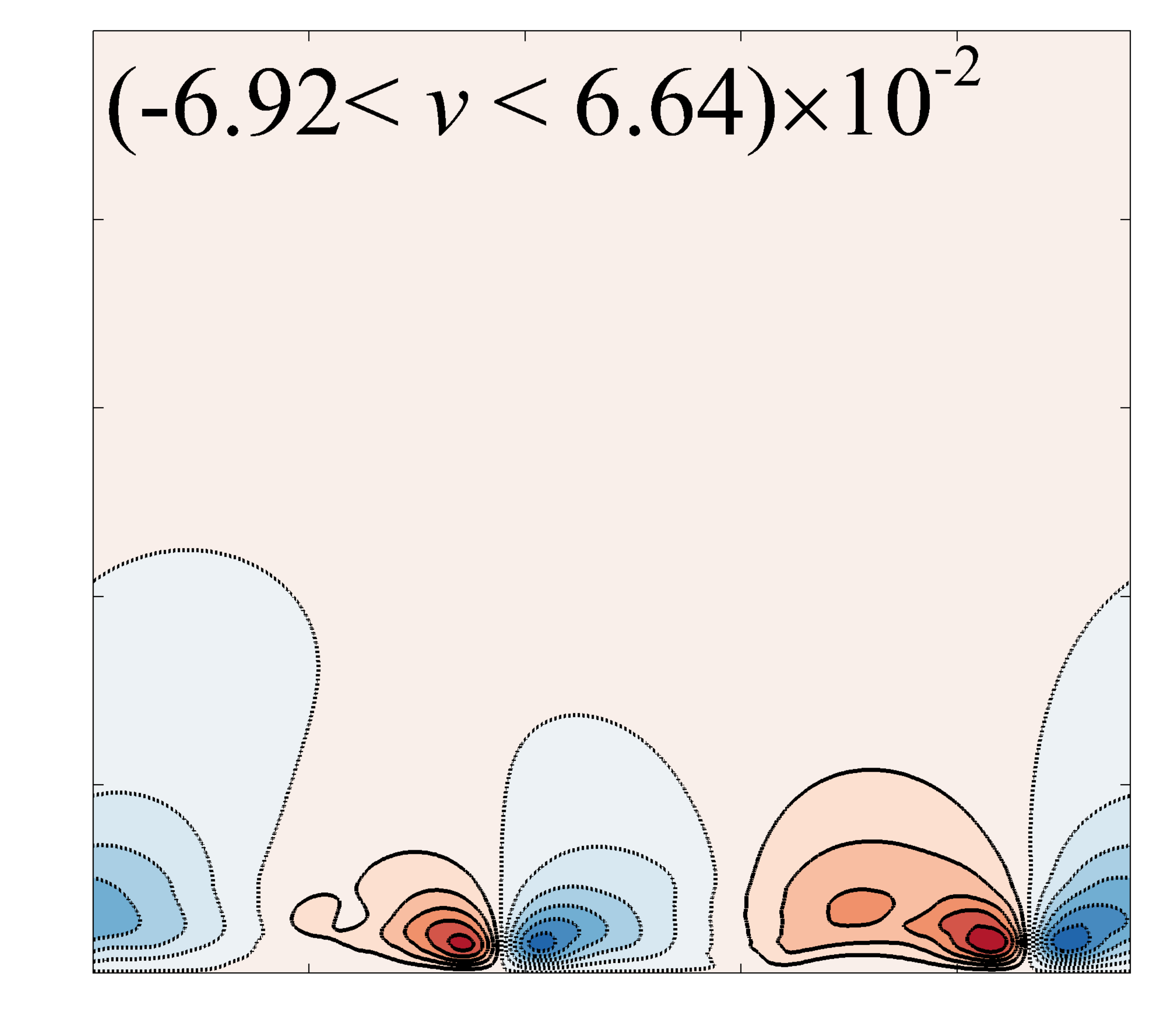}} &
\makecell{\vspace{10mm} \footnotesize{(c)}  \\  \vspace{15mm} }  & \makecell{\includegraphics[width=0.208\textwidth]{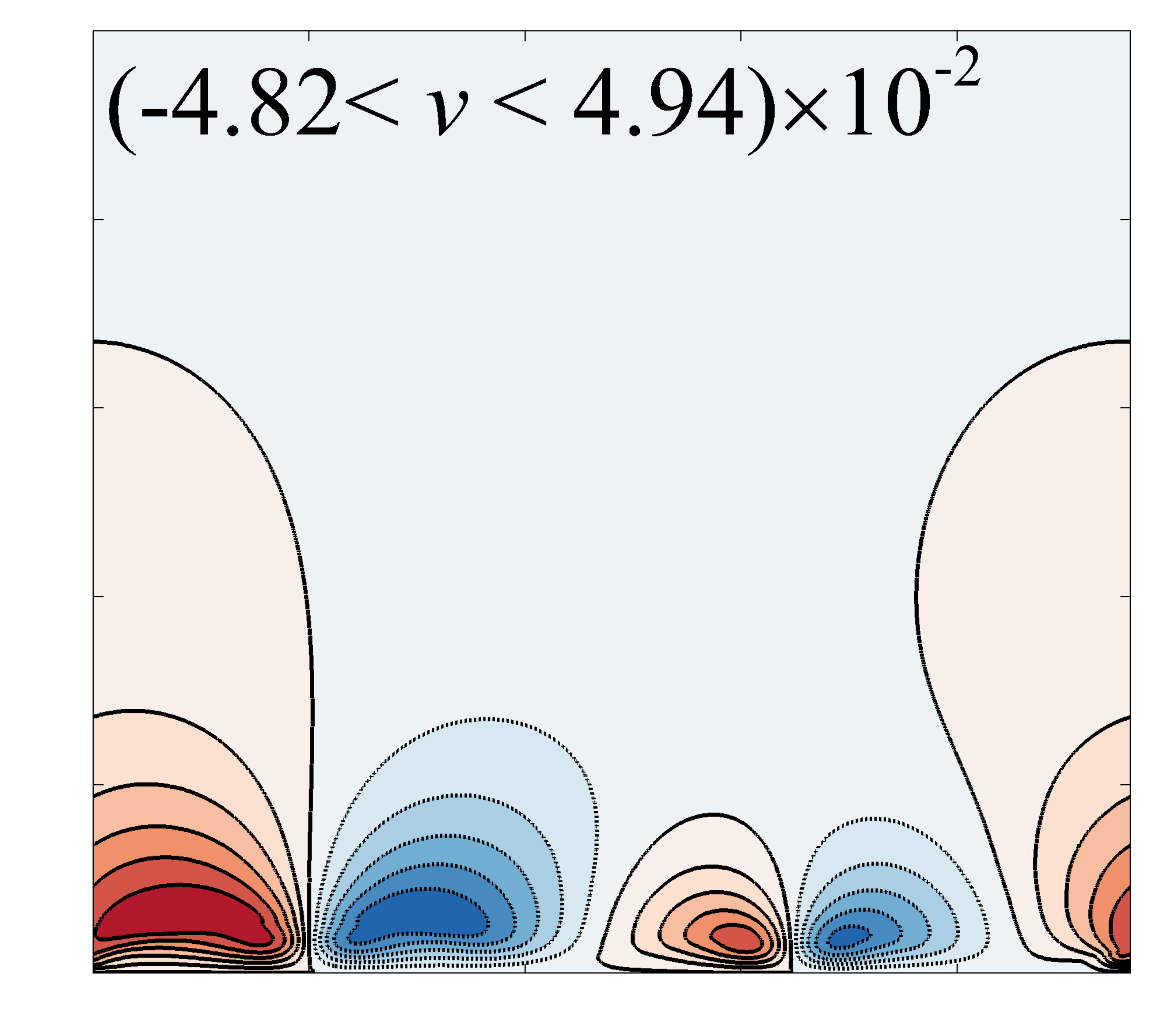}} & 
\makecell{\vspace{10mm} \footnotesize{(d)}  \\  \vspace{15mm} } & \makecell{\includegraphics[width=0.208\textwidth]{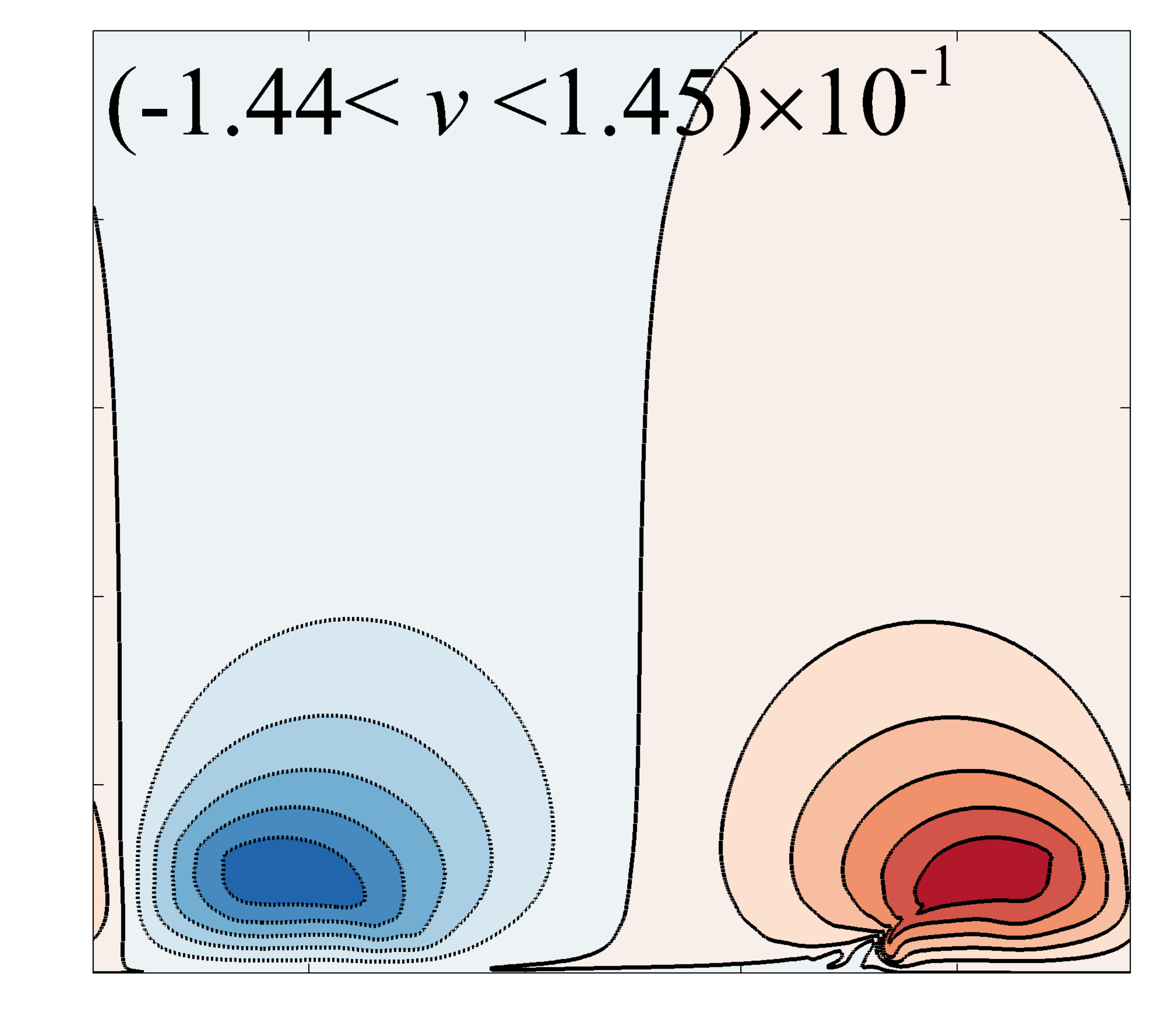}} \\
 \makecell{\vspace{10mm} \footnotesize{(e)}  \\  \vspace{15mm} \rotatebox{90}{\footnotesize{$y$}}} & \makecell{\includegraphics[width=0.208\textwidth]{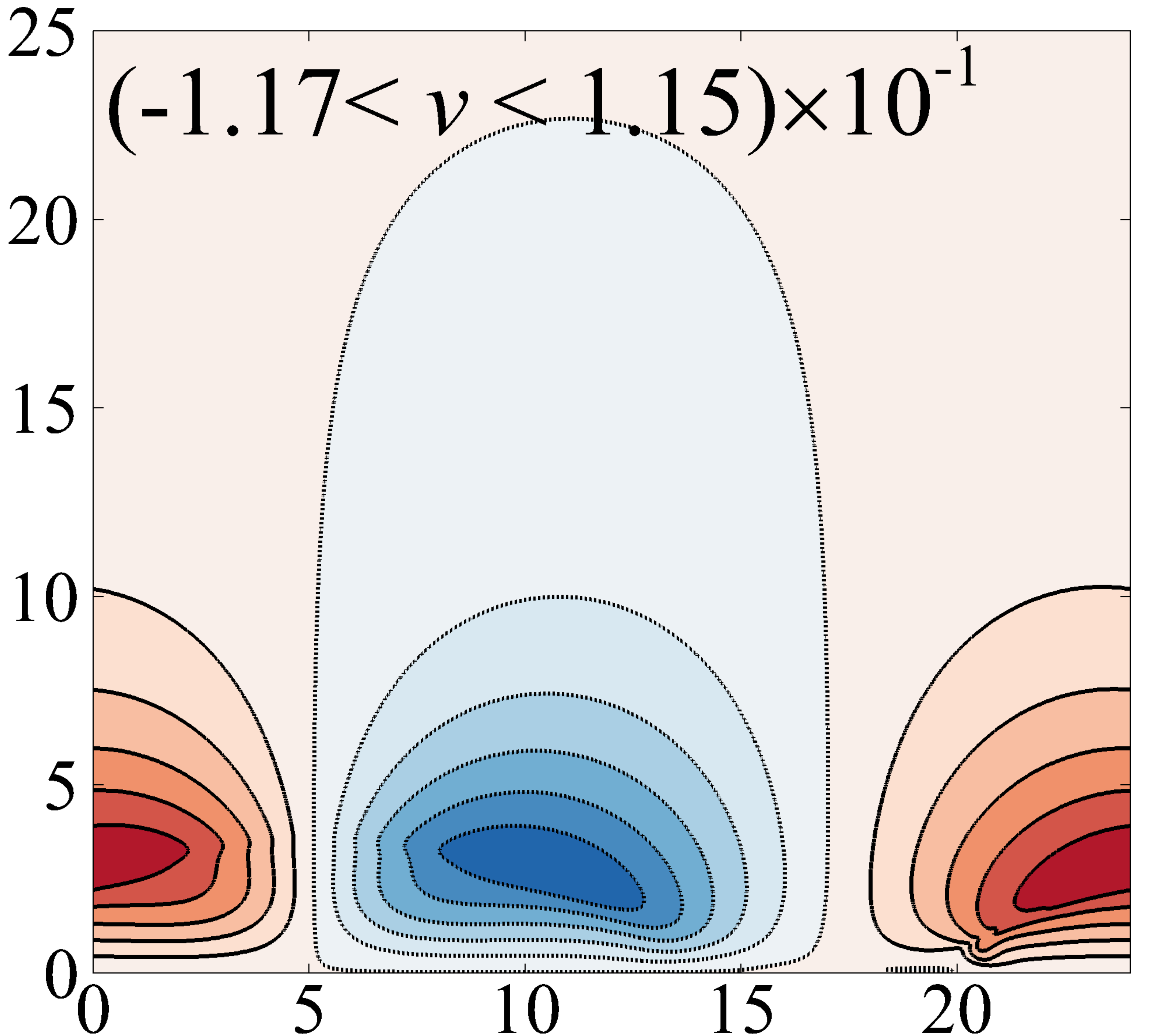}} & 
\makecell{\vspace{10mm} \footnotesize{(f)}  \\  \vspace{15mm} } & \makecell{\includegraphics[width=0.208\textwidth]{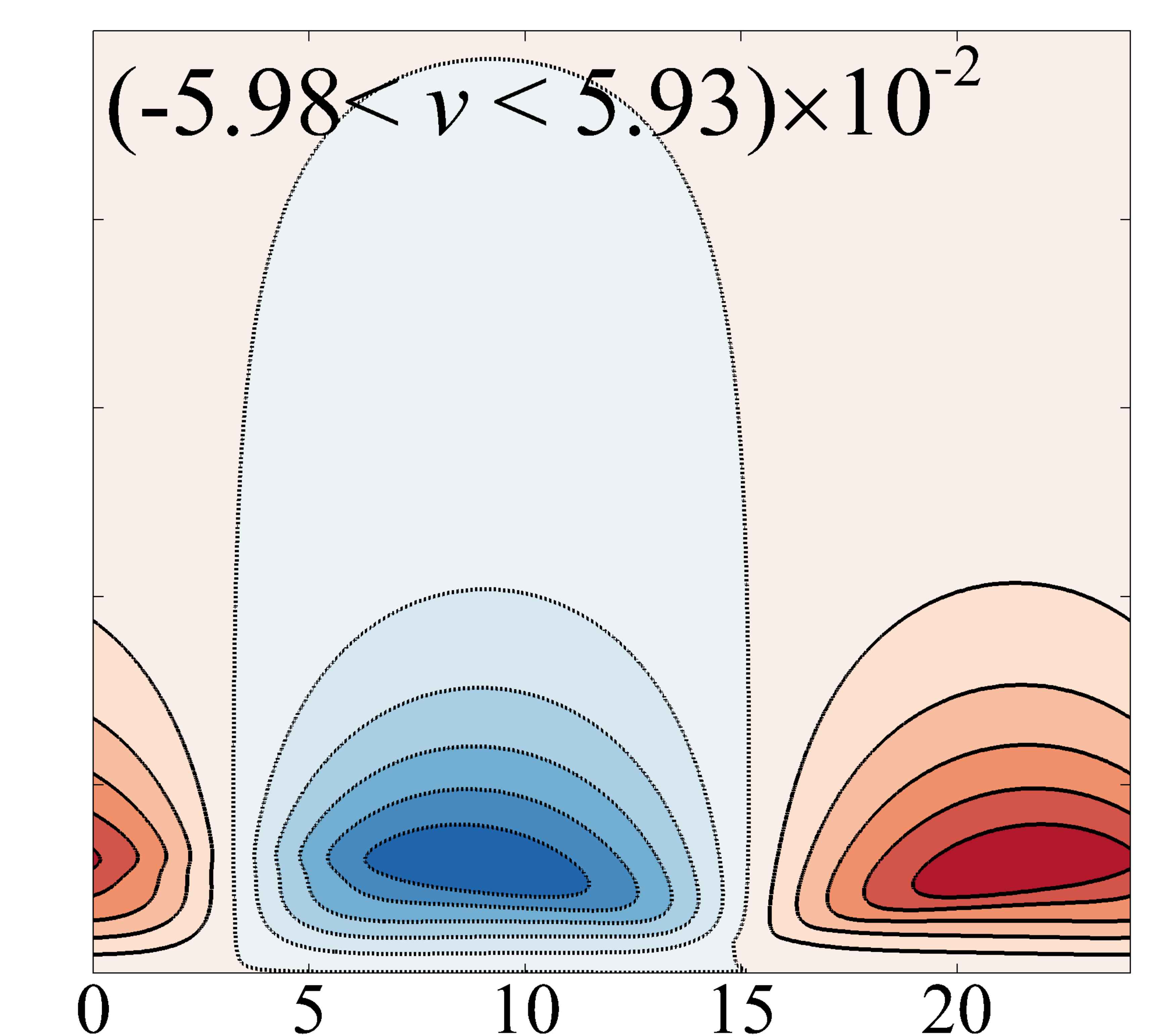}} &
\makecell{\vspace{10mm} \footnotesize{(g)}  \\  \vspace{15mm} }  & \makecell{\includegraphics[width=0.208\textwidth]{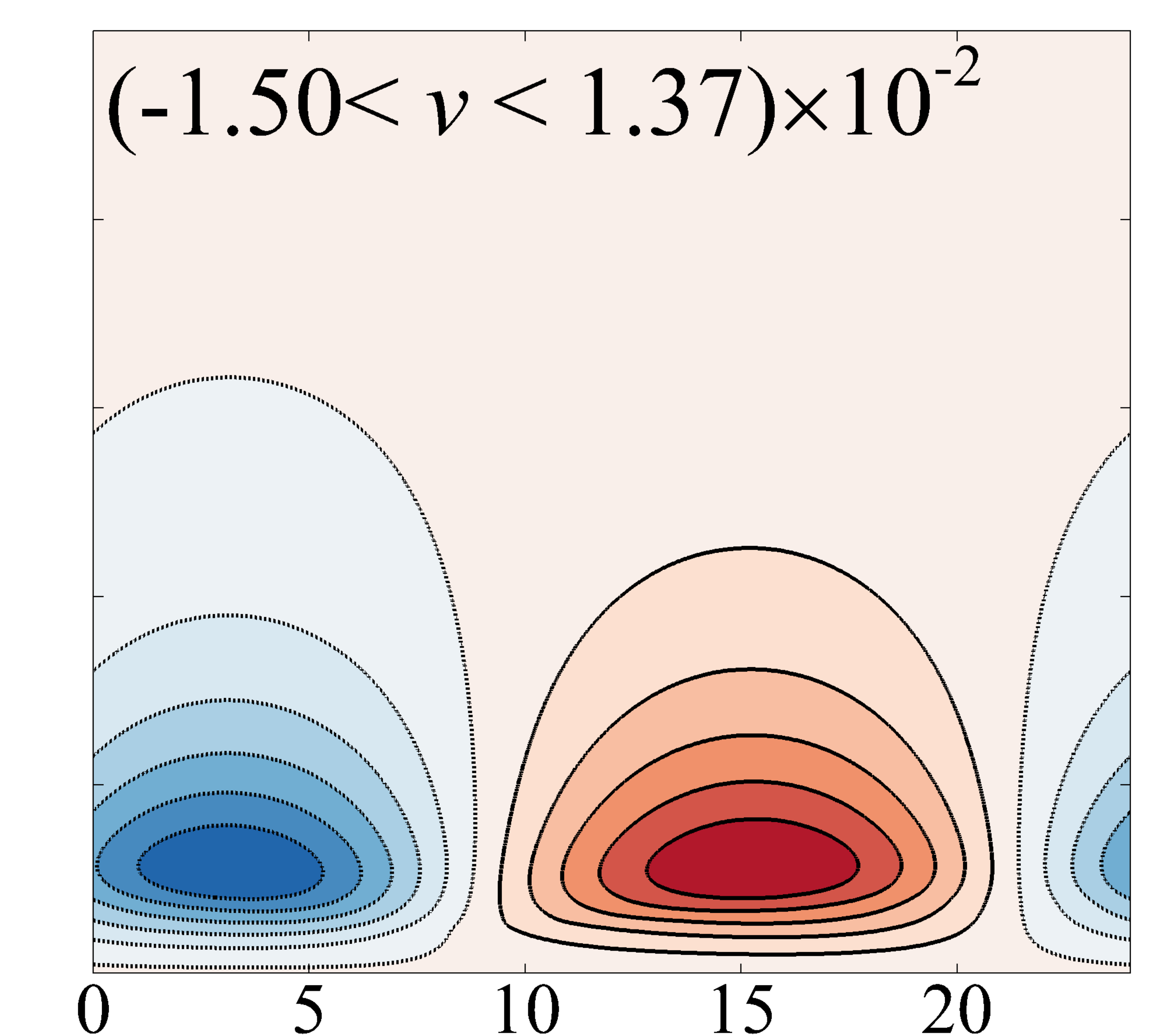}} & 
\makecell{\vspace{10mm} \footnotesize{(h)}  \\  \vspace{15mm} } & \makecell{\includegraphics[width=0.208\textwidth]{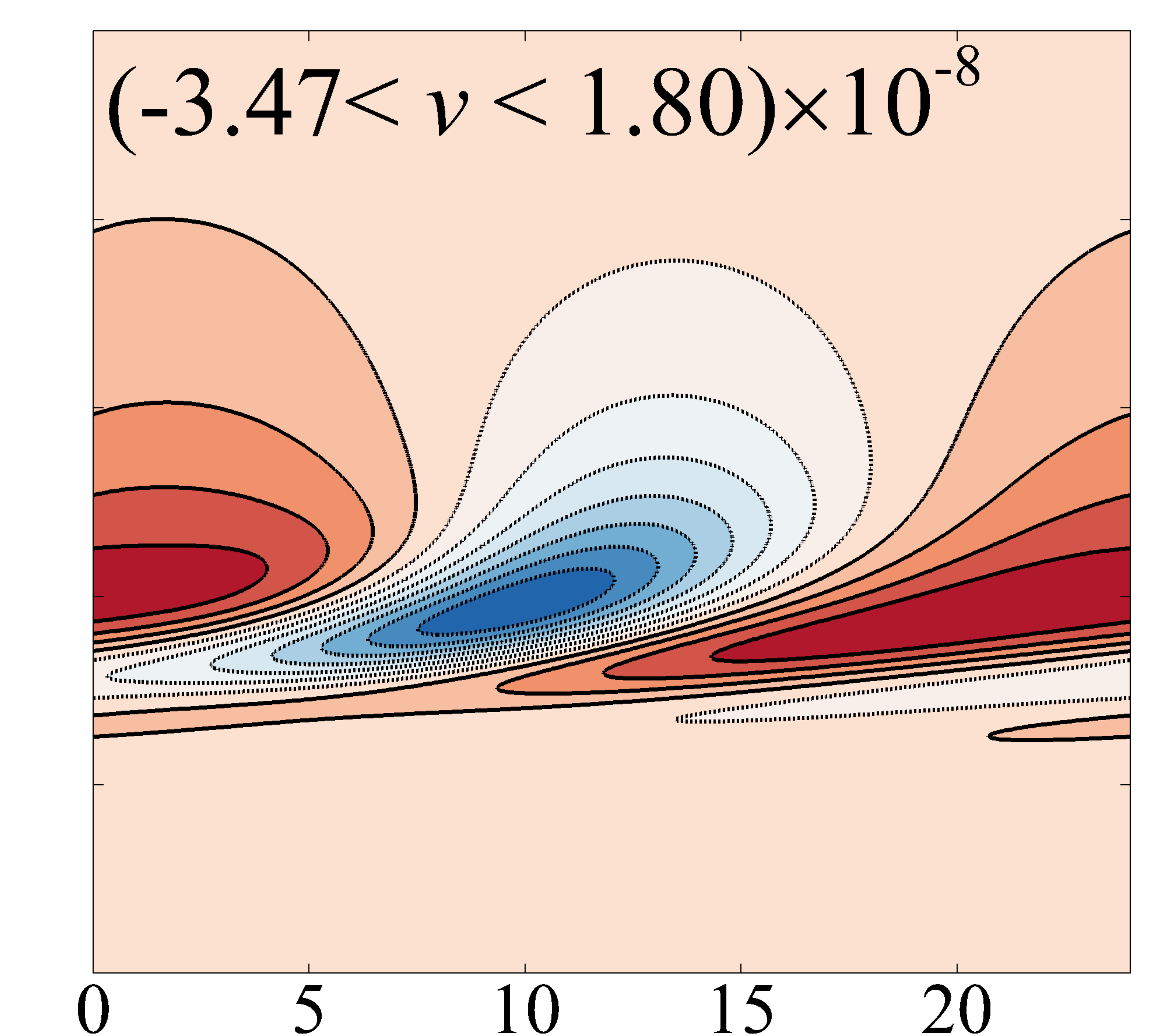}} \\
 & \hspace{16mm} \footnotesize{$x$} &  & \hspace{16mm} \footnotesize{$x$} &  & \hspace{16mm} \footnotesize{$x$} &  & \hspace{16mm} \footnotesize{$x$} \\
\end{tabular}
\addtolength{\tabcolsep}{+2pt}
\addtolength{\extrarowheight}{+10pt}
\end{center}
    \caption{Temporal evolution at $\rrc=0.293$, $\HsubD=28.28$, $n=2$, $\EzeroR = 5.48\times10^{-5}$; $\hat{v}$-velocity contours. Solid lines (red flooding) positive; dotted lines (blue flooding) negative. This case decays in an $n=1$ domain, but undergoes second-stage growth in an $n=2$ domain because it restructures to an arched \TS\ wave after the coalescence of the two individual perturbation repetitions.}
    \label{fig:2e4_extra}
\end{figure}


The $\hat{v}$-velocity fields are depicted in \fig\ \ref{fig:2e4_extra} for $\EzeroR=5.48\times10^{-5}$, $n=2$ at $\rrc=0.293$. Recall that with $n=1$, $\EzeroR=1.10\times10^{-5}$ attains second-stage growth, whereas $\EzeroR=5.48\times10^{-5}$ is too highly energised and rapidly decays, as the flow field does not resemble an arched \TS\ wave, \eg\ \fig\ \ref{fig:arch_high}(a). The two repetitions of the distorted \TS\ wave shown in \fig\ \ref{fig:2e4_extra}(a), \ref{fig:2e4_extra}(b) are not yet interacting. The interaction between the two wavelengths is shown in \fig\ \ref{fig:2e4_extra}(c), where one repetition becomes dominant, and will shortly subsume the other, \fig\ \ref{fig:2e4_extra}(d). In \fig\ \ref{fig:2e4_extra}(e), the wave has re-formed into a single repetition of the arched \TS\ wave structure. The arched \TS\ wave then undergoes nonlinear second-stage growth, as it slowly relaxes back to a conventional \TS\ wave, \fig\ \ref{fig:2e4_extra}(g). It finally decays to a field resembling the long time solution of a linear transient growth computation. However, unlike a linear optimal, this process will still have stored perturbation energy in a sheet of negative $\hat{u}$-velocity, visible when comparing the energy measures shown in \figs\ \ref{fig:amp_large_time}(a), \ref{fig:amp_large_time}(b).


\begin{figure}
\begin{center}
\addtolength{\extrarowheight}{-10pt}
\addtolength{\tabcolsep}{-2pt}
\begin{tabular}{ llll }
\makecell{\vspace{26mm} \footnotesize{(a)} \\  \vspace{33mm} \rotatebox{90}{\footnotesize{$\Euv$}}} & \makecell{\includegraphics[width=0.458\textwidth]{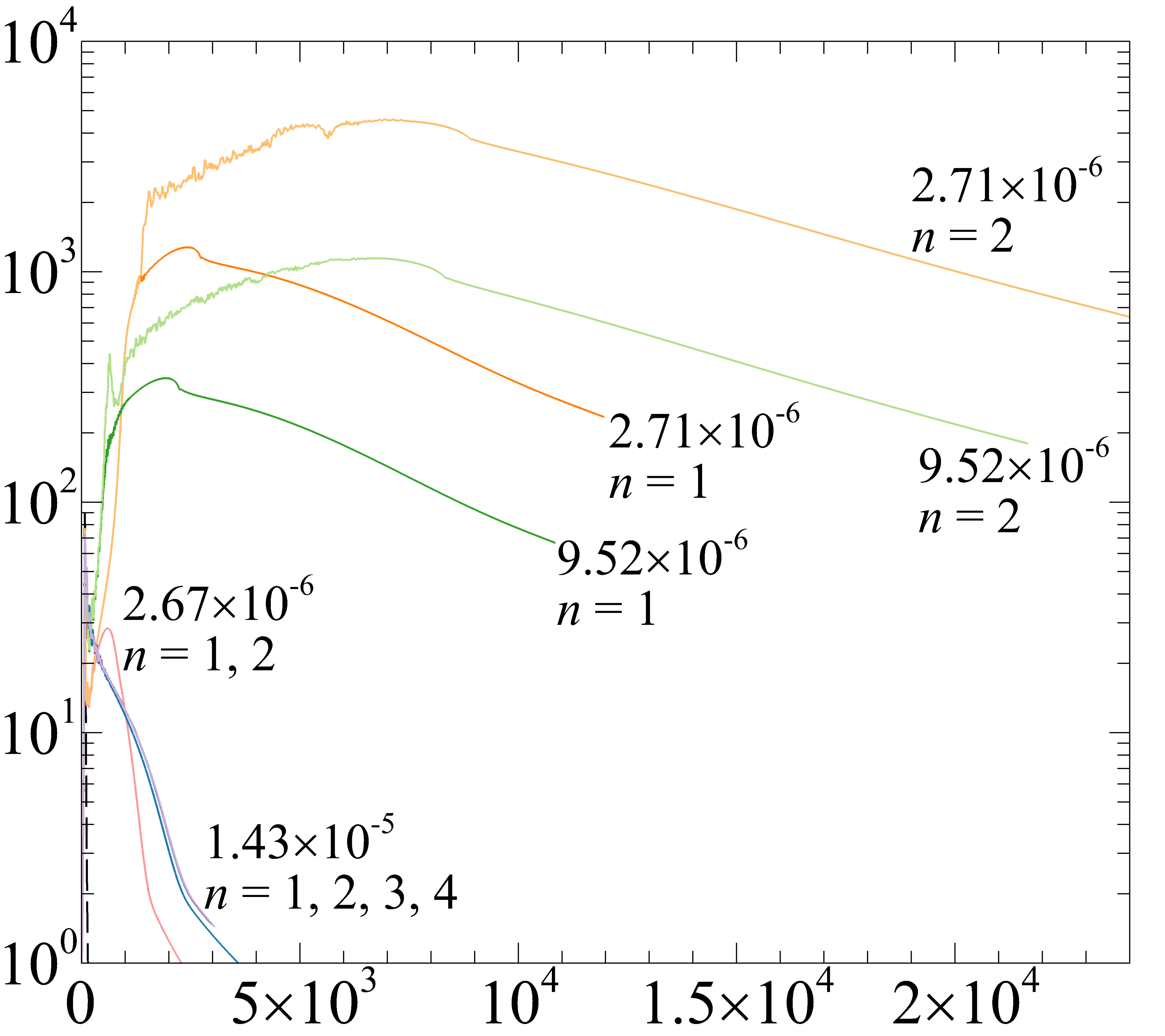}} &
\makecell{\vspace{26mm} \footnotesize{(b)} \\  \vspace{33mm} \rotatebox{90}{\footnotesize{$\Euv$}}}
 & \makecell{\includegraphics[width=0.458\textwidth]{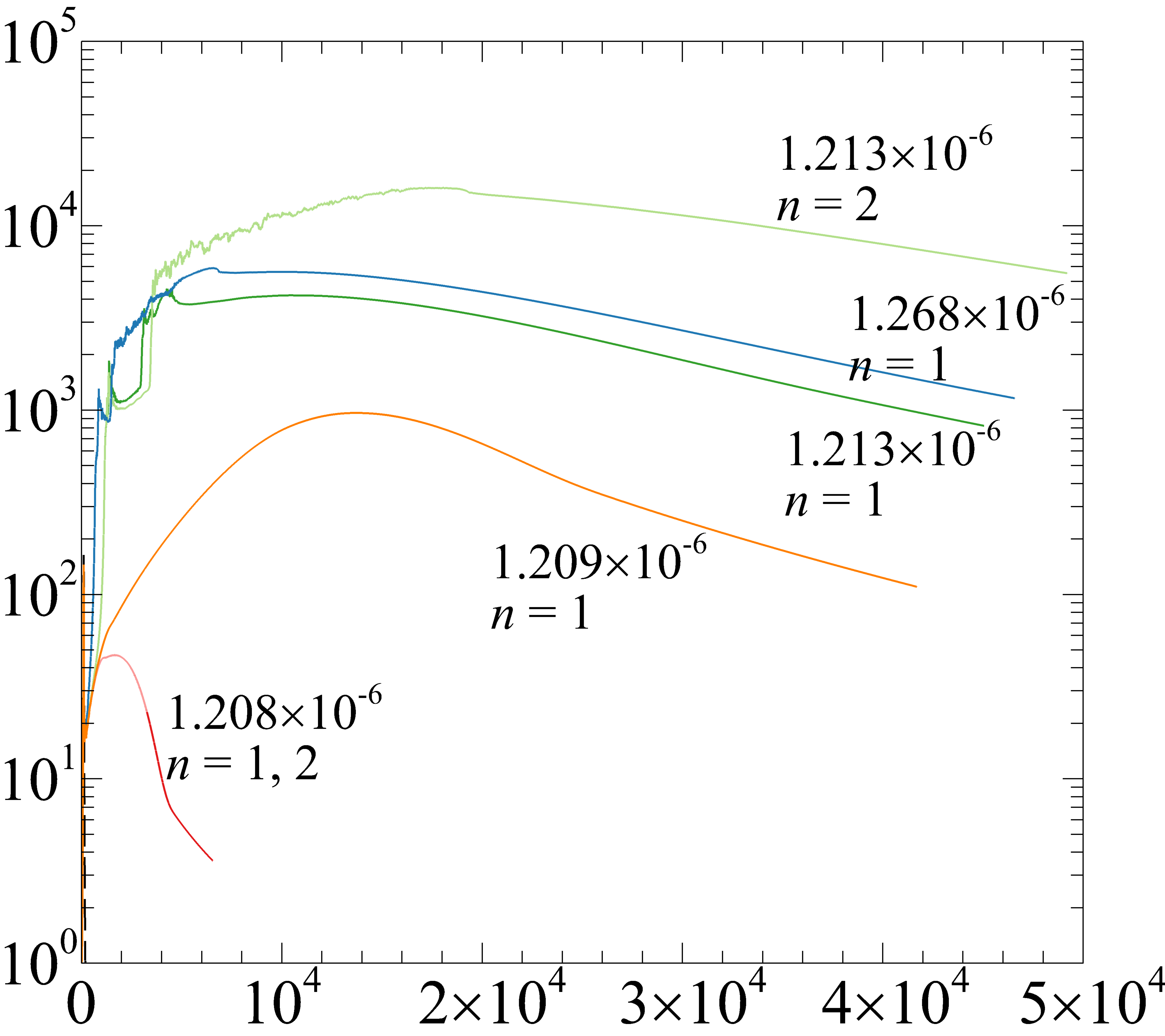}} \\
 & \hspace{36mm} \footnotesize{$t$} & & \hspace{36mm} \footnotesize{$t$} \\
\end{tabular}
\addtolength{\tabcolsep}{+2pt}
\addtolength{\extrarowheight}{+10pt}
\end{center}
    \caption{Energy time histories, varying the initial energy and domain length via repetitions $n$ of $\lxopt$. (a) $\rrc=0.585$, $\Gmax=89.9630$, $\ELD=2.6919\times10^{-6}$, maximum nonlinear gain observed for $\Ezero>\ELD$ is $\approx 4\times10^3$ ($n=2$). (b) $\rrc=1.463$, $\Gmax=166.4092$, $\ELD=1.2096\times10^{-6}$, maximum nonlinear gain observed for $\Ezero>\ELD$ is $\approx 2 \times 10^4$ ($n=2$). All curves are rescaled to unit initial energy.  $\Ezero<\ELD$ are unable to take advantage of the extra domain length, and still rapidly decay.}
    \label{fig:higher_Re}
\end{figure}


The energy growth at larger Reynolds numbers is depicted in \fig\ \ref{fig:higher_Re}. These illustrate the length of time over which high energy states are maintained when $\EzeroR>\ELD$. At $\rrc=0.585$, $n=1$, $\Ezero=2.67\times10^{-6}<\ELD$ rapidly decays, while $\Ezero=2.71\times10^{-6}>\ELD$ maintains large energies for the order of $10^4$ time units, particularly so when $n=2$. This is even clearer at $\rrc=1.463$, with very large amounts of growth, and a very slow decay, when $\EzeroR=1.213\times10^{-6}>\ELD$. A case $\EzeroR=1.209\times10^{-6}$ just slightly below $\ELD=1.2096\times10^{-6}$ provides a clearer indication of the additional growth due to reaching the turbulent attractor, compared to the underlying nonlinear second-stage growth (to a local maximum). Of additional interest is that it takes a far greater time to relaminarize turbulent states in larger domains. The oscillations appear to be less energetic, or otherwise damped out more rapidly, in the $n=1$ domains. Lastly, all $\rrc=0.585$ and $\rrc=1.463$ cases show that $\EzeroR<\ELD$ cannot take advantage of the extra space afforded in $n=2$ domains, and decay following the $n=1$ curves, such that $\ELD$ does not depend on domain length. Note that at $\rrc=1.463$ the wavenumbers in $n=1$ and $n=2$ domains are outside the neutral curve. 

\begin{figure}
\begin{center}
\addtolength{\extrarowheight}{-10pt}
\addtolength{\tabcolsep}{-2pt}
\begin{tabular}{ llll }
\makecell{\vspace{26mm} \footnotesize{(a)} \\  \vspace{33mm} \rotatebox{90}{\footnotesize{$y$}}} & \makecell{\includegraphics[width=0.458\textwidth]{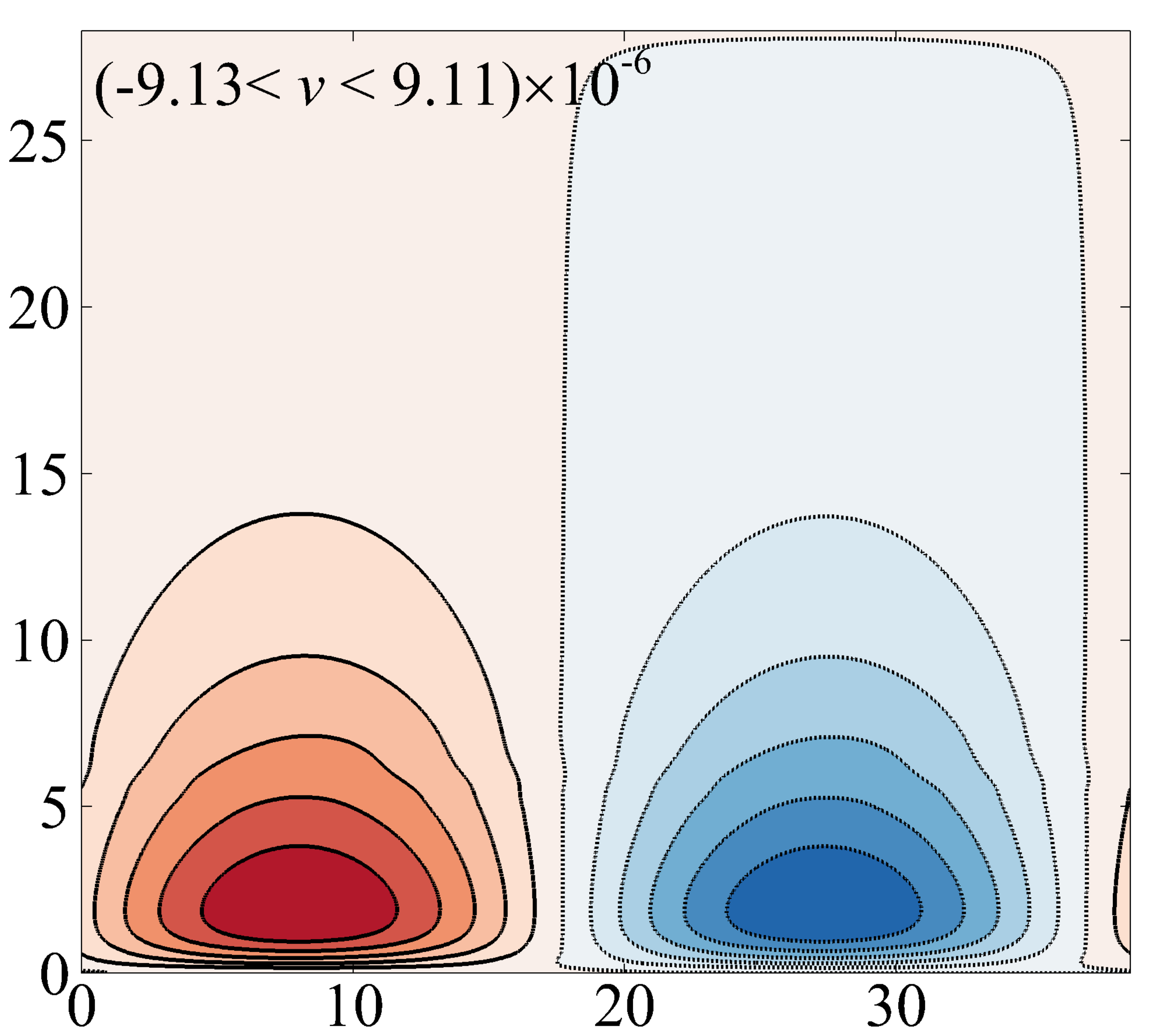}} &
\makecell{\vspace{26mm} \footnotesize{(b)} \\  \vspace{33mm} \rotatebox{90}{\footnotesize{$y$}}}
 & \makecell{\includegraphics[width=0.458\textwidth]{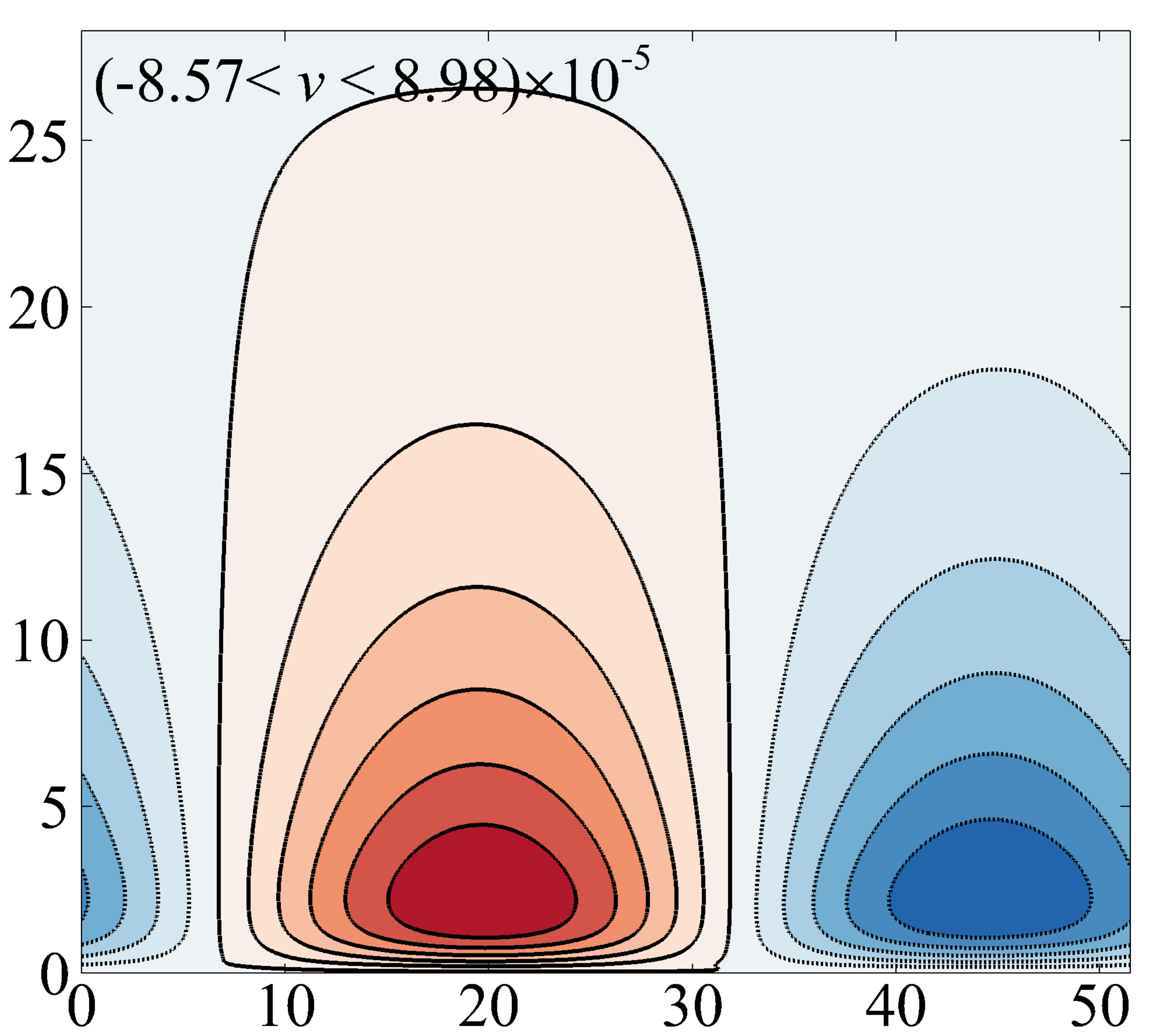}} \\
 & \hspace{36mm} \footnotesize{$x$} & & \hspace{36mm} \footnotesize{$x$} \\
\end{tabular}
\addtolength{\tabcolsep}{+2pt}
\addtolength{\extrarowheight}{+10pt}
\end{center}
    \caption{Contours of $\hat{v}$-velocity at $\rrc=0.585$, $\EzeroR=1.43\times10^{-5}$, $\HsubD=28.28$ at $t\approx 2.8\times10^3$. (a) $n=3$. (b) $n=4$. Solid lines (red flooding) positive; dotted lines (blue flooding) negative. Although the $n=3$ and $n=4$ cases coalesce, without the \TS\ wave having an arched appearance, they decay monotonically.}
    \label{fig:4e4_extra}
\end{figure}

One final influence of the domain length is considered. At $\rrc=0.585$, $\EUD=1.09646\times10^{-5}$ when $n=1$, \fig\ \ref{fig:amp_delin}(b). Over-energised cases, with $\Ezero=1.43\times10^{-5}>\EUD$ and in longer domains ($n=2$ through $n=4$), are shown in \fig\ \ref{fig:higher_Re}(a). These all appear to decay coincidentally with the $n=1$ case, seemingly implying that $\EUD$ has not significantly changed with increasingly domain length, at $\rrc=0.585$. Comparatively, at $\rrc=0.293$ with $n=2$ (\fig\ \ref{fig:amp_large_time}) second-stage growth is observed (akin to cases with $\ELD \leq \Ezero \leq \EUD$), in multiple over-energised situations, via the restructuring depicted in \fig\ \ref{fig:2e4_extra}. This would imply that at $\rrc=0.293$, $\EUD$ has changed noticeably with increasing domain length. At $\rrc=0.585$, with a larger initial energy, the vortex merging process may occur too rapidly, unlike the $\rrc=0.293$, $n=2$ cases. At $\rrc=0.585$ the $n=3$ and $n=4$ cases reformed into the simpler conventional flat bottomed \TS\ wave structure, shown part way through their decay in \fig\ \ref{fig:4e4_extra}, rather than arched \TS\ waves capable of nonlinear second-stage growth. This issue may also be exacerbated by the wavelength restrictions imposed by the periodic boundary conditions, recalling the $\rrc=0.293$, $n=3$ case indicated that a mismatch in wavelength between \TS\ wave instances can also prevent growth. Overall, results in longer domain do not contradict the fact that $\EzeroR=1.43\times10^{-5}$ does not incite sustained turbulence at $\rrc=0.585$, so that separatrix 2 is still clearly defined. However, they do indicate that $\EUD$ can be very difficult to accurately determine, as consistent behaviour was not observed across all Reynolds numbers tested. As a final note, the investigations at $\rrc=0.585$, $n=3$ and $n=4$ also highlight that the energy growth is due to the form of the merged structure, and not coalescence, as the cases monotonically decay after the linear peak, during which time they are merging.

\section{Conclusions}\label{sec:conc}
The present work has numerically illustrated a subcritical route to turbulence driven by purely \qtwod\ mechanisms, in a laminar Q2D exponential boundary layer. This system approximates a magnetohydrodynamic duct flow under a strong transverse magnetic field. It was shown that the linear optimals form an excellent approximation of the nonlinear optimals, when tested for small (linear $\tauOpt$) target times. The transition process then has two stages. First, linear transient growth, via the Orr mechanism. This was followed by a second stage of substantial nonlinear growth, able to propel the flow across the laminar-turbulent basin boundary. However, only linear optimals with specific initial energies $\ELD \leq \Ezero \leq \EUD$ were capable of following this route to a temporary turbulent state, before later relaminarizing. The lower bound, $\ELD$, defines the minimal seed energy capable of transition. The upper bound, $\EUD$, represents an initial perturbation too highly energised, which chaotically distorts the \TS\ wave, inducing rapid dissipation, rather than transitioning to turbulence. 

The additional nonlinear growth which leads to the existence of the delineation energy $\ELD$ (separating states which rapidly relaminarize, and those which temporarily maintain turbulence) is linked to the formation of an arched \TS\ wave, which forms when a conventional \TS\ wave becomes pinched close to the wall. The arched \TS\ wave still provides significant nonlinear growth when $\Ezero < \ELD$, but does not transition because the optimal is too far (measured in an energy norm) from the boundary of the turbulent attractor. While closer to the basin boundary at $\Ezero>\EUD$, distortion of the conventional \TS\ wave prevents the arch from forming. If the arch forms, the relaxing of the arched \TS\ wave into its conventional counterpart eventually results in the decay of the perturbation. However, during this process, perturbation energy is stored in a streamwise sheet of negative velocity, which effectively becomes a modulation to the original base flow. This modulated base flow may prove easier to re-excite if targeted by flow control methods. Overall, this \qtwod\ system was found to be highly sensitive to the energy and structure of the initiating perturbation, with only the optimal initial field capable of transition for tests in shorter domains. 


Larger domain lengths were also investigated. Firstly, this showed that successive vortex merging may be capable of increasing the upper delineating energy $\EUD$, by allowing distorting structures which would naturally rapidly decay, to instead coalesce into an arched \TS\ wave structure, capable of sustaining turbulence over longer times. However, for sufficiently large initial energy, even very long domains still indicated the existence of high energy states which only rapidly decay after the initial linear growth. Perturbations with energy below the lower delineating energy $\ELD$ could not make use of the merging process, and still decayed in longer domains. Perturbations with $\Ezero>\ELD$, which are sufficient to transition to turbulence, made use of the longer domains by pairwise coalescence of \TS\ wave repetitions, achieving up to an order of magnitude of additional growth (compared to the shorter domains). The largest nonlinear gains are therefore achieved with $\Ezero>\ELD$ and in longer domains. The comparison between the nonlinear growth of the linear optimal and the linear growth of the linear optimal is striking at larger Reynolds numbers. The nonlinear gains achieved, at Reynolds numbers approximately $40\%$ below and above critical, were $\approx 4\times10^3$ and $\approx 2\times10^4$, respectively, compared to the optimised linear gains of $89.96$ and $166.4$, respectively. Furthermore, it appeared to take noticeably longer for turbulent oscillations to become subdued in longer domains. 

The prospect of subcritical transitions is promising for the feasibility of self-cooled liquid metal reactor ducts. However, the fact that all Reynolds numbers are scaled on the boundary layer thickness must be kept in mind. Although a sidewall Reynolds number of $10^5$ provided both very large growth, and slow relaminarization, at a realistic magnetic field strength, the corresponding Reynolds number based on the half duct height would be around $10^7$. This is well beyond what is currently expected for reactor operation, which range from $10^4$ to $10^6$ \citep{Mas2011qualification, Smolentsev2008characterization, Vetcha2012study}. Furthermore, no assessment of the sensitivity to wall properties on the formation of the arched \TS\ wave has been performed, which given the thermal, electrical and slip issues considered in magnetohydrodynamic coolant duct flows \citep{Buhler1995influence, Buhler1996instabilities, Buhler1998laminar, Smolentsev2009duct}, provides an important avenue for future work for self-cooled reactor designs.

Lastly, further investigation is warranted from a theoretical point of view. Although subcritical turbulent transitions were obtained, it is curious that all turbulent flow fields relaminarized. It would be worth exploring whether the turbulent states are in a true basin of attraction. The Q2D turbulent states may be unstable, such that a small deviation from their trajectory drives them out of the basin, causing relaminarization. However, it cannot be excluded that the behaviour originates from the numerical method, or choice of periodic boundary conditions.


\begin{acknowledgments}
The author is grateful for discussions with Ashley Willis regarding the iterative approach applied to the nonlinear transient growth scheme. C.J.C.\ receives an Australian Government Research Training Program (RTP) Scholarship. A.P.\ is supported by Wolfson Research Merit Award Scheme grant WM140032 from the Royal Society. This research was supported by the Australian Government via the Australian Research Council (Discovery Grants DP150102920 and DP180102647), the National Computational Infrastructure (NCI) and Pawsey Supercomputing Centre (PSC), and Monash University via the MonARCH cluster. 
\end{acknowledgments}


\begin{thebibliography}{63}%
\makeatletter
\providecommand \@ifxundefined [1]{%
 \@ifx{#1\undefined}
}%
\providecommand \@ifnum [1]{%
 \ifnum #1\expandafter \@firstoftwo
 \else \expandafter \@secondoftwo
 \fi
}%
\providecommand \@ifx [1]{%
 \ifx #1\expandafter \@firstoftwo
 \else \expandafter \@secondoftwo
 \fi
}%
\providecommand \natexlab [1]{#1}%
\providecommand \enquote  [1]{``#1''}%
\providecommand \bibnamefont  [1]{#1}%
\providecommand \bibfnamefont [1]{#1}%
\providecommand \citenamefont [1]{#1}%
\providecommand \href@noop [0]{\@secondoftwo}%
\providecommand \href [0]{\begingroup \@sanitize@url \@href}%
\providecommand \@href[1]{\@@startlink{#1}\@@href}%
\providecommand \@@href[1]{\endgroup#1\@@endlink}%
\providecommand \@sanitize@url [0]{\catcode `\\12\catcode `\$12\catcode
  `\&12\catcode `\#12\catcode `\^12\catcode `\_12\catcode `\%12\relax}%
\providecommand \@@startlink[1]{}%
\providecommand \@@endlink[0]{}%
\providecommand \url  [0]{\begingroup\@sanitize@url \@url }%
\providecommand \@url [1]{\endgroup\@href {#1}{\urlprefix }}%
\providecommand \urlprefix  [0]{URL }%
\providecommand \Eprint [0]{\href }%
\providecommand \doibase [0]{https://doi.org/}%
\providecommand \selectlanguage [0]{\@gobble}%
\providecommand \bibinfo  [0]{\@secondoftwo}%
\providecommand \bibfield  [0]{\@secondoftwo}%
\providecommand \translation [1]{[#1]}%
\providecommand \BibitemOpen [0]{}%
\providecommand \bibitemStop [0]{}%
\providecommand \bibitemNoStop [0]{.\EOS\space}%
\providecommand \EOS [0]{\spacefactor3000\relax}%
\providecommand \BibitemShut  [1]{\csname bibitem#1\endcsname}%
\let\auto@bib@innerbib\@empty
\bibitem [{\citenamefont {Lindborg}(1999)}]{Lindborg1999atmospheric}%
  \BibitemOpen
  \bibfield  {author} {\bibinfo {author} {\bibfnamefont {E.}~\bibnamefont
  {Lindborg}},\ }\bibfield  {title} {\bibinfo {title} {Can the atmospheric
  kinetic energy spectrum be explained by two-dimensional turbulence?},\ }\href
  {https://doi.org/10.1017/S0022112099004851} {\bibfield  {journal} {\bibinfo
  {journal} {J.\ Fluid Mech.}\ }\textbf {\bibinfo {volume} {388}},\ \bibinfo
  {pages} {259} (\bibinfo {year} {1999})}\BibitemShut {NoStop}%
\bibitem [{\citenamefont {{Poth\'{e}rat}}\ and\ \citenamefont
  {Schweitzer}(2011)}]{Potherat2011shallow}%
  \BibitemOpen
  \bibfield  {author} {\bibinfo {author} {\bibfnamefont {A.}~\bibnamefont
  {{Poth\'{e}rat}}}\ and\ \bibinfo {author} {\bibfnamefont {J.}~\bibnamefont
  {Schweitzer}},\ }\bibfield  {title} {\bibinfo {title} {A shallow water model
  for magnetohydrodynamic flows with turbulent {H}artmann layers},\ }\href
  {https://doi.org/10.1063/1.3592326} {\bibfield  {journal} {\bibinfo
  {journal} {Phys.\ Fluids}\ }\textbf {\bibinfo {volume} {23}},\ \bibinfo
  {pages} {055108} (\bibinfo {year} {2011})}\BibitemShut {NoStop}%
\bibitem [{\citenamefont {Sommeria}\ and\ \citenamefont
  {Moreau}(1982)}]{Sommeria1982why}%
  \BibitemOpen
  \bibfield  {author} {\bibinfo {author} {\bibfnamefont {J.}~\bibnamefont
  {Sommeria}}\ and\ \bibinfo {author} {\bibfnamefont {R.}~\bibnamefont
  {Moreau}},\ }\bibfield  {title} {\bibinfo {title} {Why, how, and when, {MHD}
  turbulence becomes two-dimensional},\ }\href
  {https://doi.org/10.1017/S0022112082001177} {\bibfield  {journal} {\bibinfo
  {journal} {J.\ Fluid Mech.}\ }\textbf {\bibinfo {volume} {118}},\ \bibinfo
  {pages} {507} (\bibinfo {year} {1982})}\BibitemShut {NoStop}%
\bibitem [{\citenamefont {Thess}\ and\ \citenamefont
  {Zikanov}(2007)}]{thess2007_jfm}%
  \BibitemOpen
  \bibfield  {author} {\bibinfo {author} {\bibfnamefont {A.}~\bibnamefont
  {Thess}}\ and\ \bibinfo {author} {\bibfnamefont {O.}~\bibnamefont
  {Zikanov}},\ }\bibfield  {title} {\bibinfo {title} {Transition from
  two-dimensional to three-dimensional magnetohydrodynamic turbulence},\ }\href
  {https://doi.org/doi:10.1017/S0022112007005277} {\bibfield  {journal}
  {\bibinfo  {journal} {J. Fluid Mech.}\ }\textbf {\bibinfo {volume} {579}},\
  \bibinfo {pages} {383} (\bibinfo {year} {2007})}\BibitemShut {NoStop}%
\bibitem [{\citenamefont {Klein}\ and\ \citenamefont
  {Poth\'erat}(2010)}]{klein2010_prl}%
  \BibitemOpen
  \bibfield  {author} {\bibinfo {author} {\bibfnamefont {R.}~\bibnamefont
  {Klein}}\ and\ \bibinfo {author} {\bibfnamefont {A.}~\bibnamefont
  {Poth\'erat}},\ }\bibfield  {title} {\bibinfo {title} {Appearance of
  three-dimensionality in wall bounded {MHD} flows},\ }\href
  {https://doi.org/10.1103/PhysRevLett.104.034502} {\bibfield  {journal}
  {\bibinfo  {journal} {Phys. Rev. Lett.}\ }\textbf {\bibinfo {volume} {104}},\
  \bibinfo {pages} {034502} (\bibinfo {year} {2010})}\BibitemShut {NoStop}%
\bibitem [{\citenamefont {{Poth\'{e}rat}}\ and\ \citenamefont
  {Klein}(2014)}]{Potherat2014why}%
  \BibitemOpen
  \bibfield  {author} {\bibinfo {author} {\bibfnamefont {A.}~\bibnamefont
  {{Poth\'{e}rat}}}\ and\ \bibinfo {author} {\bibfnamefont {R.}~\bibnamefont
  {Klein}},\ }\bibfield  {title} {\bibinfo {title} {Why, how and when {MHD}
  turbulence at low {$\mathit{R_m}$} becomes three-dimensional},\ }\href
  {https://doi.org/10.1017/jfm.2014.620} {\bibfield  {journal} {\bibinfo
  {journal} {J.\ Fluid Mech.}\ }\textbf {\bibinfo {volume} {761}},\ \bibinfo
  {pages} {168} (\bibinfo {year} {2014})}\BibitemShut {NoStop}%
\bibitem [{\citenamefont {Smolentsev}\ \emph {et~al.}(2008)\citenamefont
  {Smolentsev}, \citenamefont {Moreau},\ and\ \citenamefont
  {Abdou}}]{Smolentsev2008characterization}%
  \BibitemOpen
  \bibfield  {author} {\bibinfo {author} {\bibfnamefont {S.}~\bibnamefont
  {Smolentsev}}, \bibinfo {author} {\bibfnamefont {R.}~\bibnamefont {Moreau}},\
  and\ \bibinfo {author} {\bibfnamefont {M.}~\bibnamefont {Abdou}},\ }\bibfield
   {title} {\bibinfo {title} {Characterization of key magnetohydrodynamic
  phenomena in {PbLi} flows for the {US} {DCLL} blanket},\ }\href
  {https://doi.org/10.1016/j.fusengdes.2008.07.023} {\bibfield  {journal}
  {\bibinfo  {journal} {Fusion Eng.\ Des.}\ }\textbf {\bibinfo {volume} {83}},\
  \bibinfo {pages} {771} (\bibinfo {year} {2008})}\BibitemShut {NoStop}%
\bibitem [{\citenamefont {Kl{\"{u}}ber}\ \emph {et~al.}(2019)\citenamefont
  {Kl{\"{u}}ber}, \citenamefont {B{\"{u}}hler},\ and\ \citenamefont
  {Mistrangelo}}]{Kluber2019numerical}%
  \BibitemOpen
  \bibfield  {author} {\bibinfo {author} {\bibfnamefont {V.}~\bibnamefont
  {Kl{\"{u}}ber}}, \bibinfo {author} {\bibfnamefont {L.}~\bibnamefont
  {B{\"{u}}hler}},\ and\ \bibinfo {author} {\bibfnamefont {C.}~\bibnamefont
  {Mistrangelo}},\ }\bibfield  {title} {\bibinfo {title} {Numerical simulations
  of {3D} magnetohydrodynamic flows in dual-coolant lead lithium blankets},\
  }\href {https://doi.org/10.1016/j.fusengdes.2019.01.055} {\bibfield
  {journal} {\bibinfo  {journal} {Fusion Eng.\ Des.}\ }\textbf {\bibinfo
  {volume} {146}},\ \bibinfo {pages} {684} (\bibinfo {year}
  {2019})}\BibitemShut {NoStop}%
\bibitem [{\citenamefont {Barleon}\ \emph {et~al.}(2000)\citenamefont
  {Barleon}, \citenamefont {Burr}, \citenamefont {Mack},\ and\ \citenamefont
  {Stieglitz}}]{Barleon2000heat}%
  \BibitemOpen
  \bibfield  {author} {\bibinfo {author} {\bibfnamefont {L.}~\bibnamefont
  {Barleon}}, \bibinfo {author} {\bibfnamefont {U.}~\bibnamefont {Burr}},
  \bibinfo {author} {\bibfnamefont {K.~J.}\ \bibnamefont {Mack}},\ and\
  \bibinfo {author} {\bibfnamefont {R.}~\bibnamefont {Stieglitz}},\ }\bibfield
  {title} {\bibinfo {title} {Heat transfer in liquid metal cooled fusion
  blankets},\ }\href {https://doi.org/10.1016/S0920-3796(00)00212-X} {\bibfield
   {journal} {\bibinfo  {journal} {Fusion Eng.\ Des.}\ }\textbf {\bibinfo
  {volume} {51-52}},\ \bibinfo {pages} {723} (\bibinfo {year}
  {2000})}\BibitemShut {NoStop}%
\bibitem [{\citenamefont {Burr}\ \emph {et~al.}(2000)\citenamefont {Burr},
  \citenamefont {Barleon}, \citenamefont {{M{\"u}ller}},\ and\ \citenamefont
  {Tsinober}}]{Burr2000turbulent}%
  \BibitemOpen
  \bibfield  {author} {\bibinfo {author} {\bibfnamefont {U.}~\bibnamefont
  {Burr}}, \bibinfo {author} {\bibfnamefont {L.}~\bibnamefont {Barleon}},
  \bibinfo {author} {\bibfnamefont {U.}~\bibnamefont {{M{\"u}ller}}},\ and\
  \bibinfo {author} {\bibfnamefont {A.}~\bibnamefont {Tsinober}},\ }\bibfield
  {title} {\bibinfo {title} {Turbulent transport of momentum and heat in
  magnetohydrodynamic rectangular duct flow with strong sidewall jets},\ }\href
  {https://doi.org/10.1017/S0022112099007405} {\bibfield  {journal} {\bibinfo
  {journal} {J.\ Fluid Mech.}\ }\textbf {\bibinfo {volume} {406}},\ \bibinfo
  {pages} {247} (\bibinfo {year} {2000})}\BibitemShut {NoStop}%
\bibitem [{\citenamefont {Cassels}\ \emph {et~al.}(2016)\citenamefont
  {Cassels}, \citenamefont {Hussam},\ and\ \citenamefont
  {Sheard}}]{Cassels2016heat}%
  \BibitemOpen
  \bibfield  {author} {\bibinfo {author} {\bibfnamefont {O.~G.~W.}\
  \bibnamefont {Cassels}}, \bibinfo {author} {\bibfnamefont {W.~K.}\
  \bibnamefont {Hussam}},\ and\ \bibinfo {author} {\bibfnamefont {G.~J.}\
  \bibnamefont {Sheard}},\ }\bibfield  {title} {\bibinfo {title} {Heat transfer
  enhancement using rectangular vortex promoters in confined
  quasi-two-dimensional magnetohydrodynamic flows},\ }\href
  {https://doi.org/10.1016/j.ijheatmasstransfer.2015.10.006} {\bibfield
  {journal} {\bibinfo  {journal} {Int.\ J.\ Heat Mass Transf.}\ }\textbf
  {\bibinfo {volume} {93}},\ \bibinfo {pages} {186} (\bibinfo {year}
  {2016})}\BibitemShut {NoStop}%
\bibitem [{\citenamefont {Mistrangelo}\ and\ \citenamefont
  {B{\"{u}}hler}(2009)}]{Mistrangelo2009influence}%
  \BibitemOpen
  \bibfield  {author} {\bibinfo {author} {\bibfnamefont {C.}~\bibnamefont
  {Mistrangelo}}\ and\ \bibinfo {author} {\bibfnamefont {L.}~\bibnamefont
  {B{\"{u}}hler}},\ }\bibfield  {title} {\bibinfo {title} {Influence of helium
  cooling channels on magnetohydrodynamic flows in the {HCLL} blanket},\ }\href
  {https://doi.org/10.1016/j.fusengdes.2009.01.055} {\bibfield  {journal}
  {\bibinfo  {journal} {Fusion Eng.\ Des.}\ }\textbf {\bibinfo {volume} {84}},\
  \bibinfo {pages} {1323} (\bibinfo {year} {2009})}\BibitemShut {NoStop}%
\bibitem [{\citenamefont {Mistrangelo}\ \emph {et~al.}(2014)\citenamefont
  {Mistrangelo}, \citenamefont {{B{\"u}hler}},\ and\ \citenamefont
  {Aiello}}]{Mistrangelo2014buoyant}%
  \BibitemOpen
  \bibfield  {author} {\bibinfo {author} {\bibfnamefont {C.}~\bibnamefont
  {Mistrangelo}}, \bibinfo {author} {\bibfnamefont {L.}~\bibnamefont
  {{B{\"u}hler}}},\ and\ \bibinfo {author} {\bibfnamefont {G.}~\bibnamefont
  {Aiello}},\ }\bibfield  {title} {\bibinfo {title} {Buoyant-{MHD} flows in
  {HCLL} blankets caused by spatially varying thermal loads},\ }\href
  {https://doi.org/10.1109/TPS.2014.2311510} {\bibfield  {journal} {\bibinfo
  {journal} {IEEE Trans.\ Plasma Sci.}\ }\textbf {\bibinfo {volume} {42}},\
  \bibinfo {pages} {1407} (\bibinfo {year} {2014})}\BibitemShut {NoStop}%
\bibitem [{\citenamefont {Smolentsev}\ \emph {et~al.}(2010)\citenamefont
  {Smolentsev}, \citenamefont {Wong}, \citenamefont {Malang}, \citenamefont
  {Dagher},\ and\ \citenamefont {Abdou}}]{Smolentsev2010considerations}%
  \BibitemOpen
  \bibfield  {author} {\bibinfo {author} {\bibfnamefont {S.}~\bibnamefont
  {Smolentsev}}, \bibinfo {author} {\bibfnamefont {C.}~\bibnamefont {Wong}},
  \bibinfo {author} {\bibfnamefont {S.}~\bibnamefont {Malang}}, \bibinfo
  {author} {\bibfnamefont {M.}~\bibnamefont {Dagher}},\ and\ \bibinfo {author}
  {\bibfnamefont {M.}~\bibnamefont {Abdou}},\ }\bibfield  {title} {\bibinfo
  {title} {{MHD} considerations for the {DCLL} inboard blanket and access
  ducts},\ }\href {https://doi.org/10.1016/j.fusengdes.2009.12.005} {\bibfield
  {journal} {\bibinfo  {journal} {Fusion Eng.\ Des.}\ }\textbf {\bibinfo
  {volume} {85}},\ \bibinfo {pages} {1007} (\bibinfo {year}
  {2010})}\BibitemShut {NoStop}%
\bibitem [{\citenamefont {Mistrangelo}\ and\ \citenamefont
  {B{\"{u}}hler}(2011)}]{Mistrangelo2011magnetohydrodynamic}%
  \BibitemOpen
  \bibfield  {author} {\bibinfo {author} {\bibfnamefont {C.}~\bibnamefont
  {Mistrangelo}}\ and\ \bibinfo {author} {\bibfnamefont {L.}~\bibnamefont
  {B{\"{u}}hler}},\ }\bibfield  {title} {\bibinfo {title} {Magnetohydrodynamic
  pressure drops in geometric elements forming a {HCLL} blanket mock-up},\
  }\href {https://doi.org/10.1016/j.fusengdes.2011.03.011} {\bibfield
  {journal} {\bibinfo  {journal} {Fusion Eng.\ Des.}\ }\textbf {\bibinfo
  {volume} {86}},\ \bibinfo {pages} {2304} (\bibinfo {year}
  {2011})}\BibitemShut {NoStop}%
\bibitem [{\citenamefont {Hussam}\ \emph
  {et~al.}(2012{\natexlab{a}})\citenamefont {Hussam}, \citenamefont
  {Thompson},\ and\ \citenamefont {Sheard}}]{Hussam2012enhancing}%
  \BibitemOpen
  \bibfield  {author} {\bibinfo {author} {\bibfnamefont {W.~K.}\ \bibnamefont
  {Hussam}}, \bibinfo {author} {\bibfnamefont {M.~C.}\ \bibnamefont
  {Thompson}},\ and\ \bibinfo {author} {\bibfnamefont {G.~J.}\ \bibnamefont
  {Sheard}},\ }\bibfield  {title} {\bibinfo {title} {Enhancing heat transfer in
  a high {H}artmann number magnetohydrodynamic channel flow via torsional
  oscillation of a cylindrical obstacle},\ }\href
  {https://doi.org/10.1063/1.4767515} {\bibfield  {journal} {\bibinfo
  {journal} {Phys.\ Fluids}\ }\textbf {\bibinfo {volume} {24}},\ \bibinfo
  {pages} {113601} (\bibinfo {year} {2012}{\natexlab{a}})}\BibitemShut
  {NoStop}%
\bibitem [{\citenamefont {Hamid}\ \emph
  {et~al.}(2016{\natexlab{a}})\citenamefont {Hamid}, \citenamefont {Hussam},\
  and\ \citenamefont {Sheard}}]{Hamid2016combining}%
  \BibitemOpen
  \bibfield  {author} {\bibinfo {author} {\bibfnamefont {A.~H.~A.}\
  \bibnamefont {Hamid}}, \bibinfo {author} {\bibfnamefont {W.~K.}\ \bibnamefont
  {Hussam}},\ and\ \bibinfo {author} {\bibfnamefont {G.~J.}\ \bibnamefont
  {Sheard}},\ }\bibfield  {title} {\bibinfo {title} {Combining an obstacle and
  electrically driven vortices to enhance heat transfer in a
  quasi-two-dimensional {MHD} duct flow},\ }\href
  {https://doi.org/10.1017/jfm.2016.90} {\bibfield  {journal} {\bibinfo
  {journal} {J.\ Fluid Mech.}\ }\textbf {\bibinfo {volume} {792}},\ \bibinfo
  {pages} {364} (\bibinfo {year} {2016}{\natexlab{a}})}\BibitemShut {NoStop}%
\bibitem [{\citenamefont {Hamid}\ \emph
  {et~al.}(2016{\natexlab{b}})\citenamefont {Hamid}, \citenamefont {Hussam},\
  and\ \citenamefont {Sheard}}]{Hamid2016heat}%
  \BibitemOpen
  \bibfield  {author} {\bibinfo {author} {\bibfnamefont {A.~H.~A.}\
  \bibnamefont {Hamid}}, \bibinfo {author} {\bibfnamefont {W.~K.}\ \bibnamefont
  {Hussam}},\ and\ \bibinfo {author} {\bibfnamefont {G.~J.}\ \bibnamefont
  {Sheard}},\ }\bibfield  {title} {\bibinfo {title} {Heat transfer augmentation
  of a quasi-two-dimensional {MHD} duct flow via electrically driven
  vortices},\ }\href {https://doi.org/10.1080/10407782.2016.1214518} {\bibfield
   {journal} {\bibinfo  {journal} {Numer.\ Heat Tr.\ A-Appl.}\ }\textbf
  {\bibinfo {volume} {70}},\ \bibinfo {pages} {847} (\bibinfo {year}
  {2016}{\natexlab{b}})}\BibitemShut {NoStop}%
\bibitem [{\citenamefont {Krasnov}\ \emph {et~al.}(2004)\citenamefont
  {Krasnov}, \citenamefont {Zienicke}, \citenamefont {Zikanov}, \citenamefont
  {Boeck},\ and\ \citenamefont {Thess}}]{Krasnov2004numerical}%
  \BibitemOpen
  \bibfield  {author} {\bibinfo {author} {\bibfnamefont {D.~S.}\ \bibnamefont
  {Krasnov}}, \bibinfo {author} {\bibfnamefont {E.}~\bibnamefont {Zienicke}},
  \bibinfo {author} {\bibfnamefont {O.}~\bibnamefont {Zikanov}}, \bibinfo
  {author} {\bibfnamefont {T.}~\bibnamefont {Boeck}},\ and\ \bibinfo {author}
  {\bibfnamefont {A.}~\bibnamefont {Thess}},\ }\bibfield  {title} {\bibinfo
  {title} {Numerical study of the instability of the {H}artmann layer},\ }\href
  {https://doi.org/10.1017/S0022112004008006} {\bibfield  {journal} {\bibinfo
  {journal} {J.\ Fluid Mech.}\ }\textbf {\bibinfo {volume} {504}},\ \bibinfo
  {pages} {183} (\bibinfo {year} {2004})}\BibitemShut {NoStop}%
\bibitem [{\citenamefont {Krasnov}\ \emph {et~al.}(2008)\citenamefont
  {Krasnov}, \citenamefont {Rossi}, \citenamefont {Zikanov},\ and\
  \citenamefont {Boeck}}]{Krasnov2008optimal}%
  \BibitemOpen
  \bibfield  {author} {\bibinfo {author} {\bibfnamefont {D.}~\bibnamefont
  {Krasnov}}, \bibinfo {author} {\bibfnamefont {M.}~\bibnamefont {Rossi}},
  \bibinfo {author} {\bibfnamefont {O.}~\bibnamefont {Zikanov}},\ and\ \bibinfo
  {author} {\bibfnamefont {T.}~\bibnamefont {Boeck}},\ }\bibfield  {title}
  {\bibinfo {title} {Optimal growth and transition to turbulence in channel
  flow with spanwise magnetic field},\ }\href
  {https://doi.org/10.1017/S002211200700924X} {\bibfield  {journal} {\bibinfo
  {journal} {J.\ Fluid Mech.}\ }\textbf {\bibinfo {volume} {596}},\ \bibinfo
  {pages} {73} (\bibinfo {year} {2008})}\BibitemShut {NoStop}%
\bibitem [{\citenamefont {Moresco}\ and\ \citenamefont
  {Alboussi{\'{e}}re}(2004)}]{Moresco2004experimental}%
  \BibitemOpen
  \bibfield  {author} {\bibinfo {author} {\bibfnamefont {P.}~\bibnamefont
  {Moresco}}\ and\ \bibinfo {author} {\bibfnamefont {T.}~\bibnamefont
  {Alboussi{\'{e}}re}},\ }\bibfield  {title} {\bibinfo {title} {Experimental
  study of the instability of the {H}artmann layer},\ }\href
  {https://doi.org/10.1017/S0022112004007992} {\bibfield  {journal} {\bibinfo
  {journal} {J.\ Fluid Mech.}\ }\textbf {\bibinfo {volume} {504}},\ \bibinfo
  {pages} {167} (\bibinfo {year} {2004})}\BibitemShut {NoStop}%
\bibitem [{\citenamefont {M{\"{u}}ck}\ \emph {et~al.}(2000)\citenamefont
  {M{\"{u}}ck}, \citenamefont {G{\"{u}}nther}, \citenamefont {M{\"{u}}ller},\
  and\ \citenamefont {B{\"{u}}hler}}]{Muck2000flows}%
  \BibitemOpen
  \bibfield  {author} {\bibinfo {author} {\bibfnamefont {B.}~\bibnamefont
  {M{\"{u}}ck}}, \bibinfo {author} {\bibfnamefont {C.}~\bibnamefont
  {G{\"{u}}nther}}, \bibinfo {author} {\bibfnamefont {U.}~\bibnamefont
  {M{\"{u}}ller}},\ and\ \bibinfo {author} {\bibfnamefont {L.}~\bibnamefont
  {B{\"{u}}hler}},\ }\bibfield  {title} {\bibinfo {title} {Three-dimensional
  {MHD} flows in rectangular ducts with internal obstacles},\ }\href
  {https://doi.org/10.1017/S0022112000001300} {\bibfield  {journal} {\bibinfo
  {journal} {J.\ Fluid Mech.}\ }\textbf {\bibinfo {volume} {418}},\ \bibinfo
  {pages} {265} (\bibinfo {year} {2000})}\BibitemShut {NoStop}%
\bibitem [{\citenamefont {{Poth\'{e}rat}}\ \emph {et~al.}(2000)\citenamefont
  {{Poth\'{e}rat}}, \citenamefont {Sommeria},\ and\ \citenamefont
  {Moreau}}]{Potherat2000effective}%
  \BibitemOpen
  \bibfield  {author} {\bibinfo {author} {\bibfnamefont {A.}~\bibnamefont
  {{Poth\'{e}rat}}}, \bibinfo {author} {\bibfnamefont {J.}~\bibnamefont
  {Sommeria}},\ and\ \bibinfo {author} {\bibfnamefont {R.}~\bibnamefont
  {Moreau}},\ }\bibfield  {title} {\bibinfo {title} {An effective
  two-dimensional model for {MHD} flows with a transverse magnetic field},\
  }\href {https://doi.org/10.1017/S0022112000001944} {\bibfield  {journal}
  {\bibinfo  {journal} {J.\ Fluid Mech.}\ }\textbf {\bibinfo {volume} {424}},\
  \bibinfo {pages} {75} (\bibinfo {year} {2000})}\BibitemShut {NoStop}%
\bibitem [{\citenamefont {Dousset}\ and\ \citenamefont
  {Poth{\'{e}}rat}(2008)}]{Dousset2008numerical}%
  \BibitemOpen
  \bibfield  {author} {\bibinfo {author} {\bibfnamefont {V.}~\bibnamefont
  {Dousset}}\ and\ \bibinfo {author} {\bibfnamefont {A.}~\bibnamefont
  {Poth{\'{e}}rat}},\ }\bibfield  {title} {\bibinfo {title} {Numerical
  simulations of a cylinder wake under a strong axial magnetic field},\ }\href
  {https://doi.org/10.1063/1.2831153} {\bibfield  {journal} {\bibinfo
  {journal} {Phys.\ Fluids}\ }\textbf {\bibinfo {volume} {20}},\ \bibinfo
  {pages} {7104} (\bibinfo {year} {2008})}\BibitemShut {NoStop}%
\bibitem [{\citenamefont {Kanaris}\ \emph {et~al.}(2013)\citenamefont
  {Kanaris}, \citenamefont {Albets}, \citenamefont {Grigoriadis},\ and\
  \citenamefont {Kassinos}}]{Kanaris2013numerical}%
  \BibitemOpen
  \bibfield  {author} {\bibinfo {author} {\bibfnamefont {N.}~\bibnamefont
  {Kanaris}}, \bibinfo {author} {\bibfnamefont {X.}~\bibnamefont {Albets}},
  \bibinfo {author} {\bibfnamefont {D.}~\bibnamefont {Grigoriadis}},\ and\
  \bibinfo {author} {\bibfnamefont {S.}~\bibnamefont {Kassinos}},\ }\bibfield
  {title} {\bibinfo {title} {Three-dimensional numerical simulations of
  magnetohydrodynamic flow around a confined circular cylinder under low,
  moderate, and strong magnetic fields},\ }\href
  {https://doi.org/10.1063/1.4811398} {\bibfield  {journal} {\bibinfo
  {journal} {Phys.\ Fluids}\ }\textbf {\bibinfo {volume} {25}},\ \bibinfo
  {pages} {074102} (\bibinfo {year} {2013})}\BibitemShut {NoStop}%
\bibitem [{\citenamefont {Cassels}\ \emph {et~al.}(2018)\citenamefont
  {Cassels}, \citenamefont {Vo}, \citenamefont {Poth{\'e}rat},\ and\
  \citenamefont {Sheard}}]{Cassels2019from3D}%
  \BibitemOpen
  \bibfield  {author} {\bibinfo {author} {\bibfnamefont {O.~G.~W.}\
  \bibnamefont {Cassels}}, \bibinfo {author} {\bibfnamefont {T.}~\bibnamefont
  {Vo}}, \bibinfo {author} {\bibfnamefont {A.}~\bibnamefont {Poth{\'e}rat}},\
  and\ \bibinfo {author} {\bibfnamefont {G.~J.}\ \bibnamefont {Sheard}},\
  }\bibfield  {title} {\bibinfo {title} {From three-dimensional to
  quasi-two-dimensional: transient growth in magnetohydrodynamic duct flows},\
  }\href {https://doi.org/10.1017/jfm.2018.863} {\bibfield  {journal} {\bibinfo
   {journal} {J.\ Fluid Mech.}\ }\textbf {\bibinfo {volume} {861}},\ \bibinfo
  {pages} {382} (\bibinfo {year} {2019})}\BibitemShut {NoStop}%
\bibitem [{\citenamefont {{Poth\'{e}rat}}(2007)}]{Potherat2007quasi}%
  \BibitemOpen
  \bibfield  {author} {\bibinfo {author} {\bibfnamefont {A.}~\bibnamefont
  {{Poth\'{e}rat}}},\ }\bibfield  {title} {\bibinfo {title}
  {Quasi-two-dimensional perturbations in duct flows under transverse magnetic
  field},\ }\href {https://doi.org/10.1063/1.2747233} {\bibfield  {journal}
  {\bibinfo  {journal} {Phys.\ Fluids}\ }\textbf {\bibinfo {volume} {19}},\
  \bibinfo {pages} {074104} (\bibinfo {year} {2007})}\BibitemShut {NoStop}%
\bibitem [{\citenamefont {Vo}\ \emph {et~al.}(2017)\citenamefont {Vo},
  \citenamefont {Poth{\'e}rat},\ and\ \citenamefont {Sheard}}]{Vo2017linear}%
  \BibitemOpen
  \bibfield  {author} {\bibinfo {author} {\bibfnamefont {T.}~\bibnamefont
  {Vo}}, \bibinfo {author} {\bibfnamefont {A.}~\bibnamefont {Poth{\'e}rat}},\
  and\ \bibinfo {author} {\bibfnamefont {G.~J.}\ \bibnamefont {Sheard}},\
  }\bibfield  {title} {\bibinfo {title} {Linear stability of horizontal,
  laminar fully developed, quasi-two-dimensional liquid metal duct flow under a
  transverse magnetic field heated from below},\ }\href
  {https://doi.org/10.1103/PhysRevFluids.2.033902} {\bibfield  {journal}
  {\bibinfo  {journal} {Phys.\ Rev.\ Fluids}\ }\textbf {\bibinfo {volume}
  {2}},\ \bibinfo {pages} {033902} (\bibinfo {year} {2017})}\BibitemShut
  {NoStop}%
\bibitem [{\citenamefont {Krasnov}\ \emph {et~al.}(2010)\citenamefont
  {Krasnov}, \citenamefont {Zikanov}, \citenamefont {Rossi},\ and\
  \citenamefont {Boeck}}]{Krasnov2010optimal}%
  \BibitemOpen
  \bibfield  {author} {\bibinfo {author} {\bibfnamefont {D.}~\bibnamefont
  {Krasnov}}, \bibinfo {author} {\bibfnamefont {O.}~\bibnamefont {Zikanov}},
  \bibinfo {author} {\bibfnamefont {M.}~\bibnamefont {Rossi}},\ and\ \bibinfo
  {author} {\bibfnamefont {T.}~\bibnamefont {Boeck}},\ }\bibfield  {title}
  {\bibinfo {title} {Optimal linear growth in magnetohydrodynamic duct flow},\
  }\href {https://doi.org/10.1017/S0022112010000273} {\bibfield  {journal}
  {\bibinfo  {journal} {J.\ Fluid Mech.}\ }\textbf {\bibinfo {volume} {653}},\
  \bibinfo {pages} {273} (\bibinfo {year} {2010})}\BibitemShut {NoStop}%
\bibitem [{\citenamefont {Roberts}(1967)}]{Roberts1967introduction}%
  \BibitemOpen
  \bibfield  {author} {\bibinfo {author} {\bibfnamefont {P.~H.}\ \bibnamefont
  {Roberts}},\ }\href@noop {} {\emph {\bibinfo {title} {An Introduction to
  Magnetohydrodynamics}}}\ (\bibinfo  {publisher} {Longmans, Green New York},\
  \bibinfo {year} {1967})\BibitemShut {NoStop}%
\bibitem [{\citenamefont {Levin}\ \emph {et~al.}(2005)\citenamefont {Levin},
  \citenamefont {Davidsson},\ and\ \citenamefont
  {Henningson}}]{Levin2005transition}%
  \BibitemOpen
  \bibfield  {author} {\bibinfo {author} {\bibfnamefont {O.}~\bibnamefont
  {Levin}}, \bibinfo {author} {\bibfnamefont {E.~N.}\ \bibnamefont
  {Davidsson}},\ and\ \bibinfo {author} {\bibfnamefont {D.~S.}\ \bibnamefont
  {Henningson}},\ }\bibfield  {title} {\bibinfo {title} {Transition thresholds
  in the asymptotic suction boundary layer},\ }\href
  {https://doi.org/doi.org/10.1063/1.2136900} {\bibfield  {journal} {\bibinfo
  {journal} {Phys.\ Fluids}\ }\textbf {\bibinfo {volume} {17}},\ \bibinfo
  {pages} {4104} (\bibinfo {year} {2005})}\BibitemShut {NoStop}%
\bibitem [{\citenamefont {Albrecht}\ \emph {et~al.}(2006)\citenamefont
  {Albrecht}, \citenamefont {Grundmann}, \citenamefont {Mutschke},\ and\
  \citenamefont {Gerbeth}}]{Albrecht2006stability}%
  \BibitemOpen
  \bibfield  {author} {\bibinfo {author} {\bibfnamefont {T.}~\bibnamefont
  {Albrecht}}, \bibinfo {author} {\bibfnamefont {R.}~\bibnamefont {Grundmann}},
  \bibinfo {author} {\bibfnamefont {G.}~\bibnamefont {Mutschke}},\ and\
  \bibinfo {author} {\bibfnamefont {G.}~\bibnamefont {Gerbeth}},\ }\bibfield
  {title} {\bibinfo {title} {On the stability of the boundary layer subject to
  a wall-parallel {L}orentz force},\ }\href {https://doi.org/10.1063/1.2353401}
  {\bibfield  {journal} {\bibinfo  {journal} {Phys.\ Fluids}\ }\textbf
  {\bibinfo {volume} {18}},\ \bibinfo {pages} {098103} (\bibinfo {year}
  {2006})}\BibitemShut {NoStop}%
\bibitem [{\citenamefont {Pringle}\ \emph {et~al.}(2012)\citenamefont
  {Pringle}, \citenamefont {Willis},\ and\ \citenamefont
  {Kerswell}}]{Pringle2012minimal}%
  \BibitemOpen
  \bibfield  {author} {\bibinfo {author} {\bibfnamefont {C.~C.~T.}\
  \bibnamefont {Pringle}}, \bibinfo {author} {\bibfnamefont {A.~P.}\
  \bibnamefont {Willis}},\ and\ \bibinfo {author} {\bibfnamefont {R.~R.}\
  \bibnamefont {Kerswell}},\ }\bibfield  {title} {\bibinfo {title} {Minimal
  seeds for shear flow turbulence: using nonlinear transient growth to touch
  the edge of chaos},\ }\href {https://doi.org/10.1017/jfm.2012.192} {\bibfield
   {journal} {\bibinfo  {journal} {J.\ Fluid Mech.}\ }\textbf {\bibinfo
  {volume} {702}},\ \bibinfo {pages} {415} (\bibinfo {year}
  {2012})}\BibitemShut {NoStop}%
\bibitem [{\citenamefont {Kerswell}\ \emph {et~al.}(2014)\citenamefont
  {Kerswell}, \citenamefont {Pringle},\ and\ \citenamefont
  {Willis}}]{Kerswell2014optimization}%
  \BibitemOpen
  \bibfield  {author} {\bibinfo {author} {\bibfnamefont {R.~R.}\ \bibnamefont
  {Kerswell}}, \bibinfo {author} {\bibfnamefont {C.~C.~T.}\ \bibnamefont
  {Pringle}},\ and\ \bibinfo {author} {\bibfnamefont {A.~P.}\ \bibnamefont
  {Willis}},\ }\bibfield  {title} {\bibinfo {title} {An optimization approach
  for analysing nonlinear stability with transition to turbulence in fluids as
  an exemplar},\ }\href {https://doi.org/10.1088/0034-4885/77/8/085901}
  {\bibfield  {journal} {\bibinfo  {journal} {Rep.\ Prog.\ Phys.}\ }\textbf
  {\bibinfo {volume} {77}},\ \bibinfo {pages} {085901} (\bibinfo {year}
  {2014})}\BibitemShut {NoStop}%
\bibitem [{\citenamefont {Duguet}\ \emph {et~al.}(2009)\citenamefont {Duguet},
  \citenamefont {Schlatter},\ and\ \citenamefont
  {Henningson}}]{Duguet2009localized}%
  \BibitemOpen
  \bibfield  {author} {\bibinfo {author} {\bibfnamefont {Y.}~\bibnamefont
  {Duguet}}, \bibinfo {author} {\bibfnamefont {P.}~\bibnamefont {Schlatter}},\
  and\ \bibinfo {author} {\bibfnamefont {D.~S.}\ \bibnamefont {Henningson}},\
  }\bibfield  {title} {\bibinfo {title} {Localized edge states in plane Couette
  flow},\ }\href {https://doi.org/10.1063/1.3265962} {\bibfield  {journal}
  {\bibinfo  {journal} {Phys.\ Fluids}\ }\textbf {\bibinfo {volume} {21}},\
  \bibinfo {pages} {1701} (\bibinfo {year} {2009})}\BibitemShut {NoStop}%
\bibitem [{\citenamefont {Duguet}\ \emph {et~al.}(2013)\citenamefont {Duguet},
  \citenamefont {Monokrousos}, \citenamefont {Brandt},\ and\ \citenamefont
  {Henningson}}]{Duguet2013minimal}%
  \BibitemOpen
  \bibfield  {author} {\bibinfo {author} {\bibfnamefont {Y.}~\bibnamefont
  {Duguet}}, \bibinfo {author} {\bibfnamefont {A.}~\bibnamefont {Monokrousos}},
  \bibinfo {author} {\bibfnamefont {L.}~\bibnamefont {Brandt}},\ and\ \bibinfo
  {author} {\bibfnamefont {D.~S.}\ \bibnamefont {Henningson}},\ }\bibfield
  {title} {\bibinfo {title} {Minimal transition thresholds in plane Couette
  flow},\ }\href {https://doi.org/10.1063/1.4817328} {\bibfield  {journal}
  {\bibinfo  {journal} {Phys.\ Fluids}\ }\textbf {\bibinfo {volume} {25}},\
  \bibinfo {pages} {4103} (\bibinfo {year} {2013})}\BibitemShut {NoStop}%
\bibitem [{\citenamefont {Farano}\ \emph {et~al.}(2016)\citenamefont {Farano},
  \citenamefont {Cherubini}, \citenamefont {Robinet},\ and\ \citenamefont
  {Palma}}]{Farano2016subcritical}%
  \BibitemOpen
  \bibfield  {author} {\bibinfo {author} {\bibfnamefont {M.}~\bibnamefont
  {Farano}}, \bibinfo {author} {\bibfnamefont {S.}~\bibnamefont {Cherubini}},
  \bibinfo {author} {\bibfnamefont {J.-C.}\ \bibnamefont {Robinet}},\ and\
  \bibinfo {author} {\bibfnamefont {P.~D.}\ \bibnamefont {Palma}},\ }\bibfield
  {title} {\bibinfo {title} {Subcritical transition scenarios via linear and
  nonlinear localized optimal perturbations in plane Poiseuille flow},\ }\href
  {https://doi.org/10.1088/0169-5983/48/6/061409} {\bibfield  {journal}
  {\bibinfo  {journal} {Fluid Dyn.\ Res.}\ }\textbf {\bibinfo {volume} {48}},\
  \bibinfo {pages} {1409} (\bibinfo {year} {2016})}\BibitemShut {NoStop}%
\bibitem [{\citenamefont {Zammert}\ and\ \citenamefont
  {Eckhardt}(2019)}]{Zammert2019transition}%
  \BibitemOpen
  \bibfield  {author} {\bibinfo {author} {\bibfnamefont {S.}~\bibnamefont
  {Zammert}}\ and\ \bibinfo {author} {\bibfnamefont {B.}~\bibnamefont
  {Eckhardt}},\ }\bibfield  {title} {\bibinfo {title} {Transition to turbulence
  when the {T}ollmien--{S}chlichting and bypass routes coexist},\ }\href
  {https://doi.org/10.1017/jfm.2019.724} {\bibfield  {journal} {\bibinfo
  {journal} {J.\ Fluid Mech.}\ }\textbf {\bibinfo {volume} {880}},\ \bibinfo
  {pages} {R2} (\bibinfo {year} {2019})}\BibitemShut {NoStop}%
\bibitem [{\citenamefont {Duguet}\ \emph {et~al.}(2012)\citenamefont {Duguet},
  \citenamefont {Schlatter}, \citenamefont {Henningson},\ and\ \citenamefont
  {Eckhardt}}]{Duguet2012self}%
  \BibitemOpen
  \bibfield  {author} {\bibinfo {author} {\bibfnamefont {Y.}~\bibnamefont
  {Duguet}}, \bibinfo {author} {\bibfnamefont {P.}~\bibnamefont {Schlatter}},
  \bibinfo {author} {\bibfnamefont {D.~S.}\ \bibnamefont {Henningson}},\ and\
  \bibinfo {author} {\bibfnamefont {B.}~\bibnamefont {Eckhardt}},\ }\bibfield
  {title} {\bibinfo {title} {Self-sustained localized structures in a boundary
  layer flow},\ }\href {https://doi.org/10.1103/PhysRevLett.108.044501}
  {\bibfield  {journal} {\bibinfo  {journal} {Phys.\ Rev.\ Lett.}\ }\textbf
  {\bibinfo {volume} {108}},\ \bibinfo {pages} {4501} (\bibinfo {year}
  {2012})}\BibitemShut {NoStop}%
\bibitem [{\citenamefont {Cherubini}\ \emph {et~al.}(2011)\citenamefont
  {Cherubini}, \citenamefont {Palma}, \citenamefont {Robinet},\ and\
  \citenamefont {Bottaro}}]{Cherubini2011minimal}%
  \BibitemOpen
  \bibfield  {author} {\bibinfo {author} {\bibfnamefont {S.}~\bibnamefont
  {Cherubini}}, \bibinfo {author} {\bibfnamefont {P.~D.}\ \bibnamefont
  {Palma}}, \bibinfo {author} {\bibfnamefont {J.-C.}\ \bibnamefont {Robinet}},\
  and\ \bibinfo {author} {\bibfnamefont {A.}~\bibnamefont {Bottaro}},\
  }\bibfield  {title} {\bibinfo {title} {The minimal seed of turbulent
  transition in the boundary layer},\ }\href
  {https://doi.org/10.1017/jfm.2011.412} {\bibfield  {journal} {\bibinfo
  {journal} {J.\ Fluid Mech.}\ }\textbf {\bibinfo {volume} {689}},\ \bibinfo
  {pages} {221} (\bibinfo {year} {2011})}\BibitemShut {NoStop}%
\bibitem [{\citenamefont {Beneitez}\ \emph {et~al.}(2019)\citenamefont
  {Beneitez}, \citenamefont {Duguet}, \citenamefont {Schlatter},\ and\
  \citenamefont {Henningson}}]{Beneitez2019edge}%
  \BibitemOpen
  \bibfield  {author} {\bibinfo {author} {\bibfnamefont {M.}~\bibnamefont
  {Beneitez}}, \bibinfo {author} {\bibfnamefont {Y.}~\bibnamefont {Duguet}},
  \bibinfo {author} {\bibfnamefont {P.}~\bibnamefont {Schlatter}},\ and\
  \bibinfo {author} {\bibfnamefont {D.~S.}\ \bibnamefont {Henningson}},\
  }\bibfield  {title} {\bibinfo {title} {Edge tracking in spatially developing
  boundary layer flows},\ }\href {https://doi.org/10.1017/jfm.2019.763}
  {\bibfield  {journal} {\bibinfo  {journal} {J.\ Fluid Mech.}\ }\textbf
  {\bibinfo {volume} {881}},\ \bibinfo {pages} {164} (\bibinfo {year}
  {2019})}\BibitemShut {NoStop}%
\bibitem [{\citenamefont {Vavaliaris}\ \emph {et~al.}(2020)\citenamefont
  {Vavaliaris}, \citenamefont {Beneitez},\ and\ \citenamefont
  {Henningson}}]{Vavaliaris2020optimal}%
  \BibitemOpen
  \bibfield  {author} {\bibinfo {author} {\bibfnamefont {C.}~\bibnamefont
  {Vavaliaris}}, \bibinfo {author} {\bibfnamefont {M.}~\bibnamefont
  {Beneitez}},\ and\ \bibinfo {author} {\bibfnamefont {D.~S.}\ \bibnamefont
  {Henningson}},\ }\bibfield  {title} {\bibinfo {title} {Optimal perturbations
  and transitions thresholds in boundary layer shear flows},\ }\bibfield
  {journal} {\bibinfo  {journal} {Phys.\ Rev.\ Fluids}\ }\textbf
  {\bibinfo {volume} {5}},\ \bibinfo {pages} {062401(R)} (\bibinfo {year} {2020})\BibitemShut
  {NoStop}%
\bibitem [{\citenamefont {Khapko}\ \emph {et~al.}(2014)\citenamefont {Khapko},
  \citenamefont {Duguet}, \citenamefont {Kreilos}, \citenamefont {Schlatter},
  \citenamefont {Eckhardt},\ and\ \citenamefont
  {Henningson}}]{Khapko2014complexity}%
  \BibitemOpen
  \bibfield  {author} {\bibinfo {author} {\bibfnamefont {T.}~\bibnamefont
  {Khapko}}, \bibinfo {author} {\bibfnamefont {Y.}~\bibnamefont {Duguet}},
  \bibinfo {author} {\bibfnamefont {T.}~\bibnamefont {Kreilos}}, \bibinfo
  {author} {\bibfnamefont {P.}~\bibnamefont {Schlatter}}, \bibinfo {author}
  {\bibfnamefont {B.}~\bibnamefont {Eckhardt}},\ and\ \bibinfo {author}
  {\bibfnamefont {D.~S.}\ \bibnamefont {Henningson}},\ }\bibfield  {title}
  {\bibinfo {title} {Complexity of localised coherent structures in a
  boundary-layer flow},\ }\href {https://doi.org/10.1140/epje/i2014-14032-3}
  {\bibfield  {journal} {\bibinfo  {journal} {Eur.\ Phys.\ J.\ E}\ }\textbf
  {\bibinfo {volume} {37}},\ \bibinfo {pages} {1} (\bibinfo {year}
  {2014})}\BibitemShut {NoStop}%
\bibitem [{\citenamefont {Cherubini}\ \emph {et~al.}(2015)\citenamefont
  {Cherubini}, \citenamefont {Palma},\ and\ \citenamefont
  {Robinet}}]{Cherubini2015nonlinear}%
  \BibitemOpen
  \bibfield  {author} {\bibinfo {author} {\bibfnamefont {S.}~\bibnamefont
  {Cherubini}}, \bibinfo {author} {\bibfnamefont {P.~D.}\ \bibnamefont
  {Palma}},\ and\ \bibinfo {author} {\bibfnamefont {J.-C.}\ \bibnamefont
  {Robinet}},\ }\bibfield  {title} {\bibinfo {title} {Nonlinear optimals in the
  asymptotic suction boundary layer: {T}ransition thresholds and symmetry
  breaking},\ }\href {https://doi.org/10.1063/1.4916017} {\bibfield  {journal}
  {\bibinfo  {journal} {Phys.\ Fluids}\ }\textbf {\bibinfo {volume} {27}},\
  \bibinfo {pages} {4108} (\bibinfo {year} {2015})}\BibitemShut {NoStop}%
\bibitem [{\citenamefont {Budanur}\ \emph {et~al.}(2020)\citenamefont
  {Budanur}, \citenamefont {Marensi}, \citenamefont {Willis},\ and\
  \citenamefont {Hof}}]{Budanur2020upper}%
  \BibitemOpen
  \bibfield  {author} {\bibinfo {author} {\bibfnamefont {N.~B.}\ \bibnamefont
  {Budanur}}, \bibinfo {author} {\bibfnamefont {E.}~\bibnamefont {Marensi}},
  \bibinfo {author} {\bibfnamefont {A.~P.}\ \bibnamefont {Willis}},\ and\
  \bibinfo {author} {\bibfnamefont {B.}~\bibnamefont {Hof}},\ }\bibfield
  {title} {\bibinfo {title} {Upper edge of chaos and the energetics of
  transition in pipe flow},\ }\href
  {https://doi.org/10.1103/PhysRevFluids.5.023903} {\bibfield  {journal}
  {\bibinfo  {journal} {Phys.\ Rev.\ Fluids}\ }\textbf {\bibinfo {volume}
  {5}},\ \bibinfo {pages} {023903} (\bibinfo {year} {2020})}\BibitemShut
  {NoStop}%
\bibitem [{\citenamefont {Schmid}\ and\ \citenamefont
  {Henningson}(2001)}]{Schmid2001stability}%
  \BibitemOpen
  \bibfield  {author} {\bibinfo {author} {\bibfnamefont {P.~J.}\ \bibnamefont
  {Schmid}}\ and\ \bibinfo {author} {\bibfnamefont {D.~S.}\ \bibnamefont
  {Henningson}},\ }\href@noop {} {\emph {\bibinfo {title} {Stability and
  Transition in Shear Flows}}}\ (\bibinfo  {publisher} {Springer-Verlag New
  York},\ \bibinfo {year} {2001})\BibitemShut {NoStop}%
\bibitem [{\citenamefont {Butler}\ and\ \citenamefont
  {Farrell}(1992)}]{Butler1992optimal}%
  \BibitemOpen
  \bibfield  {author} {\bibinfo {author} {\bibfnamefont {K.~M.}\ \bibnamefont
  {Butler}}\ and\ \bibinfo {author} {\bibfnamefont {B.~F.}\ \bibnamefont
  {Farrell}},\ }\bibfield  {title} {\bibinfo {title} {Three-dimensional optimal
  perturbations in viscous shear flow},\ }\href
  {https://doi.org/10.1063/1.858386} {\bibfield  {journal} {\bibinfo  {journal}
  {Phys.\ Fluids A}\ }\textbf {\bibinfo {volume} {4}},\ \bibinfo {pages} {1637}
  (\bibinfo {year} {1992})}\BibitemShut {NoStop}%
\bibitem [{\citenamefont {Hussam}\ \emph
  {et~al.}(2012{\natexlab{b}})\citenamefont {Hussam}, \citenamefont
  {Thompson},\ and\ \citenamefont {Sheard}}]{Hussam2012optimal}%
  \BibitemOpen
  \bibfield  {author} {\bibinfo {author} {\bibfnamefont {W.~K.}\ \bibnamefont
  {Hussam}}, \bibinfo {author} {\bibfnamefont {M.~C.}\ \bibnamefont
  {Thompson}},\ and\ \bibinfo {author} {\bibfnamefont {G.~J.}\ \bibnamefont
  {Sheard}},\ }\bibfield  {title} {\bibinfo {title} {Optimal transient
  disturbances behind a circular cylinder in a quasi-two-dimensional
  magnetohydodynamic duct flow},\ }\href {https://doi.org/10.1063/1.3686809}
  {\bibfield  {journal} {\bibinfo  {journal} {Phys.\ Fluids}\ }\textbf
  {\bibinfo {volume} {24}},\ \bibinfo {pages} {024105} (\bibinfo {year}
  {2012}{\natexlab{b}})}\BibitemShut {NoStop}%
\bibitem [{\citenamefont {Sheard}\ \emph {et~al.}(2009)\citenamefont {Sheard},
  \citenamefont {Fitzgerald},\ and\ \citenamefont
  {Ryan}}]{Sheard2009cylinders}%
  \BibitemOpen
  \bibfield  {author} {\bibinfo {author} {\bibfnamefont {G.~J.}\ \bibnamefont
  {Sheard}}, \bibinfo {author} {\bibfnamefont {M.~J.}\ \bibnamefont
  {Fitzgerald}},\ and\ \bibinfo {author} {\bibfnamefont {K.}~\bibnamefont
  {Ryan}},\ }\bibfield  {title} {\bibinfo {title} {Cylinders with square
  cross-section: wake instabilities with incidence angle variation},\ }\href
  {https://doi.org/10.1017/S0022112009006879} {\bibfield  {journal} {\bibinfo
  {journal} {J.\ Fluid Mech.}\ }\textbf {\bibinfo {volume} {630}},\ \bibinfo
  {pages} {43} (\bibinfo {year} {2009})}\BibitemShut {NoStop}%
\bibitem [{\citenamefont {Karniadakis}\ \emph {et~al.}(1991)\citenamefont
  {Karniadakis}, \citenamefont {Israeli},\ and\ \citenamefont
  {Orszag}}]{Karniadakis1991high}%
  \BibitemOpen
  \bibfield  {author} {\bibinfo {author} {\bibfnamefont {G.~E.}\ \bibnamefont
  {Karniadakis}}, \bibinfo {author} {\bibfnamefont {M.}~\bibnamefont
  {Israeli}},\ and\ \bibinfo {author} {\bibfnamefont {S.~A.}\ \bibnamefont
  {Orszag}},\ }\bibfield  {title} {\bibinfo {title} {High-order splitting
  methods for the incompressible {N}avier-{S}tokes equations},\ }\href
  {https://doi.org/10.1016/0021-9991(91)90007-8} {\bibfield  {journal}
  {\bibinfo  {journal} {J.\ Comput.\ Phys.}\ }\textbf {\bibinfo {volume}
  {97}},\ \bibinfo {pages} {414} (\bibinfo {year} {1991})}\BibitemShut
  {NoStop}%
\bibitem [{\citenamefont {Reddy}\ \emph {et~al.}(1993)\citenamefont {Reddy},
  \citenamefont {Schmidt},\ and\ \citenamefont
  {Henningson}}]{Reddy1993pseudospectra}%
  \BibitemOpen
  \bibfield  {author} {\bibinfo {author} {\bibfnamefont {S.~C.}\ \bibnamefont
  {Reddy}}, \bibinfo {author} {\bibfnamefont {P.~J.}\ \bibnamefont {Schmidt}},\
  and\ \bibinfo {author} {\bibfnamefont {D.~S.}\ \bibnamefont {Henningson}},\
  }\bibfield  {title} {\bibinfo {title} {Pseudospectra of the
  {O}rr--{S}ommerfeld operator},\ }\href {https://doi.org/10.1137/0153002}
  {\bibfield  {journal} {\bibinfo  {journal} {SIAM J.\ Appl.\ Math.}\ }\textbf
  {\bibinfo {volume} {53}},\ \bibinfo {pages} {15} (\bibinfo {year}
  {1993})}\BibitemShut {NoStop}%
\bibitem [{\citenamefont {Trefethen}\ \emph {et~al.}(1993)\citenamefont
  {Trefethen}, \citenamefont {Trefethen}, \citenamefont {Reddy},\ and\
  \citenamefont {Driscoll}}]{Trefethen1993hydrodynamic}%
  \BibitemOpen
  \bibfield  {author} {\bibinfo {author} {\bibfnamefont {L.~N.}\ \bibnamefont
  {Trefethen}}, \bibinfo {author} {\bibfnamefont {A.~E.}\ \bibnamefont
  {Trefethen}}, \bibinfo {author} {\bibfnamefont {S.~C.}\ \bibnamefont
  {Reddy}},\ and\ \bibinfo {author} {\bibfnamefont {T.~A.}\ \bibnamefont
  {Driscoll}},\ }\bibfield  {title} {\bibinfo {title} {Hydrodynamic stability
  without eigenvalues},\ }\href {https://doi.org/10.1063/1.4791605} {\bibfield
  {journal} {\bibinfo  {journal} {Science}\ }\textbf {\bibinfo {volume}
  {261}},\ \bibinfo {pages} {578} (\bibinfo {year} {1993})}\BibitemShut
  {NoStop}%
\bibitem [{\citenamefont {Barkley}\ \emph {et~al.}(2008)\citenamefont
  {Barkley}, \citenamefont {Blackburn},\ and\ \citenamefont
  {Sherwin}}]{Barkley2008direct}%
  \BibitemOpen
  \bibfield  {author} {\bibinfo {author} {\bibfnamefont {D.}~\bibnamefont
  {Barkley}}, \bibinfo {author} {\bibfnamefont {H.~M.}\ \bibnamefont
  {Blackburn}},\ and\ \bibinfo {author} {\bibfnamefont {S.~J.}\ \bibnamefont
  {Sherwin}},\ }\bibfield  {title} {\bibinfo {title} {Direct optimal growth
  analysis for timesteppers},\ }\href {https://doi.org/10.1002/fld.1824}
  {\bibfield  {journal} {\bibinfo  {journal} {Int.\ J.\ Numer.\ Methods
  Fluids}\ }\textbf {\bibinfo {volume} {57}},\ \bibinfo {pages} {1435}
  (\bibinfo {year} {2008})}\BibitemShut {NoStop}%
\bibitem [{\citenamefont {Blackburn}\ \emph {et~al.}(2008)\citenamefont
  {Blackburn}, \citenamefont {Barkley},\ and\ \citenamefont
  {Sherwin}}]{Blackburn2008convective}%
  \BibitemOpen
  \bibfield  {author} {\bibinfo {author} {\bibfnamefont {H.~M.}\ \bibnamefont
  {Blackburn}}, \bibinfo {author} {\bibfnamefont {D.}~\bibnamefont {Barkley}},\
  and\ \bibinfo {author} {\bibfnamefont {S.~J.}\ \bibnamefont {Sherwin}},\
  }\bibfield  {title} {\bibinfo {title} {Convective instability and transient
  growth in flow over a backward-facing step},\ }\href
  {https://doi.org/10.1017/S0022112008001109} {\bibfield  {journal} {\bibinfo
  {journal} {J.\ Fluid Mech.}\ }\textbf {\bibinfo {volume} {603}},\ \bibinfo
  {pages} {271} (\bibinfo {year} {2008})}\BibitemShut {NoStop}%
\bibitem [{\citenamefont {Pringle}\ \emph {et~al.}(2015)\citenamefont
  {Pringle}, \citenamefont {Willis},\ and\ \citenamefont
  {Kerswell}}]{Pringle2015fully}%
  \BibitemOpen
  \bibfield  {author} {\bibinfo {author} {\bibfnamefont {C.~C.~T.}\
  \bibnamefont {Pringle}}, \bibinfo {author} {\bibfnamefont {A.~P.}\
  \bibnamefont {Willis}},\ and\ \bibinfo {author} {\bibfnamefont {R.~R.}\
  \bibnamefont {Kerswell}},\ }\bibfield  {title} {\bibinfo {title} {Fully
  localised nonlinear energy growth optimals in pipe flow},\ }\href
  {https://doi.org/10.1063/1.4922183} {\bibfield  {journal} {\bibinfo
  {journal} {Phys.\ Fluids}\ }\textbf {\bibinfo {volume} {27}},\ \bibinfo
  {pages} {064102} (\bibinfo {year} {2015})}\BibitemShut {NoStop}%
\bibitem [{\citenamefont {Pringle}\ and\ \citenamefont
  {Kerswell}(2010)}]{Pringle2010using}%
  \BibitemOpen
  \bibfield  {author} {\bibinfo {author} {\bibfnamefont {C.~C.~T.}\
  \bibnamefont {Pringle}}\ and\ \bibinfo {author} {\bibfnamefont {R.~R.}\
  \bibnamefont {Kerswell}},\ }\bibfield  {title} {\bibinfo {title} {Using
  nonlinear transient growth to construct the minimal seed for shear flow
  turbulence},\ }\href {https://doi.org/10.1103/PhysRevLett.105.154502}
  {\bibfield  {journal} {\bibinfo  {journal} {Phys.\ Rev.\ Lett.}\ }\textbf
  {\bibinfo {volume} {105}},\ \bibinfo {pages} {154502} (\bibinfo {year}
  {2010})}\BibitemShut {NoStop}%
\bibitem [{Sup()}]{Supvideos2020}%
  \BibitemOpen
  \href@noop {} {\bibinfo {title} {See supplemental material at}},\ \bibinfo
  {howpublished} {URL},\ \bibinfo {note} {for comparisons of linear and
  nonlinear evolution; short and long time histories}\BibitemShut {NoStop}%
\bibitem [{\citenamefont {de~les Valls}\ \emph {et~al.}(2011)\citenamefont
  {de~les Valls}, \citenamefont {Batet}, \citenamefont {de~Medina},
  \citenamefont {Fradera},\ and\ \citenamefont
  {Sedano}}]{Mas2011qualification}%
  \BibitemOpen
  \bibfield  {author} {\bibinfo {author} {\bibfnamefont {E.~M.}\ \bibnamefont
  {de~les Valls}}, \bibinfo {author} {\bibfnamefont {L.}~\bibnamefont {Batet}},
  \bibinfo {author} {\bibfnamefont {V.}~\bibnamefont {de~Medina}}, \bibinfo
  {author} {\bibfnamefont {J.}~\bibnamefont {Fradera}},\ and\ \bibinfo {author}
  {\bibfnamefont {L.~A.}\ \bibnamefont {Sedano}},\ }\bibfield  {title}
  {\bibinfo {title} {Qualification of {MHD} effects in dual-coolant {DEMO}
  blanket and approaches to their modelling},\ }\href
  {https://doi.org/10.1016/j.fusengdes.2011.03.071} {\bibfield  {journal}
  {\bibinfo  {journal} {Fusion Eng.\ Des.}\ }\textbf {\bibinfo {volume} {86}},\
  \bibinfo {pages} {2326} (\bibinfo {year} {2011})}\BibitemShut {NoStop}%
\bibitem [{\citenamefont {Vetcha}(2012)}]{Vetcha2012study}%
  \BibitemOpen
  \bibfield  {author} {\bibinfo {author} {\bibfnamefont {N.}~\bibnamefont
  {Vetcha}},\ }\emph {\bibinfo {title} {Study of instability and transition in
  {MHD} flows as applied to liquid metal blankets}},\ \href@noop {} {\bibinfo
  {type} {Doctor of {P}hilosophy}},\ \bibinfo  {school} {University of
  California}, \bibinfo {address} {Los Angeles} (\bibinfo {year}
  {2012})\BibitemShut {NoStop}%
\bibitem [{\citenamefont {B{\"u}hler}(1995)}]{Buhler1995influence}%
  \BibitemOpen
  \bibfield  {author} {\bibinfo {author} {\bibfnamefont {L.}~\bibnamefont
  {B{\"u}hler}},\ }\bibfield  {title} {\bibinfo {title} {The influence of small
  cracks in insulating coatings on the flow structure and pressure drop in
  {MHD} channel flows},\ }\href {https://doi.org/10.1016/0920-3796(95)90180-9}
  {\bibfield  {journal} {\bibinfo  {journal} {Fusion Eng.\ Des.}\ }\textbf
  {\bibinfo {volume} {27}},\ \bibinfo {pages} {650} (\bibinfo {year}
  {1995})}\BibitemShut {NoStop}%
\bibitem [{\citenamefont {B{\"u}hler}(1996)}]{Buhler1996instabilities}%
  \BibitemOpen
  \bibfield  {author} {\bibinfo {author} {\bibfnamefont {L.}~\bibnamefont
  {B{\"u}hler}},\ }\bibfield  {title} {\bibinfo {title} {Instabilities in
  quasi-two-dimensional magnetohydrodynamic flows},\ }\href
  {https://doi.org/10.1017/S0022112096008269} {\bibfield  {journal} {\bibinfo
  {journal} {J.\ Fluid Mech.}\ }\textbf {\bibinfo {volume} {326}},\ \bibinfo
  {pages} {125} (\bibinfo {year} {1996})}\BibitemShut {NoStop}%
\bibitem [{\citenamefont {B{\"u}hler}(1998)}]{Buhler1998laminar}%
  \BibitemOpen
  \bibfield  {author} {\bibinfo {author} {\bibfnamefont {L.}~\bibnamefont
  {B{\"u}hler}},\ }\bibfield  {title} {\bibinfo {title} {Laminar buoyant
  magnetohydrodynamic flow in vertical rectangular ducts},\ }\href
  {https://doi.org/10.1063/1.869562} {\bibfield  {journal} {\bibinfo  {journal}
  {Phys.\ Fluids}\ }\textbf {\bibinfo {volume} {10}},\ \bibinfo {pages} {223}
  (\bibinfo {year} {1998})}\BibitemShut {NoStop}%
\bibitem [{\citenamefont {Smolentsev}(2009)}]{Smolentsev2009duct}%
  \BibitemOpen
  \bibfield  {author} {\bibinfo {author} {\bibfnamefont {S.}~\bibnamefont
  {Smolentsev}},\ }\bibfield  {title} {\bibinfo {title} {{{MHD} duct flows
  under hydrodynamic ``slip'' condition}},\ }\href
  {https://doi.org/10.1007/s00162-009-0108-7} {\bibfield  {journal} {\bibinfo
  {journal} {Theor.\ Comput.\ Fluid Dyn.}\ }\textbf {\bibinfo {volume} {23}},\
  \bibinfo {pages} {557} (\bibinfo {year} {2009})}\BibitemShut {NoStop}%
\end{thebibliography}

%


\end{document}